\documentclass[a4paper,11pt]{article}

\pdfoutput=1
\usepackage{jheppub}

\usepackage{graphicx,color,amsmath}

\newcommand{\yr}{{\, {\rm yr}}}

\newcommand{\kpc}{{\, {\rm kpc}}}

\newcommand{\cm}{{\, {\rm cm}}}

\newcommand{\eV}{{\, {\rm eV}}}

\newcommand{\MeV}{{\, {\rm MeV}}}
\newcommand{\GeV}{{\, {\rm GeV}}}

\newcommand{\DM}{{\rm DM}}

\newcommand{\OO}{{\mathcal O}}
\newcommand{\LL}{{\mathcal L}}

\newcommand{\kms}{{\, {\rm km \, s^{-1}}}}

\begin{document}

\title{Long range dark matter self-interactions and plasma instabilities}

\author{Robert Lasenby}
\emailAdd{rlasenby@stanford.edu}
\affiliation{Stanford Institute for Theoretical Physics, Stanford University, Stanford, CA 94305, USA}

\date{\today}

\abstract{
	So far, the observed effects of dark matter
	are compatible with it having purely gravitational
	interactions. However, in many models, dark matter
	has additional interactions with itself, with the
	Standard Model, and/or with additional hidden sector states.
	In this paper, we discuss models in which dark matter 
	interacts through a light vector mediator, giving rise
	to a long-ranged force between dark matter particles.
	Coherent scattering effects through this force can lead
	to the exponential growth of small perturbations,
	in analogy to electromagnetic plasma instabilities.
	These instabilities can be significant at couplings
	many orders of magnitude below those for which 
	the usual particle-by-particle constraints on dark matter
	self-interactions apply. While this possibility has
	been noted in the literature, we provide the first
	systematic study of such instabilities, including
	the case where the mediator has finite mass. The latter
	is relevant for models of kinetically mixed `dark photon'
	mediators, which represent an important target for proposed
	dark matter detection experiments.
	Our analyses are of the growth of small perturbations,
	so do not immediately provide observational constraints
	on dark matter models --- however, they do 
	motivate further study of large regions of parameter space.
	}

\maketitle


\section{Introduction}

There is overwhelming evidence that most of
the universe's matter density consists
of states other than Standard Model (SM) particles.
All of the evidence for this `dark matter' (DM)
comes from its gravitational effects on astrophysical
objects, and current observations are compatible
with gravity being the only way in which DM
interacts, either with itself or with the SM.
Nevertheless, in many models of theoretical and
experimental interest, DM must have additional interactions.
Examples include WIMP DM, whose interactions with
the SM radiation bath are responsible for
its cosmological abundance, 
and the QCD axion, which has a self-interaction
potential arising from its coupling to QCD. 
In all of these cases, DM interactions must be weak enough
that its behaviour looks effectively non-collisional, on 
the galactic scales that we have observational data for.

In this paper, we will focus mainly on DM
self-interactions.
For models such as WIMP DM, the DM number density is low
(the occupation number of particle modes in halos is $\ll 1$),
and the interactions between DM particles are
short-ranged compared to the interparticle distance.
Consequently, scattering is dominated by 
two-particle 
events. The standard DM self-interaction bounds
from systems such as the Bullet Cluster~\cite{Clowe_2006} are
based on this scenario. For DM masses
in the GeV range, these constrain the interaction
cross-section at around the barn level~\cite{Robertson_2016}, comparable
to SM nuclear cross-sections. 
The usual WIMP dark matter candidates, which have significantly heavier
masses and weak-scale interactions, are safe from such constraints.

In other models, DM self-interactions
can be long-ranged. For the simplest examples,
DM particles interact with a light mediator, whose
mass is small compared to the inverse
spacing between DM particles. Quantitatively,
\begin{align}
	n_\DM = \frac{\rho_\DM}{m_\DM} 
	&\simeq (4 \times 10^{-4} \eV)^{3} \, \frac{\MeV}{m_\DM}  \frac{\rho_\DM}{10 \GeV \cm^{-3}} \\ 
	&\simeq (0.5 {\rm \, mm})^{-3} \,  \frac{\MeV}{m_\DM}  \frac{\rho_\DM}{10 \GeV \cm^{-3}} 
\end{align}
where $10 \GeV \cm^{-3}$ corresponds to the highest
DM halo density for which we have good evidence 
(unless otherwise noted, we use natural units with
$\hbar = c = 1$).
If the mass $m$ of the mediating
particle satisfies $m \ll n_{\rm DM}^{1/3}$, 
then DM particles can interact coherently with many others.
These interactions
are coherently enhanced, and can have much larger
rates than standard $2 \rightarrow 2$ scatterings.
Of course, they are also softer (transferring less
energy and momentum), and their behaviour may
be much more complicated.
This scenario can be constrasted to
the `long-range' mediators that are used
to implement velocity-dependent
DM scattering (e.g.\ \cite{Kahlhoefer_2013}) --- in that case, we only
need the mediator mass to be $\lesssim q$, where $q$ is the typical momentum transfer
in a $2 \rightarrow 2$ collision.
For example, in a typical galactic halo, $q \sim 10^{-3} m_{\rm DM}$.

There are various theoretical motivations for
long-range hidden sector mediators.
Examples include hidden sector counterparts of the 
photon (e.g.\ in scenarios such as Twin Higgs
models~\cite{Chacko_2006,Garc_a_Garc_a_2015}),
light scalars~\cite{Damour_1990,Gradwohl_1992,Nusser_2005,Kesden_2006,Hellwing_2009},
or other forms of `modified gravity' (e.g.\ \cite{Comelli_2012}).
Such models are also important as an experimental 
target for low-threshold direct detection experiments,
as we discuss in Section~\ref{secap}.

The effects of coherent, many-many scattering processes
will depend on the form of the DM-mediator interaction.
For a spin-0 mediator $\phi$, the simplest kinds of
interactions (e.g.\ $\phi \bar \chi \chi$, for a DM
fermion $\chi$) will be universally attractive
for a given particle species, i.e.\ $\chi$ particles
and antiparticles all interact attractively.
This is basically
the scenario of `modified gravity' in the DM sector,
a number of models for which have been explored in the
literature (e.g.\ \cite{Damour_1990,Gradwohl_1992,Nusser_2005,Kesden_2006,Hellwing_2009}).

For a spin-1 mediator $A_\mu$, the simplest forms
of interactions, e.g. $A_\mu \bar \chi \gamma^\mu \chi$, 
are analogous to electromagnetism.\footnote{Higher-dimensional couplings, or couplings
to a non-conserved current, result in a non-renormalisable
theory that must be completed above some energy scale.}
If the cosmological DM abundance is net neutral,
with equal numbers of positive and negative charges, 
then (in the absence of large-scale charge separation)
there will not be coherent forces between bulk matter,
unlike in `modified gravity' scenarios. In cosmological
terms, inflation naturally sets any charge asymmetry to zero,
and any processes regenerating it later must break
the associated $U(1)$.
However, even for DM that is bulk neutral,
there can still be
significant `plasma effects', driven by coherent
many-many scatterings.
Analogous coherent EM effects are extremely important 
in SM astrophysics, driving processes such as
shocks, jets and other collective phenomena~\cite{Chiuderi_2015}.

This paper will investigate scenarios
in which DM interacts through a light vector mediator.
As we will see, coherent effects can be important even
with very small DM-mediator couplings; in some cases,
many orders of magnitude smaller than those constrained
by $2 \rightarrow 2$ scattering processes.
This raises the possibility of much stronger
constraints on such DM models, or equivalently, observational
signatures at much weaker couplings.
While the possibility of these effects has been noted
in the literature~\cite{Ackerman:mha},
we provide a systematic analysis.
In addition to considering massless vectors, we also
consider mediators with a small but finite mass,
as well as those with couplings to SM matter
as well as to DM.
These more complicated models are important 
phenomenological targets, in particular for
proposed experiments aiming to
detect light DM candidates.


\subsection{Parametrics of plasma instabilities}

Despite the simple underlying physics, plasma dynamics
--- the behaviour of charged matter interacting via a
light vector mediator --- is an immensely complicated 
subject. The great variety of different behaviours
of SM matter in astrophysical systems,
much of which is driven by the combination of EM
and gravity, illustrates this richness~\cite{Chiuderi_2015}.
Accordingly, this paper will not attempt to categorise
all of the different kinds of behaviour that 
could occur in the presence of long-range DM
interactions. Instead, we will focus
on identifying the smallest couplings for 
which coherent effects would be important
in astrophysical settings.
This should help guide future investigations
of such models, and identify regions of parameter
space where we might expect deviations
from collisionless behaviour.

In many circumstances, the first effects
to become important, as we increase the coupling
from zero, are exponentially-growing instabilities.
For a non-relativistic plasma, the fastest-growing of these
instabilities are `electrostatic' ones, in which
small charge separation perturbations grow,
by extracting kinetic energy from a non-Maxwellian
velocity distribution~\cite{Chen_2015}.
The simplest case is a `two-stream' instability,
in which two uniform `streams' of plasma pass
through each other at high velocity (relative to their
individual velocity dispersions).
Perturbations, in the form of longitudinal charge-density
waves, are exponentially amplified,
with growth rate $\sim \omega_p$, where
$\omega_p^2 = q^2 n / m_q$ is the `plasma frequency',
$q$ the charge of individual particles, $m_q$
their mass, and
$n$ their number density~\cite{Briggs_71}.

Taking our `plasma' to be composed of dark matter particles,
interacting through a hidden-sector vector mediator,
this two-stream scenario approximates
a collision between the DM halos of two galaxy
clusters,
such as the Bullet Cluster~\cite{Clowe_2006,Paraficz_2016}.
In that system, the central densities
of the colliding clusters' DM halos
are modelled to be $\gtrsim 0.1 \GeV \cm^{-3}$
(see Section~\ref{secastro}).
This DM density gives a plasma frequency of
\begin{equation}
	\omega_p = \sqrt{\frac{g^2 \rho_\chi}{m_\chi^2}}
	\simeq 4 \times 10^{-7} \yr^{-1} \sqrt{\frac{\rho_\chi}{0.1 \GeV \cm^{-3}}} \frac{\GeV}{m_\chi} \frac{g}{10^{-17}}
	\label{eqwp1}
\end{equation}
where $m_\chi$ is the DM mass, $\rho_\chi$ is the DM
density, and $g$ is the DM-mediator coupling.
The scale radius of the DM halos
is $\sim 100 \kpc$, and their relative velocity is 
$\sim 3000 \kms$, giving a crossing
time of $\sim 100 \kpc / (3000 \kms) \sim 3 \times 10^7 \yr$.
Comparing $\omega_p^{-1}$ to this crossing time,
we see that the exponential growth of perturbations
could occur even for $g \lll 1$.
For comparison, constraints from $2 \rightarrow 2$ 
particle scattering give
$g \lesssim 4 \times 10^{-3} (m_\chi / \GeV)^{3/4}$
(see Section~\ref{secexisting}).

This estimate shows that coherent effects may become
important at
DM-mediator couplings orders of magnitude smaller than those
at which $2 \rightarrow 2$ scatterings have significant effects.
A natural question is what the observational
consequences of coherent effects might be,
and whether they can give observational constraints
or potential signatures
for such models. As discussed at the start of this section,
plasma behaviour can be extremely complicated, and we will
not try to answer this question in generality. However,
in Section~\ref{secobscons}, we discuss some
of the first observational signatures that may arise
as we increase the DM-mediator coupling from zero. 
For cluster collisions, such as the Bullet Cluster,
we might expect DM-DM momentum transfer to alter the
post-collision DM density profile, leaving it more similar
to that of the SM gas
than the (effectively collisionless) stars. For DM subhalos moving within larger halos (e.g.\
dwarf galaxies in the Milky Way halo), potential effects
include a drag force on the subhalo, modifying its orbit,
or heating, resulting in its expansion / evaporation.
Developing a proper understanding of these potential
signatures is left to future work.

In the remainder of this paper, we analyse the
growth of small perturbations more generally and quantitatively.
In particular:

\begin{itemize}
	\item  We review the linear theory of instabilities
		in a homogeneous plasma (Section~\ref{secinst}). Standard 
		electromagnetic
		theory can be applied directly to the case
		of DM interacting via a massless hidden-sector
		mediator, simply by modifying the charges
		and masses of the particles.
	\item We extend this theory to the case 
		of a massive vector mediator (Section~\ref{secinst}). 
		We find
		that mediator masses
		$m \gtrsim \omega_p / v_{\rm th}$
		suppress the growth of perturbations,
		where $v_{\rm th}$ is the velocity dispersion of the plasma.
		For velocity distributions without multiple
		peaks (such as an anisotropic Maxwellian
		distribution), only magnetic instabilities exist,
		and these are suppressed for $m \gtrsim \OO(\omega_p)$.
	\item We identify regions in the $m_\chi, m_A, g_\chi$ parameter
		space (where $m_\chi$ is the DM mass,
		$m_A$ is the mediator mass, and $g_\chi$ is the
		DM-mediator coupling)
		where we expect small perturbations to grow
		in astrophysical systems (Section~\ref{secastro}).
		At small $m_A$, the potential constraints on
		$g_\chi$ are many orders of magnitude stronger
		than those from 
		$2 \rightarrow 2$ scattering (this is illustrated
		in Figures~\ref{figmg} and~\ref{figmag1}).
		Observational signatures that may lead to such
		constraints are discussed in Section~\ref{secobscons}.
		We also comment on the potential for improved constraints
		when the DM is so light ($m \lesssim \eV$) that it has occupation number
		$\gg 1$ in halos (Section~\ref{seclightbosonic}).
	\item We study models in which DM interacts
		non-gravitationally with both itself
		and with SM matter, through a `dark photon' vector
		mediator that kinetically mixes with the SM photon
		(Section~\ref{secap}).
		This is one of the simplest such possibilities,
		and represents an important target
		for low-threshold direct detection experiments.
		In large parts of the theoretically and experimentally
		interesting parameter space,
		we find that DM-DM plasma instabilities
		would be expected to grow exponentially in astrophysical systems
		(Figure~\ref{figqeff}). This motivates
		further investigations to determine whether such models
		are astrophysically viable.
		We also discuss why such effects may not be as 
		important for the case of millicharged DM
		(Section~\ref{secmillicharged}).
\end{itemize}


\subsection{Previous literature}
\label{secprev}

To our knowledge, the first paper to raise the possibility
of strong constraints from coherent DM-DM scattering
was~\cite{Ackerman:mha}. This considered DM interacting
through a massless, purely dark-sector vector mediator,
and focussed mainly on the constraints and signatures
arising from $2 \rightarrow 2$ scattering.
However, it also commented on the possibility of
a magnetic Weibel instability (see section~\ref{secmagnetic}),
and noted that such phenomena may provide significantly
stronger constraints than $2 \rightarrow 2$ scattering,
while leaving further investigations to future work.

In unpublished work, \cite{Mardon} identified
the electrostatic two-stream instability as being faster-growing
than the Weibel instability (again, for the case
of a massless hidden-sector mediator), and estimated
the parameter space where constraints from Bullet Cluster
type systems might exist. In addition, they conducted preliminary
simulations of such cluster collision scenarios.

\cite{Heikinheimo:2015kra} also identified
the electrostatic two-stream instability as the fastest-growing
one, and argued that since this instability would
make DM effectively collisional, `models where all of DM is charged are ruled out
by observations of cluster collisions unless the charge is
extremely small' (though they did not attempt
to quantify this).
Their main focus was a scenario in only a sub-component
of DM is self-interacting through a massless
hidden-sector mediator. In \cite{Sepp:2016tfs},
cluster collisions with such a subcomponent were simulated,
assuming that it behaves like a fluid with some effective
viscosity.

Coherent interactions involving DM have been investigated
extensively in the context of millicharged DM,
where DM is taken to have a (very small) charge
under Standard Model EM.
As discussed in Section~\ref{secmillicharged},
the most important interactions are generally `one-many', with
individual DM particles scattering off the large-scale
EM fields created by SM charges and currents.
While some of this literature (e.g.\ \cite{Stebbins:2019xjr})
claims that similar bounds apply to models
with massive mediators, such as kinetically mixed
dark photons, this is only true when the mediator
mass is extremely small
(see Section~\ref{secap}).
There have also been investigations of whether
`many-many' collective instabilities can be important
for millicharged DM~\cite{Li_2020}, but futher work
would be needed to establish whether this could occur
for couplings not excluded by other effects.

DM coupled to a light vector mediator is considered in
many other papers (e.g.\ \cite{Dai_2009,Feng_2009,Agrawal_2017,Garny_2019}), but in most cases,
they do not consider plasma effects (or if they do, only
insofar as they affect Coulomb scattering through Debye screening
--- see Section~\ref{secexisting}).
\cite{Clarke_2016} does consider plasma effects in a model with
a massive hidden-sector mediator, but focusses
on the properties of the DM plasma near Earth (in the context of 
direct detection experiments), rather than on systems
such as cluster collisions.


\section{Plasma instabilities}
\label{secinst}

In this section, we will discuss the linear theory of plasma
instabilities.
Astrophysically, the DM distributions that arise
from collisionless evolution will often be very
far from thermal equilibrium (for example, the
basically-bimodal velocity distributions in cluster
collisions). Given strong enough interactions,
such a distribution will relax towards equilibrium.\footnote{
	When gravitational interactions are important,
	there is generally no equilibrium state for a particle
	system; entropy can always be increased by
	a subset of the particles becoming
	more tightly bound~\cite{BT,Lynden_Bell_1968}.
	However, as we will discuss, the plasma instabilities
	that we are interested in will generally
	be important on scales well below those
	where gravity is important.}
For some distributions, the initial stages
of this relaxation are driven by 
 exponentially-growing
instabilities,
which grow by converting particle kinetic
energy into field energy.

\subsection{Instabilities in a uniform plasma}
\label{secuniform}

We will begin by considering perturbations of 
a spatially uniform plasma, in a freely falling frame,
with no large-scale currents or charge separation in 
the initial state.
In real astrophysical systems, spatial 
inhomogeneities and gravitational forces 
are, of course, often important. However, 
if the spatial scale of a perturbation is small,
these approximations can be good, and as we will see, the spatial scales associated
with instabilities are often small enough.
If the initial state does not feature large-scale
currents or charge separation, 
then (for large enough DM number densities that
particle number effects are not important), these must
have arisen through interactions becoming important
at earlier times. Consequently, to analyse the
earliest point at which large-scale vector fields
arise, we can start from approximately zero-field conditions.

In terms of the distribution functions
$f_s(v, x, t) = \frac{dN_s}{d^3v d^3 x}$
of the particle species $s$ in the plasma,
conservation of phase space density along particle
trajectories requires that
\begin{equation}
	\partial_t f_s + v \cdot \partial_x f_s 
	+ a_s \cdot \partial_v f_s = 0
	\label{eqf0}
\end{equation}
where $a_s = \frac{dv}{dt} = \frac{q_s}{m_s} (E + v \times B)$
is the acceleration due to the Lorentz force,
with $q_s$ and $m_s$ the charge and mass of species $s$.
The charge density, $\rho = \sum_s q_s \int d^3 v f_s$,
and the current density, $J = \sum_s q_s \int d^3 v v f_s$,
give the source term $J^\nu$ in the Proca equations
$(\partial_\mu \partial^\mu - m^2) A^\nu = J^\nu$
(where $m$ is the mass of the vector mediator, and we 
use the $+---$ signature).
Together, these equations determine the evolution of the plasma
distribution function.

For a particular configuration of charged particles,
the distribution function will be a set of delta-functions
corresponding to the positions and momenta of these particles,
and solving the abov equations will correspond to
solving the interacting $N$-body system. To simplify things,
we need to perform some kind of averaging.
One approach is to coarse-grain in position and
velocity space, so that there are many charged particles
in each phase-space `bin' --- another is to track the evolution of an
ensemble-averaged distribution function, where we average
over e.g.\ different possible starting configurations of
the particles~\cite{fitzpatrick2015plasma}. 
In either case, for the averaged solution
to be a good approximation to the particular solution,
we need the particles in the latter to be sufficiently dense in
phase space. We will return to this condition in
Section~\ref{secelectro}.

Since $E$ and $B$, and consequently
$a_s$, depend on the particle trajectories, 
the $a_s \cdot \partial_v f_s$ term in equation~\ref{eqf0}
cannot simply be replaced by $\bar a_s \cdot \partial_v \bar f$,
where the bar indicates averaging. In general,
correlations need to be taken into account. However,
the most important correlations generally arise
from close encounters between particles.
In situations where collective effects dominate, and individual
collisions
are unimportant, we can use the `collisionless Boltzmann
equation' (also referred to as the Vlasov equation)~\cite{fitzpatrick2015plasma},
\begin{equation}
	\partial_t \bar f_s + v \cdot \partial_x \bar f_s 
	+ \bar a \cdot \partial_v \bar f_s = 0
\end{equation}
(in the rest of the paper, we will usually drop the bars).
Since a typical DM particle can only undergo $\OO(1)$
hard collisions during the lifetime of the universe, without
affecting observables such as galactic halo shapes
(see Section~\ref{secexisting}), the collisionless Boltzmann
equation will often be a good approximation,
in parameter space which is not already constrained
by analyses of $2 \rightarrow 2$ collisions.

If our unperturbed state is a spatially uniform
distribution function $f_s$, with no
vector field background,
then the linear-order equations for a small perturbation
$\delta f_s, A_\mu$ are 
\begin{equation}
	(\partial_t + v \cdot \partial_x) \delta f_s = - a(A_\mu) \cdot
	\partial_v f \quad , \quad
	(\partial_\mu \partial^\mu - m^2) A^\nu = \sum_s q_s \int d^3v \, v^\nu 
	\delta f_s
\end{equation}
where $v^\nu = \gamma (1, v_i)$, with $\gamma = 1/\sqrt{1 - v^2}$.
We will work with non-relativistic velocity distributions,
so will take $\gamma \simeq 1$.
If we decompose the perturbations
into Fourier modes, so $\delta f_s(t,x)
= \delta f_s e^{- i(\omega t - k \cdot x)}$ etc,
then we can express $\delta f_s$ in terms of $A_\mu$,
giving a linear equation for $A_\mu$;
\begin{equation}
	(\omega^2 - k^2 - m^2) A^\nu = - i \sum_s 
	 q_s \int d^3 v \, v^\nu  \frac{a(A_\mu) \cdot \partial_v f}
	 {\omega - k \cdot v}
	 \equiv \Pi^{\nu \mu}(\omega,k) A_\mu
	 \label{eqvlasov1}
\end{equation}
where $\Pi^{\nu\mu}$ is the leading
order response function (the `self-energy', in the language
of thermal field theory~\cite{Le_Bellac_1996}).
For a given $k$, there will be some particular
$\omega$ for which this equation can be satisfied.
These $(k,w)$ pairs give the dispersion relation 
for perturbations.

To take the simplest possible example,
a cold, uniform plasma has $f_s = n_s \delta^3(v)$.
We can evaluate equation~\ref{eqvlasov1} via integration
by parts; if we assume that $f$ decays
fast enough at large velocities, then
\begin{equation}
	\int d^3 v \, v_i \frac{((E + v \times B) \cdot \partial_v f)}
	{\omega - k \cdot v}
	= - \int d^3 v \, \frac{f}{\omega - k \cdot v} \left(
	E + v \times B + \frac{k \cdot (E + v \times B)}{\omega - k \cdot v} v\right)_i
\end{equation}
So, for $f_s = n_s \delta^3 (v)$, we have
\begin{equation}
	(\omega^2 - k^2 - m^2) A_i = i \sum_s \frac{q_s^2 n_s}{m_s}
	\frac{E_i}{\omega}
\end{equation}
Since $E = - \nabla \phi - \partial_t A$, 
and the Lorenz condition $\partial_\mu A^\mu = 0$ gives
$\dot \phi + \nabla \cdot A = 0$, we
have 
\begin{equation}
	(\omega^2 - k^2 - m^2) A_i = 
	\left(\sum_s \omega_s^2\right) \left(A_i - \frac{k_i k \cdot A}{\omega^2}\right)
\end{equation}
where
$\omega_s \equiv q_s^2 n_s / m_s$.
For $m=0$,
this gives the usual dispersion relations for transverse
and longitudinal excitations
in a cold plasma~\cite{Raffelt_1996}.
Some more realistic examples,
such as Maxwellian velocity distributions, can also
be treated analytically (see appendix~\ref{appMaxwell}).

One of the simplest examples that gives rise to 
exponentially growing perturbations is the `two-stream'
instability. If we consider a plasma consisting of
positive and negatively charged species with the same
mass (as we expect for symmetric DM models),
this has $f_\pm = \frac{n_p}{2} ( \delta^3(v - v_0) + 
\delta^3 (v + v_0))$, 
corresponding to two cold
`streams' of plasma passing through each other.
For longitudinal perturbations, i.e.\ those 
with $k \parallel A \parallel v_0$, we have
\begin{equation}
	(\omega^2 - k^2 - m^2) A = (\omega^2 - k^2) 
	\left(\frac{\omega_p^2}{(\omega - k \cdot v_0)^2}
	+ \frac{\omega_p^2}{(\omega + k \cdot v_0)^2}\right)A
	\label{eqlongdisp1ts}
\end{equation}
where $\omega_p = \sqrt{q^2 n_p / m}$ is the plasma frequency of
each of the streams, with the charge
of the two species being $\pm q$ and their mass being $m$.
If we start with some localised perturbation
at an initial time, we can decompose it into spatial
Fourier modes with real $k$. Consequently, there
are exponentially growing perturbations if there
are solutions of the dispersion relation
with ${\rm Im} \, \omega > 0$ for real $k$.
As we discuss in Section~\ref{secelectro}, such solutions
exist for $m \le \omega_p/v_0$, and the fastest
growth rate is $\omega_i \simeq 0.5 \omega_s$
(attained for $k \simeq \omega_s / v_0$, and 
$m \ll k$).

\begin{figure}[t]
	\begin{center}
\includegraphics[width = 0.25\textwidth]{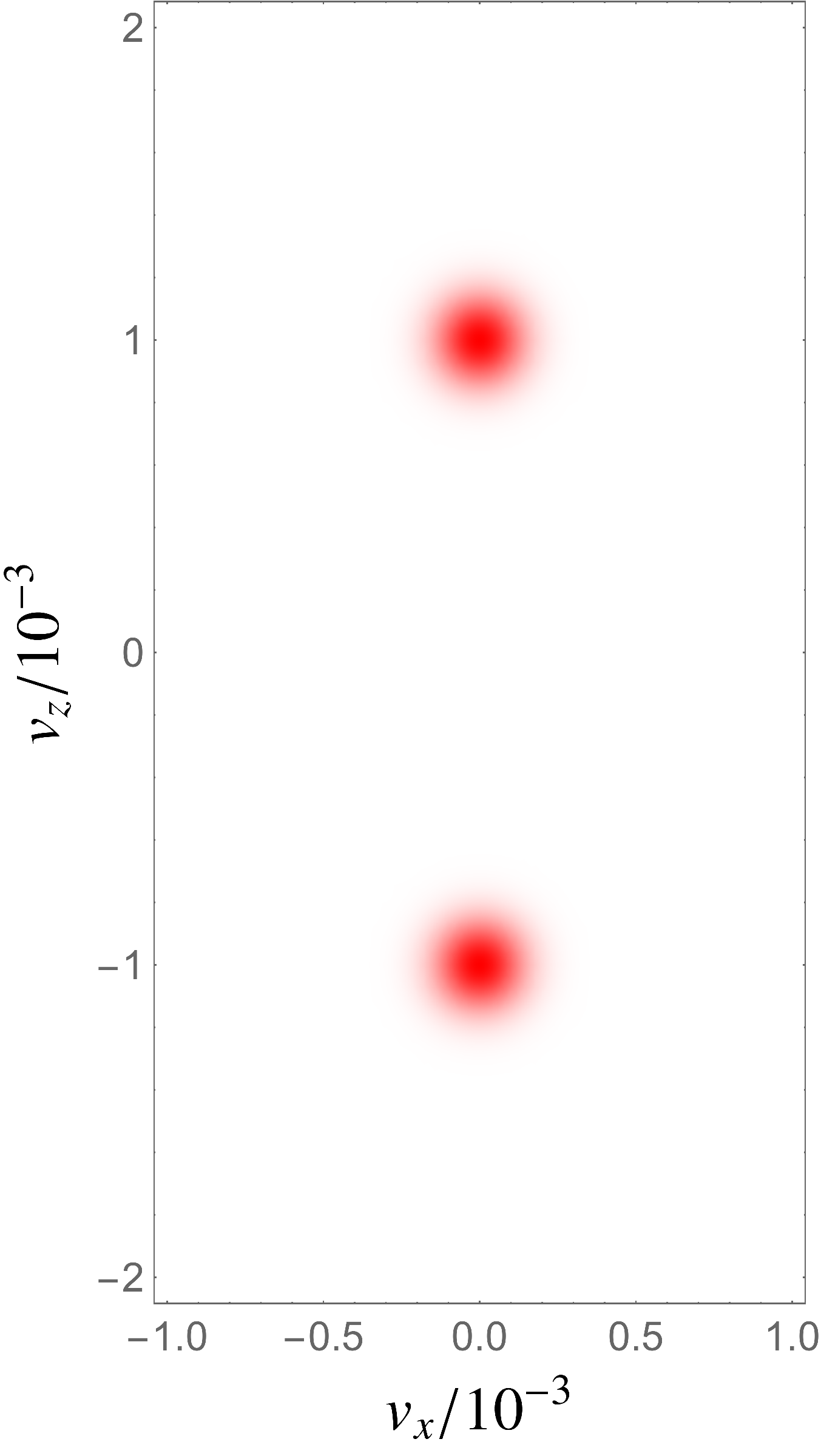}
\includegraphics[width = 0.6\textwidth]{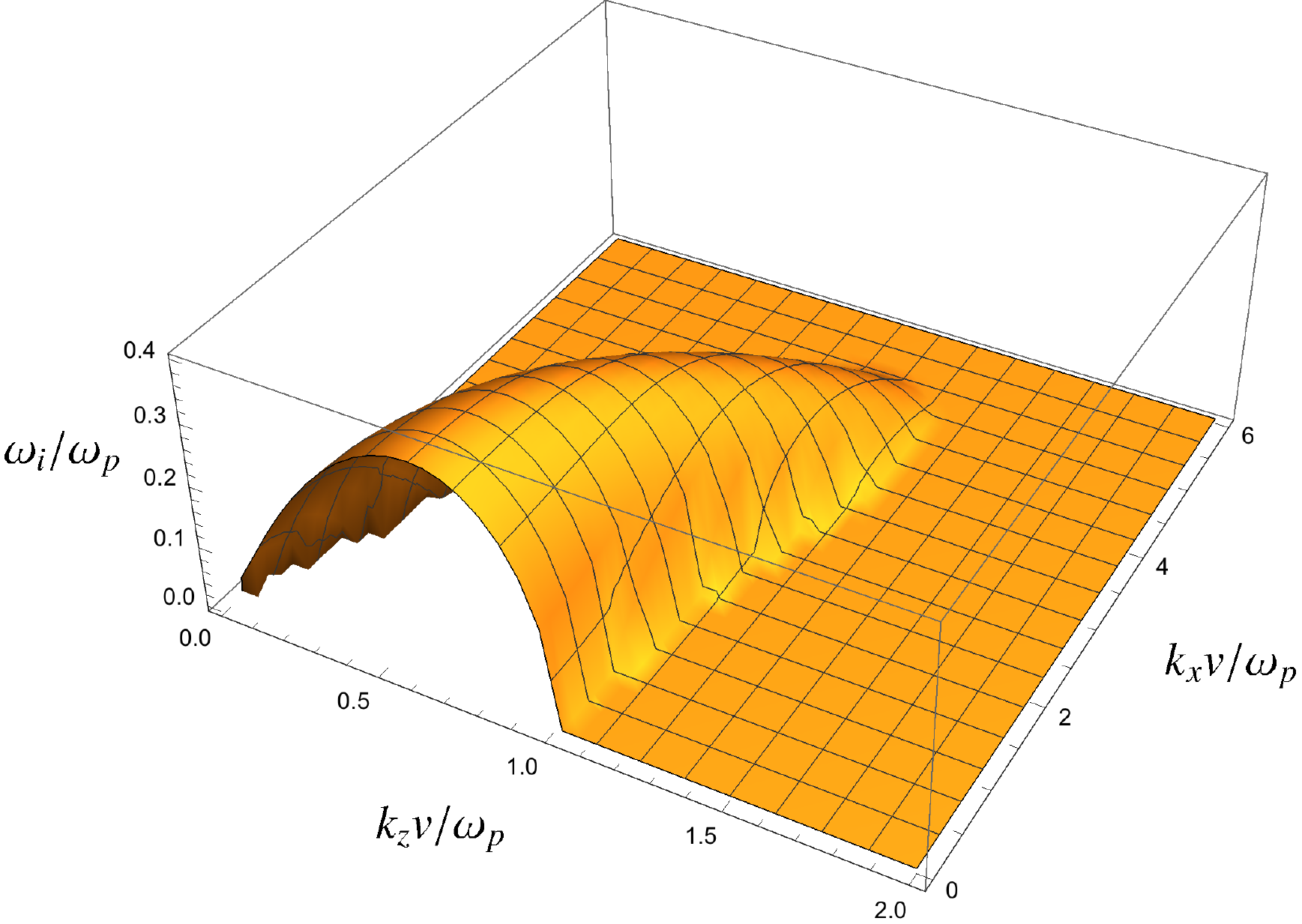}
		\caption{\emph{Left panel:} velocity distribution for a `two-stream'
		plasma background, in which two Maxwellian streams
		with velocity dispersion
		$\sigma = 10^{-4}c$ are counter-propagating,
		with a closing velocity $2 \times 10^{-3}c$
		in the $z$ direction.
		The shading indicates the phase
		space density at a given velocity.
		\emph{Right panel:} fastest growth rate for
		a perturbation with wavevector $k$, on the plasma
		background illustrated in the left-hand panel.
		The height of the surface corresponds to 
		the imaginary part of the mode's frequency.
		The plasma is assumed to consist of
		positive and negatively charged species with the
		same mass, and having the same velocity distributions.
		We take $\omega_p^2 = q^2 n / m_q$, where $n$ is the number
		density in each stream, $\pm q$ is the charge of each species,
		and $m_q$ is their mass.
		}
\label{fig2s1}
	\end{center}
\end{figure}

We will discuss the physical understanding of this
instability (and others) in detail below, but 
to start with, we can understand the parametrics
of the growth rate simply by considering the
energetics of the instability.
The vector field's energy density is 
$u_A = \frac{1}{2}(E^2 + B^2)$ (or larger
 for a massive mediator --- see equation~\ref{equmass}).
 The force exerted on a charged particle
 is $F = q(E + v \times B)$, so over a time
 $\delta t$, the magnitude of the change
 in the particle's velocity is at most 
 $\delta v \sim \frac{q \delta t}{m}(E + v \times B)$.
For a collection of particles with charges
$q_i$ and velocities $v_i$, the rate
of energy transfer from particle
KE to field energy is $P = \dot U_A = - \sum_i q_i v_i \cdot E$.

If we suppose that our perturbation is growing
exponentially, and we take $\delta t$ to be the
$e$-folding time, then we must have
$\delta P \sim P$ and $\delta U_A \sim U_A$, so $\delta P \delta t \sim U_A$. For a perturbation where $u_A$ is dominated by the electric
field, such as the longitudinal modes in
equation~\ref{eqlongdisp1ts}, we have
$|\delta P \delta t|  \lesssim \frac{q^2}{m} N (\delta t)^2 E^2$,
where $N = n V$ is the number of particles in the volume $V$.
Since $|\delta P \delta t|$ needs to be $\gtrsim u_A V \sim \frac{1}{2} E^2 V$ for the perturbation to grow,
we have 
\begin{equation}
	\frac{q^2 n}{m} (\delta t)^2 \gtrsim 1
	\quad \Rightarrow \quad
	\delta t \lesssim \left(\frac{q^2 n}{m}\right)^{-1/2}
	= \omega_p^{-1}
\end{equation}
For a magnetic-field-dominated perturbation,
the maximum growth rate is suppressed by
some power of $v/c \ll 1$ (for a non-relativistic plasma).\footnote{If
a Fourier mode of a perturbation has both $E$ and
$B$ large, then its Poynting vector
$S = E \times B$ is large, corresponding to
a propagating photon. In our scenarios, this means
that energy will be transported away from 
an initially localised perturbation, rather
than the perturbation growing.}
Consequently, the $\omega_i \sim \omega_p$ 
growth rate obtained from the two-stream
instability is as fast as perturbations can
grow, starting from a background without larger
fields present (equivalently, it is the fastest
rate at which the overall field energy can increase).

\subsection{Electrostatic instabilities}
\label{secelectro}

\begin{figure}[t]
	\begin{center}
\includegraphics[width = 0.48\textwidth]{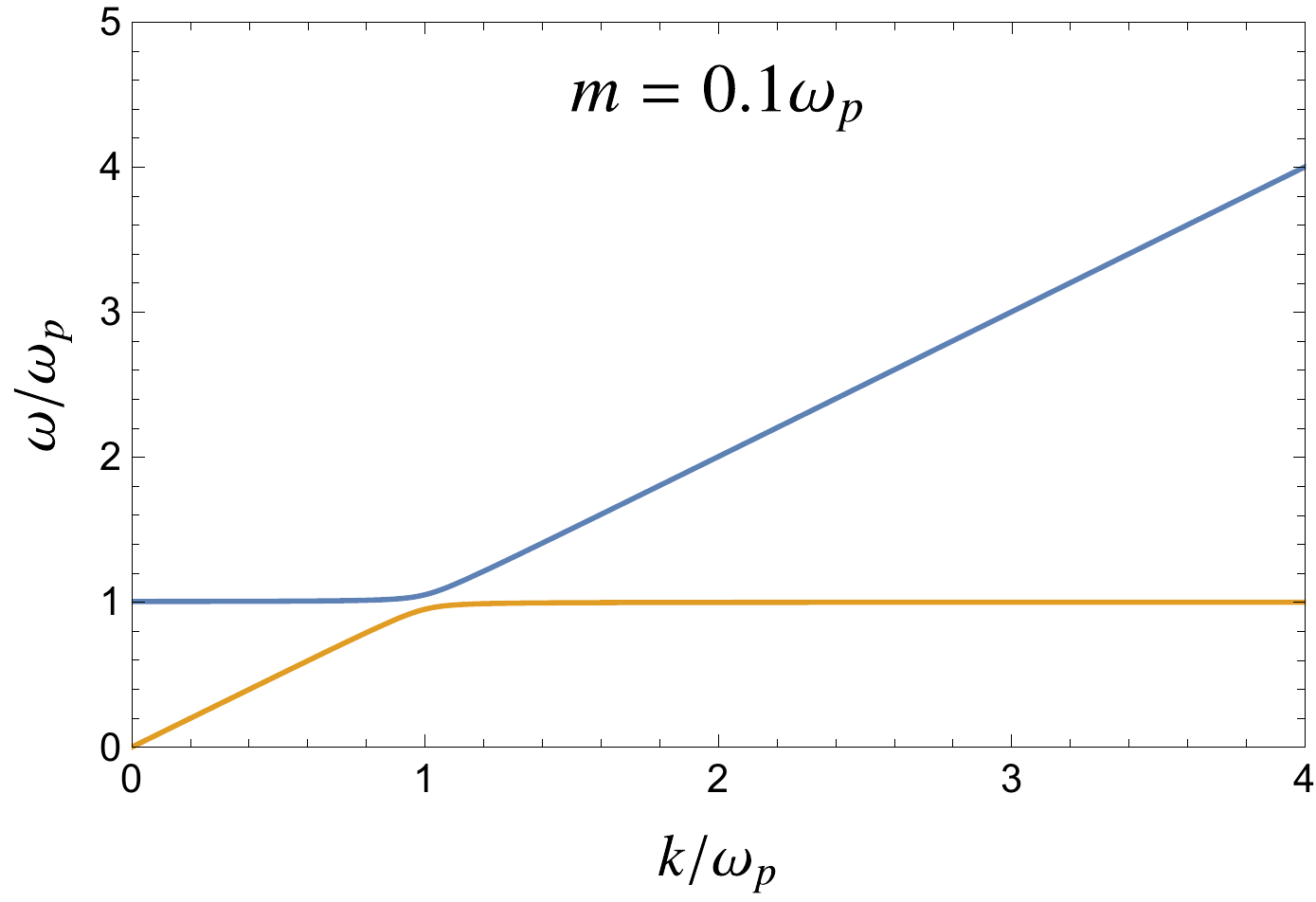}
\includegraphics[width = 0.48\textwidth]{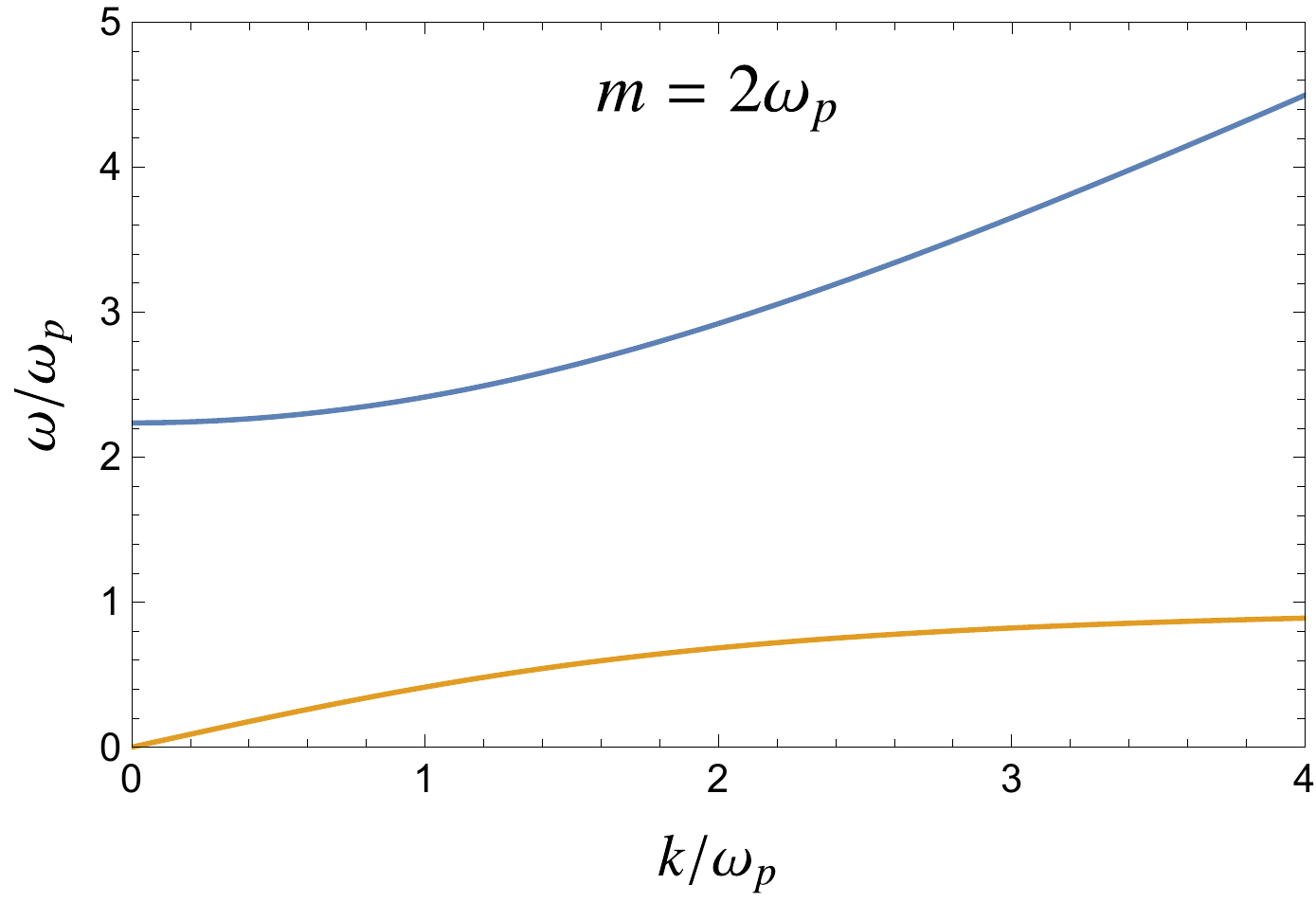}
		\caption{Dispersion relation for longitudinal
		oscillations in a cold, uniform plasma, with a massive mediator. In the left-hand panel, the mediator mass is $m = 0.1 \omega_p$.
		Longitudinal plasmons, which for a massive
		mediator have $\omega = \omega_p$, are mostly
		unaffected (apart from a small avoided crossing). At small $\omega$, there is
		also the $\phi \simeq$ const Goldstone mode
		(which would be pure-gauge for a massless mediator),
		and at high $\omega$, there are relativistic $\phi$
		excitations.
		In the right-hand panel, we have
		$m = 2 \omega_p$. Here, longitudinal plasmons
		are strongly affected for $k \lesssim m$, 
		linking with the Goldstone dispersion relation at smaller
		$k$. At higher $\omega$, the dispersion relation is
		approximately that for a free massive particle.
		}
\label{figlongdisp}
	\end{center}
\end{figure}

\begin{figure}[t]
\includegraphics[width = 0.5\textwidth]{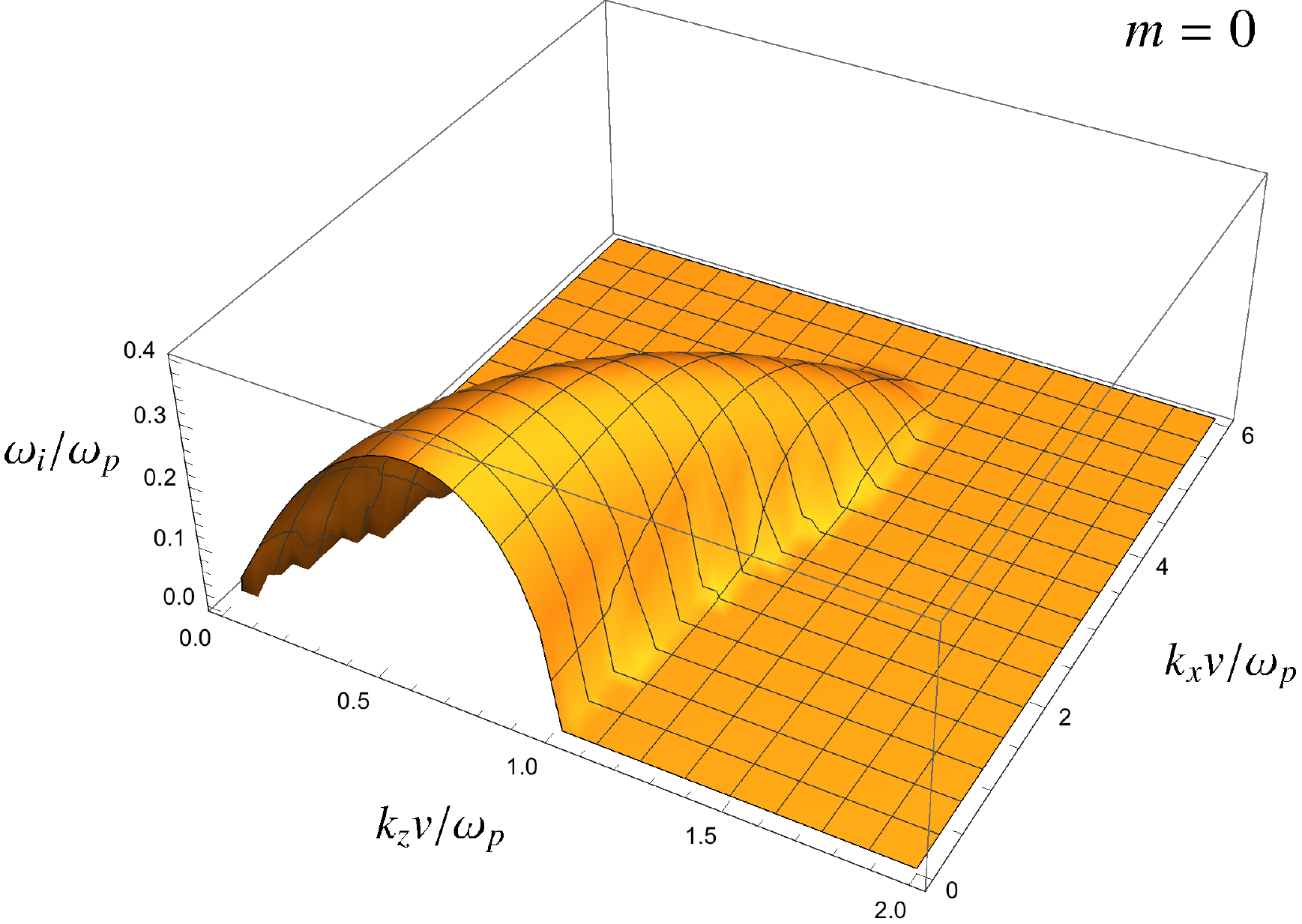}
\includegraphics[width = 0.5\textwidth]{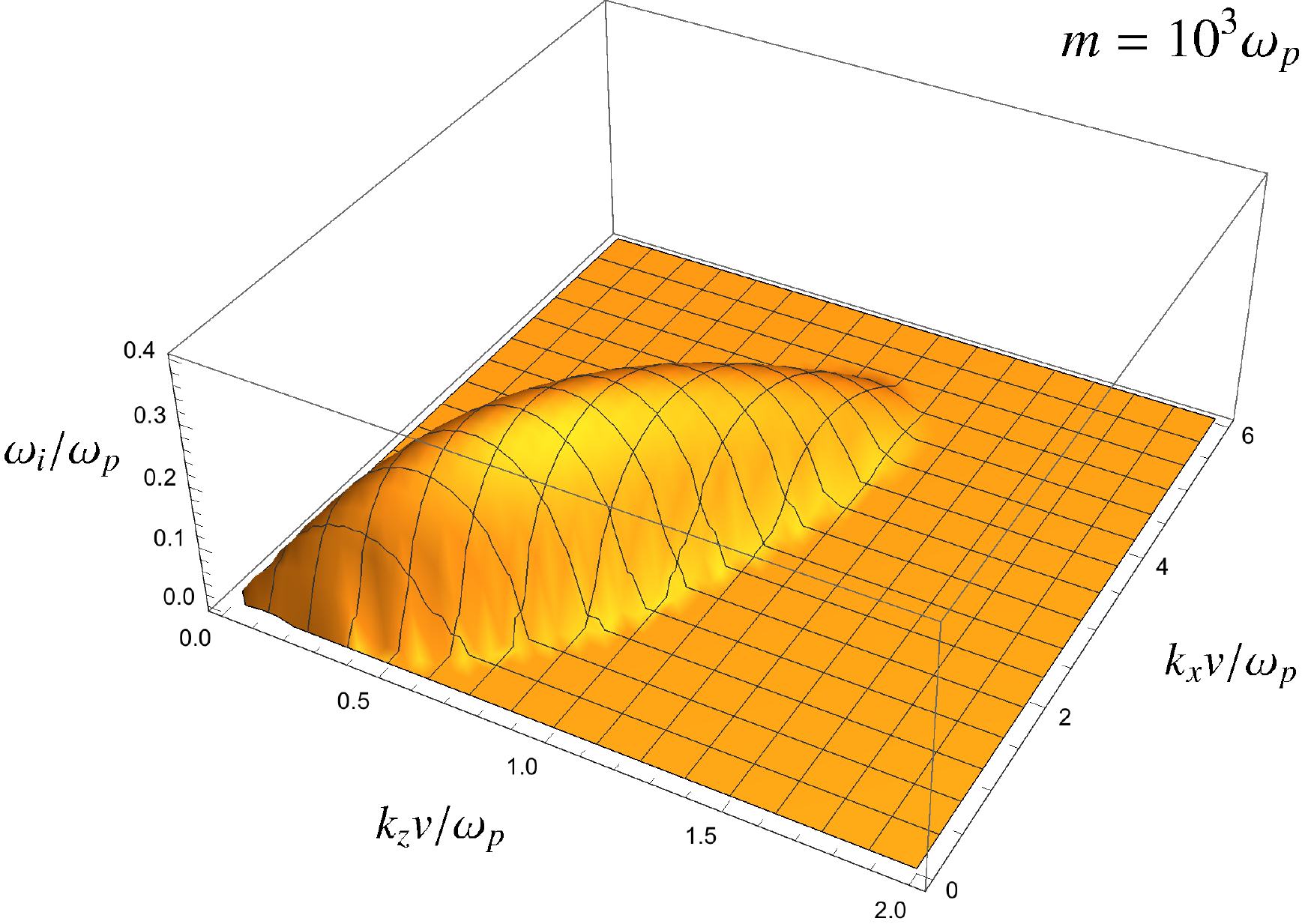}
\includegraphics[width = 0.5\textwidth]{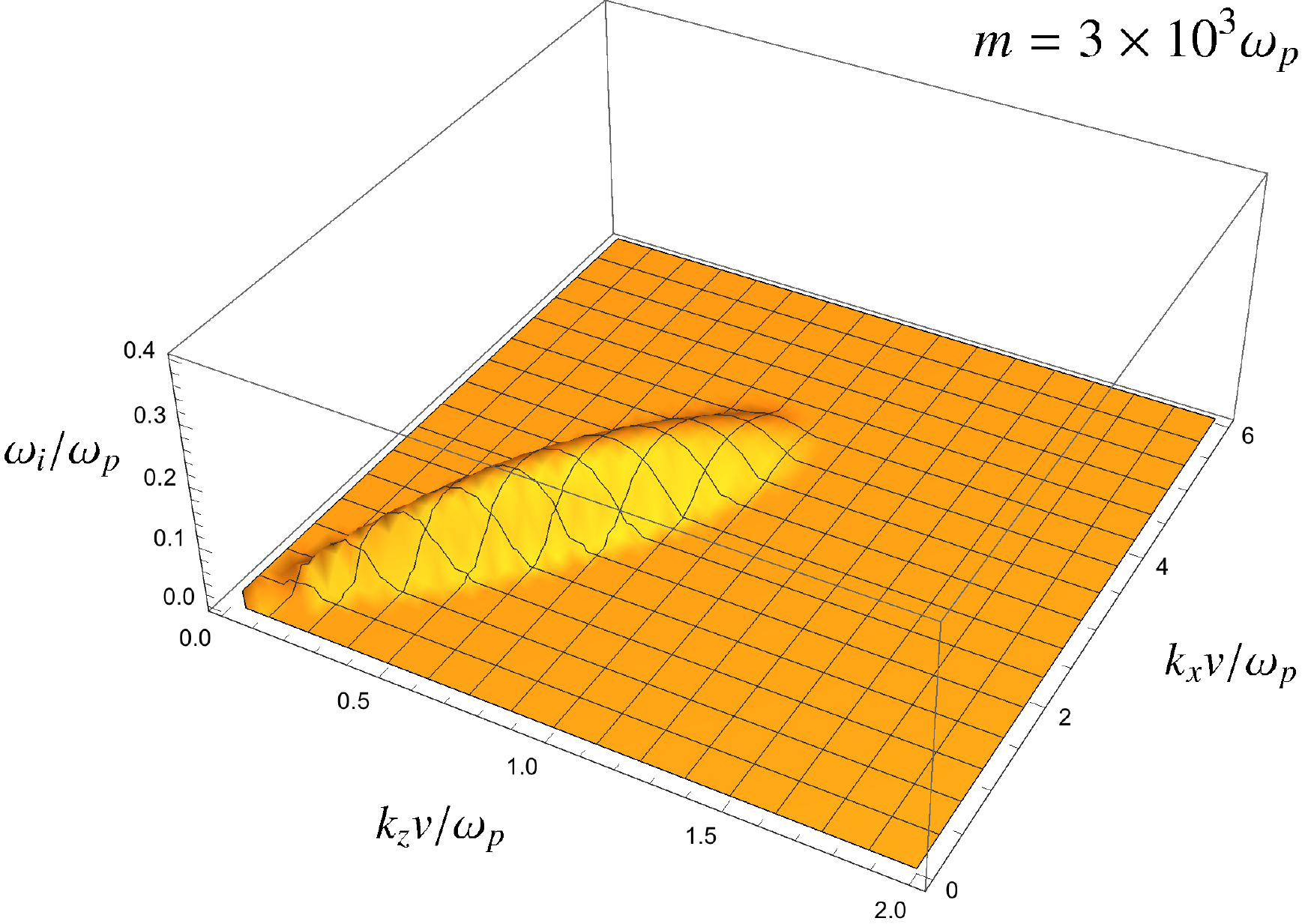}
\includegraphics[width = 0.5\textwidth]{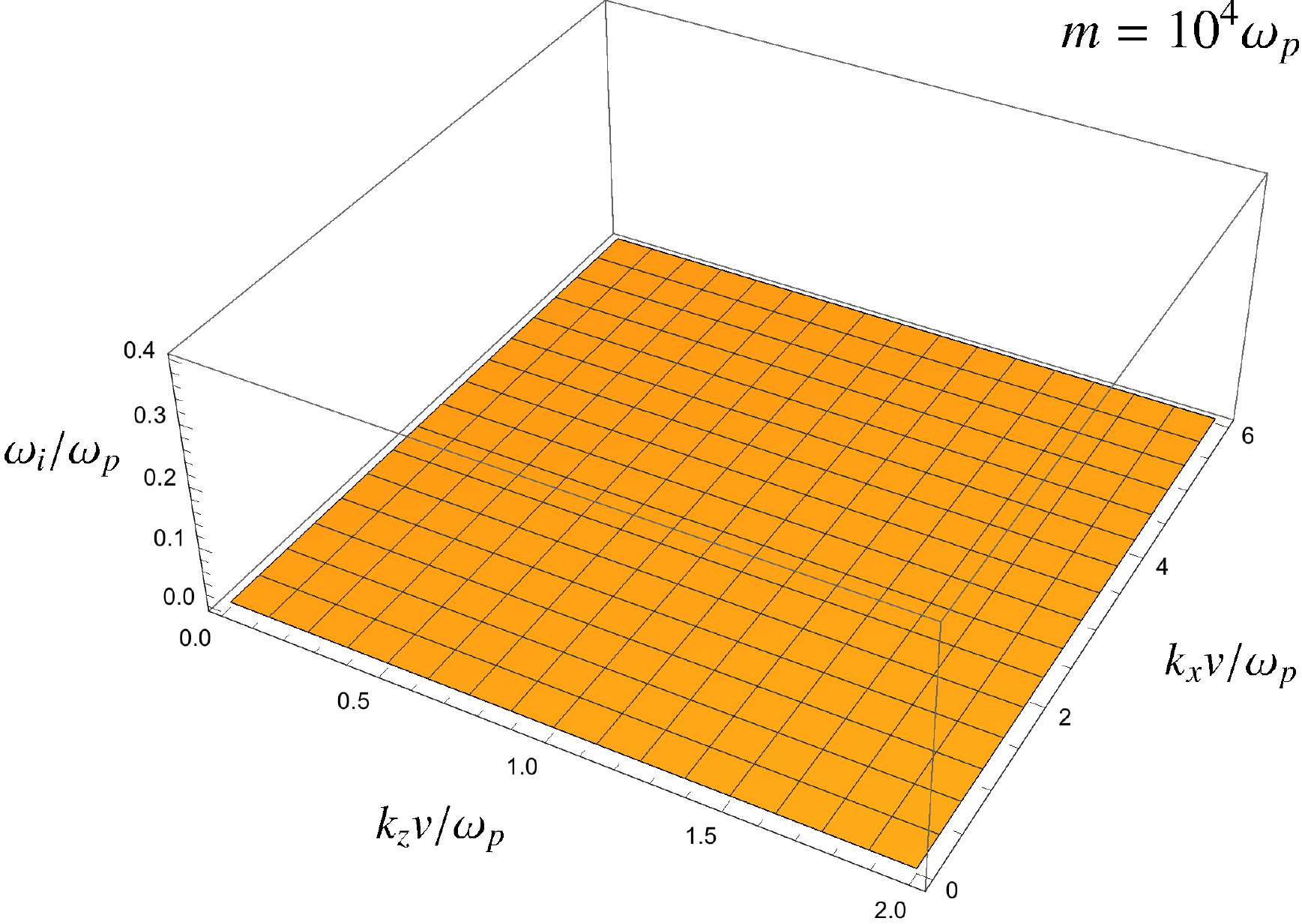}
	\caption{Fastest growth rate for a perturbation
	with wavevector $k$, on the plasma background
	illustrated in Figure~\ref{fig2s1}, for 
	a vector mediator of the indicated mass $m$.
	As this illustrates, longitudinal perturbations
	are suppressed for $m \gtrsim \omega_p / v_0 \sim 10^3 \omega_p$,
	while `oblique' perturbations with significant
	$k_x$ component exist up to $m \sim \omega_p / \sigma$
	(see Section~\ref{secelectro}).}
\label{figmm}
\end{figure}

As we will see, unstable modes in a non-relativistic
plasma generally have $|k| \gg |\omega|$;
parametrically, for the velocity distribution
of the plasma to be significant, we need 
$|k v| \gtrsim \omega$. Consequently,
instabilities in which the electric field perturbations
are much larger than the magnetic field perturbations
are dominantly electrostatic, with $|\phi| \gg |A|$
(since $B = \nabla \times A$ and $E = - \partial_t A - \nabla \phi$).
In this case, we can simply the Vlasov-Proca equations
to a one-dimensional integral. Reinstating
$c$ for clarity, we have
\begin{equation}
	(\omega^2 - c^2 k^2 - m^2 c^4)\phi 
	= -i \sum_s \frac{q_s^2}{m} \int d^3v \frac{1}{\sqrt{1 - v^2/c^2}}
	\frac{((E + v \times B) \cdot \partial_v f_s)}{\omega -
	k \cdot v}
\end{equation}
so taking $E \simeq -i k \phi$,
\begin{equation}
	(c^2 k^2 + c^4 m^2) \phi \simeq c^2 \phi \sum_s \frac{q_s^2}{m} \int d^3 v
	\frac{k \cdot \partial_v f_s}{\omega - k \cdot v}
	\label{eqlongdisp1}
\end{equation}
which gives us a scalar equation for the dispersion
relation. 
For purely longitudinal perturbations,
the full Vlasov equation has this form
(since there can be no magnetic fields), so
the behaviour of electrostatic perturbations
with a particular wavevector $k$ is equivalent
to longitudinal perturbations with the appropriate
projected velocity distribution (see appendix~\ref{appPenrose}).

For a massless mediator, longitudinal perturbations
have been extensively analysed in the plasma theory
literature~\cite{Briggs_71}.
In the symmetric, two-stream scenario,
a basic physical picture of the instability can
be obtained (following~\cite{Briggs_71}) by considering the longitudinal oscillations
of each of the streams individually. For a stream
with velocity $v$ in the lab frame,
oscillations with $\omega = \omega_p$ in the stream's
rest frame have $\omega = k \cdot v \pm \omega_p$ in the lab
frame\footnote{In the case of symmetric DM,
each stream will be composed of equal numbers of positive
and negative charges, with both species having the same velocity distribution.
If each has number density $n_\pm$ then the plasma frequency for the stream
is $\omega_p^2 = 2 q^2 n_\pm/m$. In the case of an SM plasma, we might instead have streams of electrons on top of a uniform, slowly-moving background of ions. The important quantity is simply the self-energy that the
distribution function gives rise to, which determines the dispersion relation.}, and the `slow' oscillation (with smaller frequency)
has less kinetic energy then the unperturbed beam, $\Delta {\rm KE} < 0$.
If we take $k$ such that $|k \cdot v| = \omega_p$, then
the slow oscillations for both streams 
are stationary in the lab frame. If the two oscillations
are in phase, they reinforce each other,
extracting KE from the streams to increase
the $E$ field energy, and giving rise to an instability.

This is the relevant picture for $k \cdot v \simeq \omega_p$,
where the peak instability
rate is obtained. As Figure~\ref{fig2s1}
shows, there are also growing perturbations
with small $k$, for which $\omega_i \searrow 0$ as $k \searrow 0$.
These can be thought of as arising from out-of-phase slow
oscillations in each stream, such that (in the lab frame)
the charge density perturbations almost cancel each other
out. 

A similar picture applies in the case of a massive
mediator, with the difference that, for
$m \gtrsim \omega_p$ and 
$k \lesssim m$, the dispersion
relation for longitudinal oscillations in a
stationary plasma is modified (see Figure~\ref{figlongdisp}).
Consequently, the peak instability rate is modified 
for $m \gtrsim \omega_p/v$.

Even without solving equation~\ref{eqlongdisp1}
directly, we can determine whether an instability exists
using the Penrose criteria~\cite{Penrose_1960} (see appendix~\ref{appPenrose}).
If the two streams have Maxwellian (projected) velocity distributions,
then for a massless mediator, an instability exists if $v_0 \gtrsim 1.3 \sigma$,
where $\sigma$ is the velocity dispersion for each
stream, and $2 v_0$ is the closing velocity. For a massive mediator, even for 
well-separated stream velocity distributions (i.e.\ closing
velocity $\gg$ velocity dispersions), we need
$m \lesssim 1.3 \omega_p/v_0$ in order for an instability 
to exist.

Figure~\ref{fig2s1} illustrates a particular example 
of two Maxwellian streams moving through each 
other (with parameters roughly
analogous to a cluster collision scenario), taking the closing
velocity to be $2 \times 10^{-3} c$, and the
velocity dispersion for each stream to be $\sigma_v = 10^{-4} c$.
The height of the surface corresponds to the fastest instability
rate for each $k$, which can be found 
from equation~\ref{eqlongdisp1} (in fact, the figure uses
the full Vlasov-Proca equation, rather than the longitudinal
approximation, though the results are almost identical).
As we go to more oblique $k$, the projected velocity distribution
has a smaller closing velocity, so past
the $v_p \simeq 1.3 \sigma$ threshold, where $2 v_p$
is the projected closing velocity, the instability
does not exist (though magnetic instabilities may still be 
present, as we discuss in Section~\ref{secmagnetic}).

For a massive mediator, if the closing velocity
is significantly greater than the velocity dispersions,
then going to oblique $k$ can be helpful, since
it decreases $v_p$, and so increases
$\omega_p / v_p$. This is illustrated
in Figure~\ref{figmm}, which shows the effects on increasing
$m$ on the instability rate as a function of $k$.
For $m \gtrsim \omega_p / \sigma$, the instability is
suppressed entirely.

This behaviour matches what we would expect from
basic energetic considerations.
The energy density associated with the $A$ field
is
\begin{equation}
	u_A = \frac{1}{2}(E^2 + B^2 + m^2 ((A^0)^2 + A^2))
	\label{equmass}
	\simeq \frac{1}{2} ((\nabla \phi)^2 + m^2 \phi^2)
\end{equation}
where the second equality is for an electrostatic
perturbation. Consequently, we expect the mediator mass
to become significant when $m \gtrsim k$, as we observed
above.

As mentioned in Section~\ref{secuniform}, for the 
collisionless Boltzmann equation to
be a good approximation to the evolution of a particular plasma
configuration, the particle density in phase space
needs to be sufficiently high. If the largest
wavenumber of interest is $k$, then
having many particles per $\sim k^{-3}$ volume
means that we have a sensible spatial coarse-graining.
Within each $k^{-3}$, we ideally want enough particles
to resolve the important features of the velocity distribution.
For the two-stream instability,
the relevant condition is $n_\chi \gg (\omega_\chi / \sigma)^3$.
If we impose the $2 \rightarrow 2$ Coulomb scattering
bound (equation~\ref{eqcoulomb}), and also that
$g \le 1/(4 \pi)$ (when $m_\chi$ is large enough that 
the Coulomb bound is ineffective), then
\begin{equation}
	\frac{n_\chi}{(\omega_\chi/\sigma)^3} 
	= \frac{\sigma^3 m_\chi^2}{\rho_\chi^{1/2} g^3}
	\gtrsim 10^{20} \left(\frac{\sigma}{10^3 \kms}\right)^3
	\left(\frac{0.1 \GeV\cm^{-3}}{\rho_\chi}\right)^{1/2}
	\label{eqn20}
\end{equation}
Consequently, the dark matter number density is always high
enough that, in a spatially uniform plasma, there
would be a good coarse-grained description. 
We discuss how well the approximation of spatial uniformity
applies to astrophysical systems in Section~\ref{secastro}.

The other way in which the collisionless Boltzmann equation can
break down is if collisions are important. For hard collisions,
the obvious estimate is that, if the collision rate for 
a particle is comparable to the growth time for
the instability, then the instability will be suppressed.
This can be confirmed by explicit calculations~\cite{Briggs_71}.
For Coulomb collisions, the rate of soft collisions is
enhanced by a Coulomb logarithm factor (Section~\ref{secexisting}),
but this alters things by a factor of $\OO(10)$ at most~\cite{Ackerman:mha,Agrawal_2017}.
As we noted above, since only a few hard collisions 
per DM particle during
the lifetime of universe are enough to significantly affect
e.g.\ halo shapes~\cite{Ackerman:mha,Agrawal_2017}, neglecting collisions will be usually be a good approximation
for couplings that are not already excluded by 
$2 \rightarrow 2$ scattering calculations.
As we will see in Section~\ref{secastro}, the couplings we are interested
in are often many orders of magnitude below
$2 \rightarrow 2$ exclusion limits. Consequently,
even if regions of very high DM density exist,
such as sharp central cusps in halos, 
collisions will usually be unimportant even there.


\subsubsection{Beam-plasma}
\label{secbp}

\begin{figure}[t]
	\begin{center}
\quad
\quad
\quad
\includegraphics[width = 0.3\textwidth]{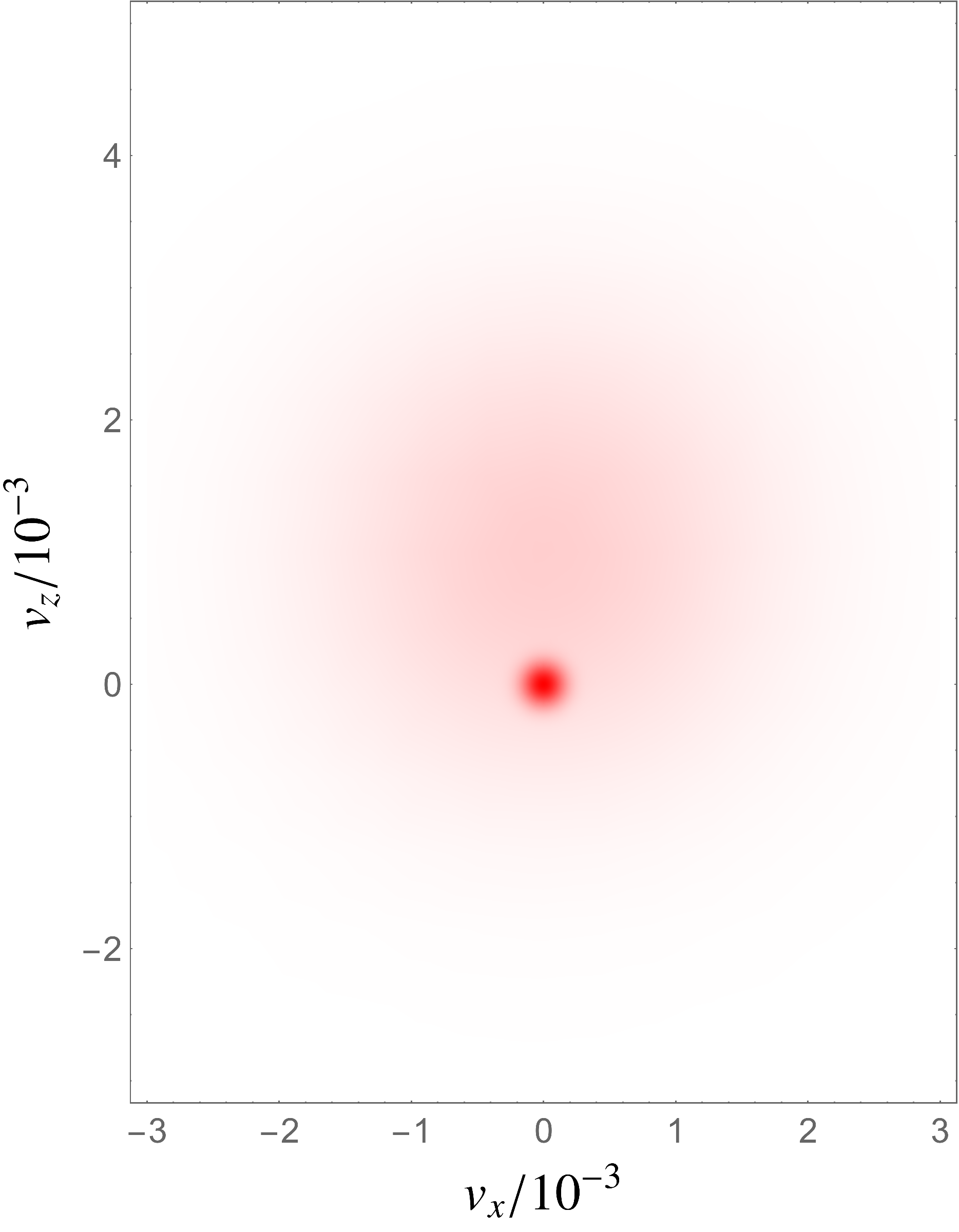}
\quad
\quad
\quad
\includegraphics[width = 0.48\textwidth]{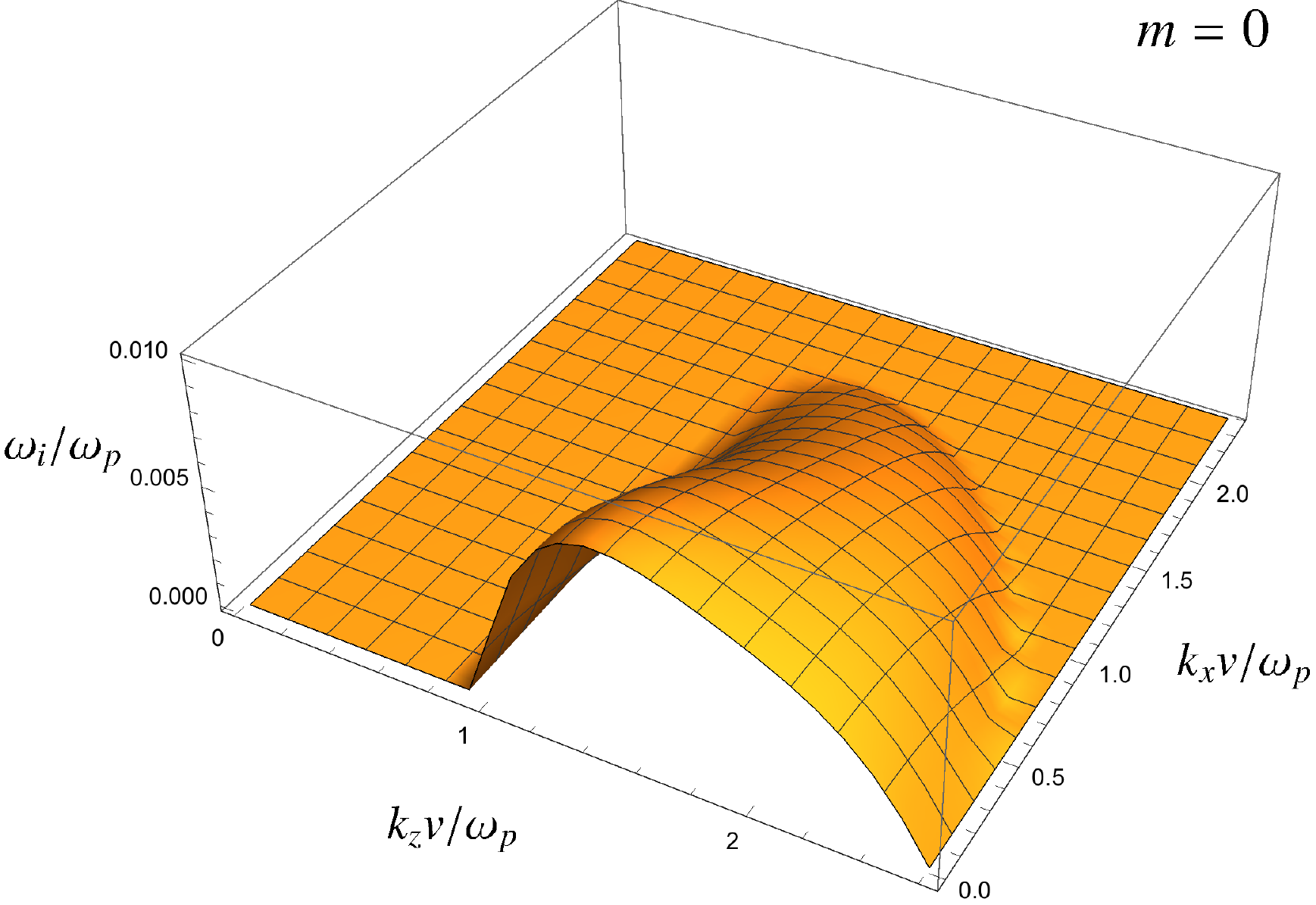}
\includegraphics[width = 0.48\textwidth]{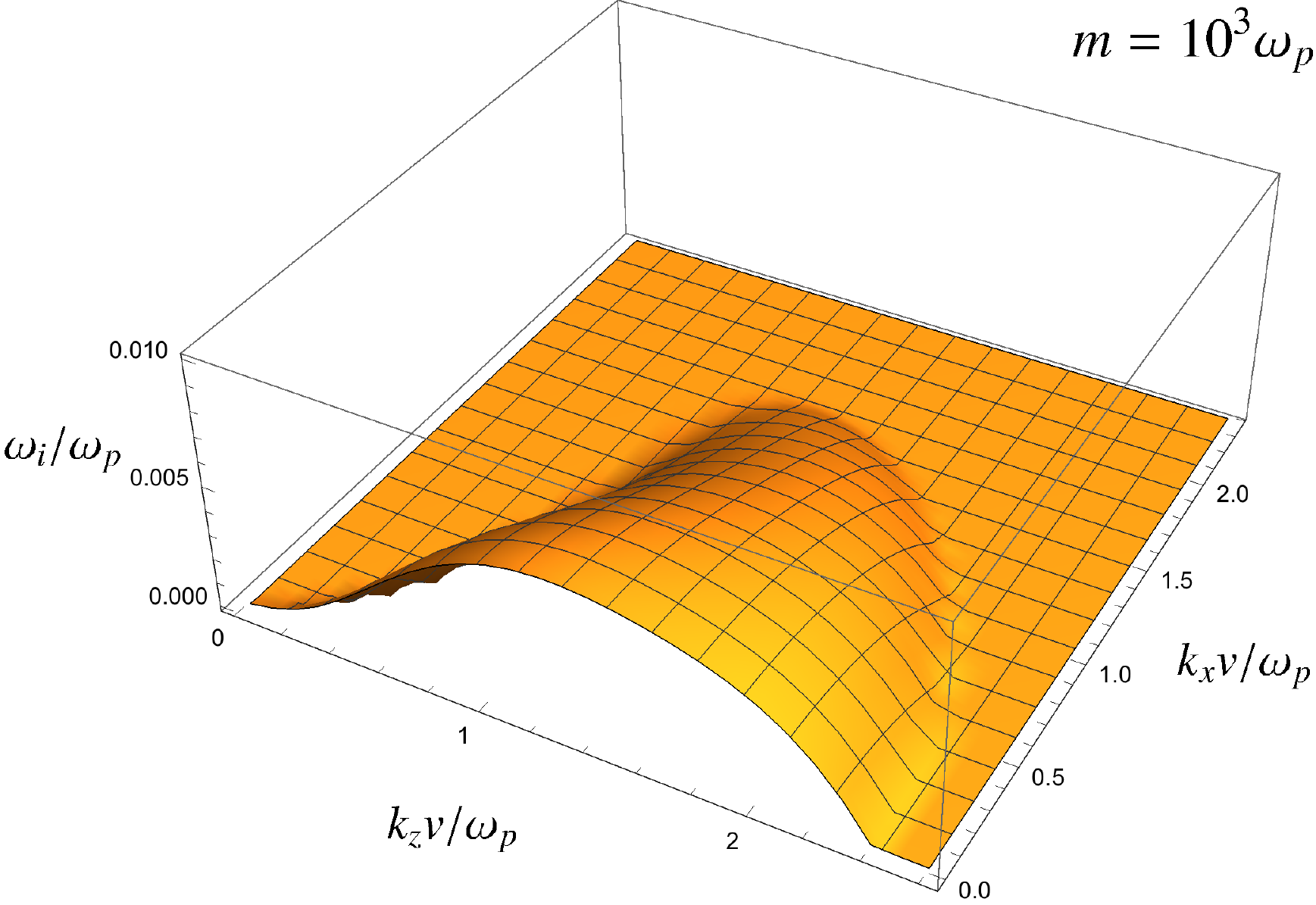}
\includegraphics[width = 0.48\textwidth]{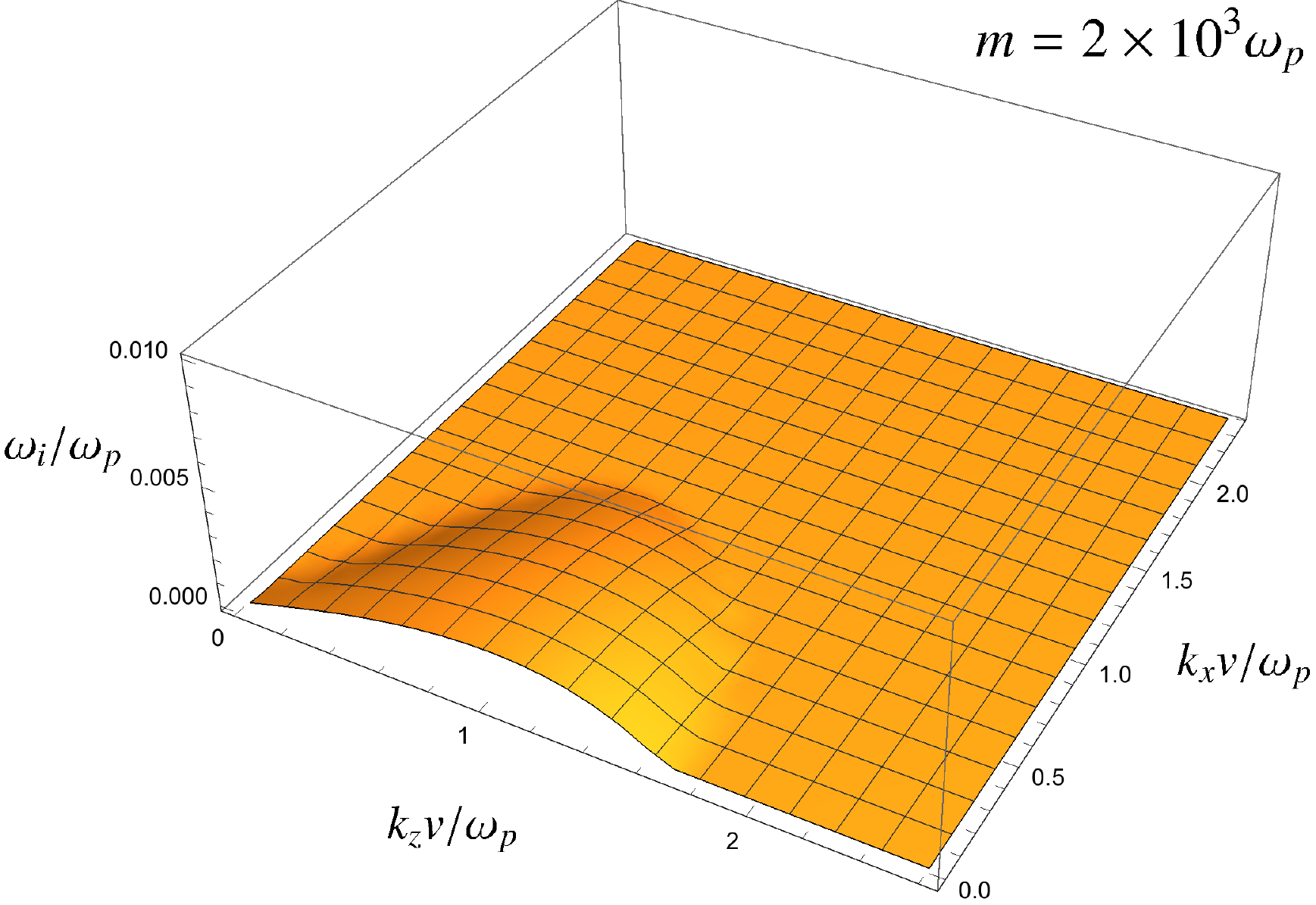}
		\caption{\emph{Top-left panel}:
		velocity distribution for
		a `beam-plasma' scenario, in which a 
		dense ($\omega_p = 1$, in arbitrary units), low velocity
		dispersion ($\sigma_p = 10^{-4}c$) Maxwellian stream
		propagates through a more dilute
		($\omega_p = 0.1$), high velocity dispersion
		($\sigma = 10^{-3}c$) at relative velocity
		$v = 10^{-3}c$. \emph{Other panels}:
		fastest growth rate for a perturbation
		with wavevector $k$, on the plasma
		background illustrated in the top-left panel,
		for a vector mediator with the masses $m$
		indicated.
		For $m \gtrsim 2.5 \times 10^3 \omega_p$, there are no unstable modes.}
	\label{figbp}
	\end{center}
\end{figure}

If we consider asymmetric streams,
as illustrated in Figure~\ref{figbp}, the behaviour
of instabilities can be somewhat different.
In the plasma literature, these are known 
as `beam-plasma' instabilities, with the `beam'
being the less dense stream and the 
`plasma' the more dense one~\cite{Briggs_71}.
For us, they can serve as a model
for astrophysical
situations such as a denser, lower-velocity-dispersion
subhalo (the `plasma') moving through a larger DM halo
(the `beam'), as discussed in 
Section~\ref{secastro}.

Setting up our problem (for a massless mediator,
our calculations will follow those in~\cite{Briggs_71}),
we will take
the denser stream to be at rest, with plasma
frequency $\omega_p$, and the less dense stream
to have relative velocity $v_0$, with plasma
frequency $\omega_1$.
If we ignore
the velocity dispersion of the streams, the dispersion
relation for a massless mediator is
\begin{equation}
	1 = \frac{\omega_p^2}{\omega^2} + \frac{\omega_1^2}{(\omega - k v_0)^2}
\end{equation}
If we assume the density ratio is small, $R \equiv \omega_1^2 / \omega_p^2 \ll 1$, 
then for $\omega \not \simeq \omega_p$, we need $|\omega - k v_0|^2 
\sim R$. Writing $\omega - k v_0 = \delta$, we 
have
\begin{equation}
	\delta^2 = R \frac{\omega^2 \omega_p^2}{\omega^2 - \omega_p^2}
\end{equation}
Consequently, for $\omega \gtrsim \omega_p$,
the modes are stable, while for 
$\omega \lesssim \omega_p$, we have $\omega_i \sim \sqrt{R} k u$.
The maximum instability rate is obtained for $\omega \simeq \omega_p$,
for which
\begin{equation}
	\delta^2 \simeq R \frac{\omega^2 \omega_p}{2 \delta} 
	\quad  \Rightarrow \quad
	\delta^3 \simeq \frac{R}{2} \omega_p^3
	\quad \Rightarrow \quad 
	\delta \simeq e^{2 \pi i n / 3} \left(\frac{R}{2}\right)^{1/3} \omega_p
\end{equation}
where $n$ is an integer. Consequently,
the maximum growth rate is $\omega_i \sim R^{1/3} \omega_p$.

Neglecting the velocity dispersions of the streams
will be a good approximation if the phase
velocity of the excitation is well outside
the velocity distributions of both streams.
In the plasma literature, this is referred to as 
the `cold-beam' case.
Since $\omega / k - v_0 = \delta / k$,
this will be true if the fractional velocity dispersion
of the beam
is $\frac{\sigma}{v_0} \lesssim \max \frac{\delta}{k v_0} \sim R^{1/3}$,
and the plasma velocity dispersion is $\ll v_0$.

The physical picture for the fastest-growing instability
is that a longitudinal oscillation in the denser stream,
which has frequency $\omega_p$, interacts with a 
$\Delta {\rm KE} < 0$ forced oscillation in the `beam',
extracting kinetic energy.
For $kv_0 \lesssim \omega_p$, we have a similar
picture as for the two-stream instability, with
partially-cancelling perturbations in the beam and the plasma.
As we take $R \rightarrow 1$, we move smoothly towards
the two-stream case.

In many circumstances, the beam's velocity dispersion
will be large enough that the `cold-beam' approximation does
not hold. For example, in the subhalo case, the velocity dispersion
of the larger halo is generally comparable to the orbital 
velocity of the subhalo through it, so $\sigma \sim v_0$.
The behaviour in this `hot-beam' scenario is rather different.

Since we need to take the beam's velocity distribution into
account, the dispersion relation for longitudinal oscillations
becomes (again, for a massless mediator)
\begin{equation}
	1 \simeq 
	\frac{\omega_p^2}{\omega^2}
	+ \frac{\omega_1^2}{k^2} \int dv \frac{f'(v)}{\omega/k - v}
	\label{eqhotdisp}
\end{equation}
(we will see that as long as the plasma's velocity dispersion
is $\ll u$, treating it as cold
will not affect the most unstable modes).
If we write $\omega /k = u_r + i u_i$, then 
we can write the integral over $v$ as an integral
over a contour slightly below the real axis,
which for $u_i$ sufficiently small is approximately
\begin{equation}
	\int dv \frac{f'(v)}{v - (u_r + i u_i)}
	= \int_{-\infty-i u_i}^{\infty - i u_i}
	dv \frac{ f'(v + i u_i)}{v - u_r}
	\simeq P \int_{-\infty}^\infty dv \frac{f'(v)}{v - u_r} 
	+ i \pi f'(u_r)
\end{equation}
For the imaginary part of equation~\ref{eqhotdisp}
to be zero,  we need
\begin{equation}
	- \frac{2 \omega_p^2 \omega_r \omega_i}{|\omega|^4} + \pi
	\frac{\omega_1^2}{k^2} f'(\omega_r/k) \simeq 0
\end{equation}
For $R \ll 1$, the real part of the RHS
of equation~\ref{eqhotdisp} is dominated
by the first term $\omega_p^2/\omega^2$, so we
need $\omega_r \simeq \omega_p$.
Consequently,
\begin{equation}
	\omega_i \simeq R \omega_p \frac{\pi}{2} \frac{\omega_p^2}{k^2} f'(\omega_r/k)
\end{equation}
For a generic beam velocity distribution, with
velocity dispersion $\sim \sigma$, the maximum
value of $f'(u)$ is $\sim 1/\sigma^2$,
so
\begin{equation}
	\max \omega_i \sim R \omega_p \frac{u^2}{\sigma^2}
	\label{eqwibp}
\end{equation}
For this to be valid, we need $f'(u_r + i u_i) \simeq
f'(u_r)$; for a simple beam velocity distribution such as a Gaussian,
this requires $u_i \ll \sigma$. From equation~\ref{eqwibp},
this holds for for $R \ll \sigma^3/v_0^3$.
This is simply the opposite of the cold-beam condition
derived above, showing that the hot-beam and cold-beam
regimes are complementary. The maximum instability rates
also have the same parametric
value at $R \sim \sigma^3/v_0^3$, consistent with this picture.

The different behaviour in the hot-beam regime corresponds
to a somewhat different physical picture of the instability,
compared to the cold-beam case. In particular, the approximation
of the counter integral used above is very similar to the
usual derivation of Landau damping~\cite{Chen_2015,fitzpatrick2015plasma}. For the latter,
the physical picture is that 
beam particles with velocities very slightly smaller
the phase velocity $\omega/k$ of a plasma oscillation
are accelerated by it, while those with velocities
very slightly larger are slowed down.
If the gradient of the velocity distribution
is negative, then there are more slightly-slower particles
than slightly-faster particles, and energy is on average transferred
from the plasma oscillation to particle KE.
If the gradient is \emph{positive}, then the opposite
occurs; energy is transferred from particle KE to 
the plasma oscillation, resulting in an instability.
This `inverse Landau damping' picture breaks down when the
instability growth rate becomes too fast,
so that `slightly-slower'
and `slightly-faster' encompass too wide a range in the velocity
distribution.

The top-right panel of Figure~\ref{figbp} shows the instability
behaviour for an example velocity distribution
with $R = 10$, and $\sigma = v$ (i.e.\ in the hot-beam regime).
While the $\omega_r$ values of the unstable modes
are not shown, these are close to $\omega_p$. 
As the figure illustrates, we need $k \cdot v \gtrsim \omega_p$
for there to be an instability, corresponding to the phase
velocity of the perturbation being in the increasing
part of the $v_z$ distribution.

For a massive mediator, the same basic physical picture
applies, once the modified dispersion relation 
of longitudinal oscillations is taken into account.
Equation~\ref{eqhotdisp}
is modified to
\begin{equation}
	1 + \frac{m^2}{k^2} \simeq 
	\frac{\omega_p^2}{\omega^2}
	+ \frac{\omega_1^2}{k^2} \int dv \frac{f'(v)}{\omega/k - v}
\end{equation}
From the real part of this equation,
we need to have $\omega \simeq \omega_p (1 + m^2/k^2)^{-1/2}$.
This corresponds to the lower branch of the dispersion
relation (for the dense plasma) in the right-hand panel of Figure~\ref{figlongdisp},
with the usual relation $\omega \simeq \omega_p$ being
modified for $m \gtrsim k$.
Parametrically, we expect instabilities
to be suppressed entirely for $m \gtrsim \frac{\omega_p}{\sigma_p}$. For longitudinal oscillations, we can confirm
this using the Penrose criteria; from 
appendix~\ref{appPenrose}, instabilities exist
for 
\begin{equation}
	m \lesssim \frac{\omega_p}{v_c} \sim \frac{\omega_p}{{\rm few} \times \sigma_p}\
	\label{eqmbp}
\end{equation}
where $v_c$ is the point at which $f_v' = 0$,
balanced between the falling $f'$ from the plasma velocity
distribution and the rising $f'$ from the beam.
The second equality applies if both the beam and plasma
have Maxwellian velocity distributions.
The basic physical picture is that, for given $k$, increasing 
$m$ pushes $\omega_r$ smaller, until $u_r = \omega_r/k$ is no
longer on the rising part of the velocity distribution
(see Figure~\ref{figveldist}). This behaviour is illustrated
in Figure~\ref{figbp}, which shows the effects of increasing
$m$ on the instability rate as a function of $k$.
Since the velocity distribution of the dense
component is isotropic, and the dense component's velocity dispersion sets the largest $k$
for an unstable mode, it does not help to
go to oblique $k$.

\subsection{Magnetic instabilities}
\label{secmagnetic}

The electrostatic instabilities discussed above
relied on velocity distributions with
sufficiently well-separated peaks. 
As well as these, there can also be `magnetic'
instabilities, in which the magnetic field
energy dominates the electric field~\cite{Weibel_1959,Fried_1959}. 
In situations where both magnetic and electrostatic
instabilities exist, the growth rate of the former is generally
suppressed relative to the latter
by $\sim v_s/c$, where $v_s$ is the stream velocity.
However, magnetic instabilities 
exist for a wider range of velocity distributions,
including those which are very close to Maxwellian.

\begin{figure}[t]
	\begin{center}
\includegraphics[width = 0.7\textwidth]{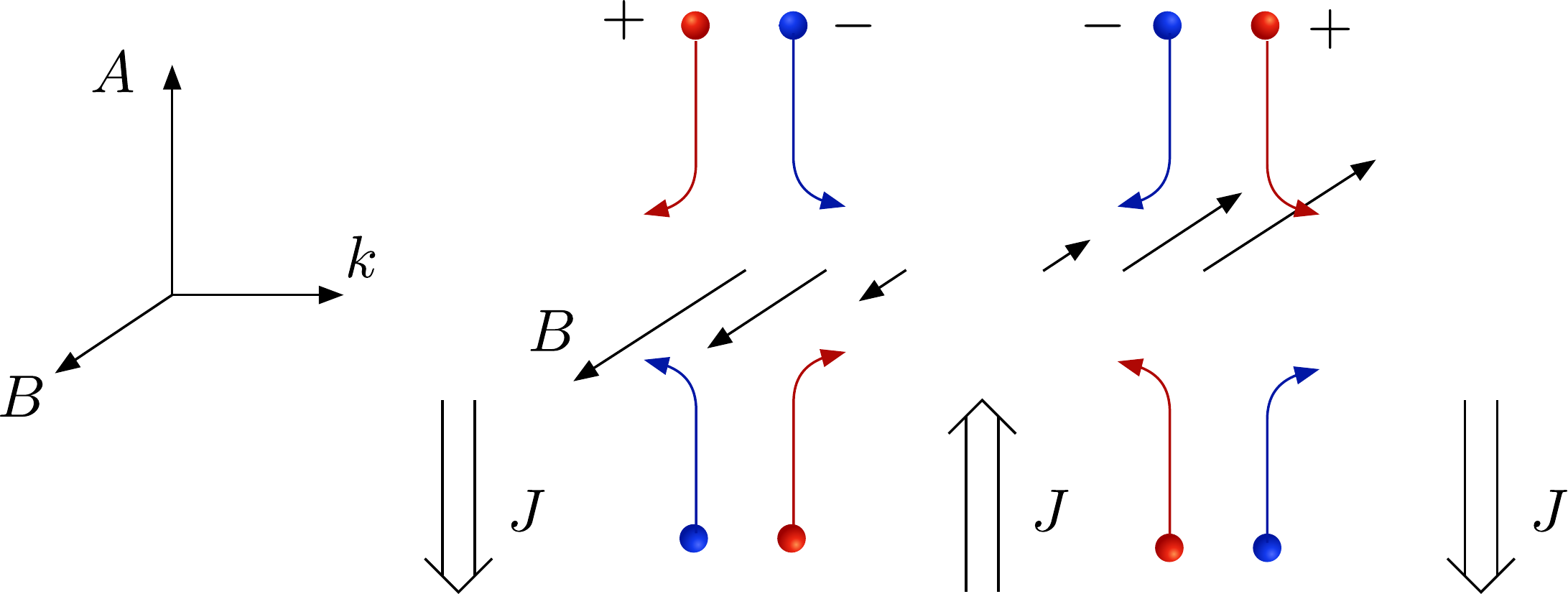}
		\caption{Illustration of a magnetic (`Weibel'~\cite{Weibel_1959})
		instability for counter-propagating streams (see
		Section~\ref{secmagnetic}).
The transverse magnetic field
		$B$ bends the trajectories of
		charged particles in the streams (which are travelling upwards
		and downwards), creating current sheets (indicated by $J$)
		\cite{Fried_1959}.
		These currents do work on the field via
		the $E \cdot J$ term (since $E$ is in the vertical direction),
		enhancing the field energy and leading to an instability.}
\label{figmagsketch}
	\end{center}
\end{figure}

The simplest examples of magnetic instabilities
can be seen in the (cold) two-stream velocity distribution
\cite{Fried_1959}.
If, instead of looking at $k \parallel v_0$
perturbations, we take $k \perp v_0$, then
the response function is
(taking the streams to be in the $z$ direction,
and $k \propto \hat x$)
\begin{equation}
	\Pi A = \begin{pmatrix}
		(\omega^2 - k^2) \frac{\omega_p^2}{\omega^2} & 0 & 0 \\
		0 & \omega_p^2 & 0 \\
		0 & 0 & (\omega^2 + k^2 v_0^2) \frac{\omega_p^2}{\omega^2}
	\end{pmatrix}
	\begin{pmatrix}
		A_x \\ A_y \\ A_z
	\end{pmatrix}
\end{equation}
For $A \perp \hat z$, we have the usual dispersion
relations for plasma oscillations (as we would
expect, since non-relativistic motion
in the $\hat z$ direction has not effect on those).
For $A \parallel \hat z$, the dispersion relation becomes
\begin{equation}
	\omega^2 - c^2 k^2 - c^4 m^2 = \omega_p^2 + k^2 v_0^2 \frac{\omega_p^2}{\omega^2}
	\label{eqmag2disp}
\end{equation}
(where we have reinstated $c$ for clarity)
which has solution
\begin{equation}
	\omega^2 = \frac{1}{2}\left(\omega_p^2 + c^2 k^2 + c^4 m^2 \pm \sqrt{(\omega_p^2 + c^2 k^2 + c^4 m^2)^2 + 4 \omega_p^2 k^2 v_0^2}\right)
\end{equation}
Expanding in small $v_0/c$, we have a solution
\begin{equation}
	\omega^2 \simeq \frac{-\omega_p^2 k^2 v_0^2}{\omega_p^2 + c^2 k^2 + c^4 m^2}
	\label{eqtts1}
\end{equation}
so there are growing modes, with $\max \omega_i \simeq \omega_p v_0/c$
(attained at $k \gtrsim \max (\omega_p,m)$).
The physical picture behind this instability is illustrated
in Figure~\ref{figmagsketch}. A transverse magnetic field
bends the trajectories of particles in the streams,
creating current sheets in the stream direction.
These currents do work on the electric field,
growing the instability.

More generally, if the velocity distribution
is separable in the $x$ and $z$ directions,
so
\begin{equation}
	\int dv_y f_\pm(v_x,v_y,v_z) = n_\pm g_x(v_x) g_z(v_z)
\end{equation}
(assuming symmetric positive and negative species)
with $\int dv_x g_x(v_x) = \int dv_z g_z(v_z) = 1$,
then
\begin{equation}
	\Pi_{zz} = \omega_p^2 \left(1 + 
	\sigma_z^2 \int dv_x \frac{g_x'(v_x)}{\omega/k - v_x}\right)
	\equiv \omega_p^2 \left(1 + \sigma_z^2 G(\omega/k) \right)
	\label{eqpzz}
\end{equation}
where $\sigma_z^2 = \int dv_z v_z^2 g_z(v_z)$ and $\omega_p^2 = 2 q_\pm^2 n_\pm / m_\pm$,
and $G$ is defined as in appendix~\ref{appPenrose}.
We also have
$\Pi_{zx}, \Pi_{xz} \propto \int dv_z v_z g_z (v_z)$,
so to simply the analysis, we can make
$\Pi$ diagonal by boosting in the $z$ direction
so that $\bar v_z = \int dv_z v_z g_z (v_z) = 0$.

For the case of a Maxwellian velocity distribution $g_x$,
we have (for $u$ with positive imaginary part)
\begin{equation}
	G(u) = - \frac{1 + \frac{u}{\sqrt{2} \sigma_x} Z \left(\frac{u}{\sqrt{2} \sigma_x}\right)}{\sigma_x^2}
\end{equation}
where $Z(x) \equiv i \sqrt{\pi} e^{-x^2} {\rm erfc}(- i x)$ is the plasma dispersion 
function~\cite{fitzpatrick2015plasma}
(inserting this into equation~\ref{eqpzz} reproduces
the dispersion relation given in the original
paper of Weibel~\cite{Weibel_1959}).\footnote{
	To understand the parametric behaviour of
	the response function, it 
	can also be useful to consider simpler
	situations, e.g.\ a top-hat velocity distribution,
	for which 
	\begin{equation}
		G(u) = \frac{1}{2 \sigma_x} \int_{-\sigma_x}^{\sigma_x}
		dv_x \frac{1}{(v_x - u)^2} = \frac{1}{u^2 - \sigma_x^2}
	\end{equation}
	}
From the dispersion relation
\begin{equation}
	\omega^2 - k^2 - m^2 = \omega_p^2 (1 + \sigma_z^2 G(\omega/k))
\end{equation}
we can find that for a Maxwellian $v_x$ distribution,
an instability exists for $k \le k_{\rm max}$, where
\begin{equation}
	k_{\rm max}^2 = \omega_p^2 \left(\frac{\sigma_z^2}{\sigma_x^2}
	- 1 \right) - m^2
\end{equation}
If the RHS is negative, then there is no instability.

In many astrophysical situations, the velocity anisotropy
will be small, $\sigma_z/\sigma_x \simeq 1$. In this
case, we can expand the plasma dispersion function
in small $\omega_i$, and obtain (for a massless mediator)
\begin{equation}
	\max \omega_{i} \simeq \sqrt{\frac{8}{27 \pi}} \omega_p
	\sigma_z \left(\frac{\sigma_z^2}{\sigma_x^2} - 1 \right)^{3/2}
\end{equation}
which is attained at 
$k \sim \left(\frac{\sigma_z^2}{\sigma_x^2} - 1\right)^{1/2}
\omega_p / \sqrt{3}$.
This expansion is valid when $\frac{\sigma_z^2}{\sigma_x^2} - 1 \ll 1$. 
For a mediator appreciably lighter than the threshold mass
$\omega_p\left(\frac{\sigma_z^2}{\sigma_x^2} - 1\right)^{1/2}$,
the same $\omega_i$ expression will hold approximately.

Comparing these results to the electrostatic
instabilities we discussed previously, the mass range in
which instabilities existed for a two-stream velocity distribution
was $m \lesssim \omega/v_{\rm th}$.
Consequently, in a scenario such as a cluster collision, 
the magnetic instability is present for a smaller range
of mediator masses (though it can be present for other velocity
distributions where electrostatic instabilities do not exist).
When both instabilities are present, the 
parametric growth rate $\omega_i \sim \omega_p v/c$, as compared
to $\omega_i \sim \omega_p$ for an electrostatic instability,
means that magnetic instabilities grow slower in non-relativistic
plasmas.
This is illustrated in Figure~\ref{fig2s1},
for which the computation includes instabilities of all types.
While for some $k$ (e.g.\ $k \parallel x$) the fastest-growing
instability will be magnetic, the closing velocity
of $\sim 10^{-3}c$ means that these growth rates
are too small to be visible on the plot.


\section{Astrophysical systems}
\label{secastro}

\begin{figure}[t]
	\begin{center}
\includegraphics[width = 0.7\textwidth]{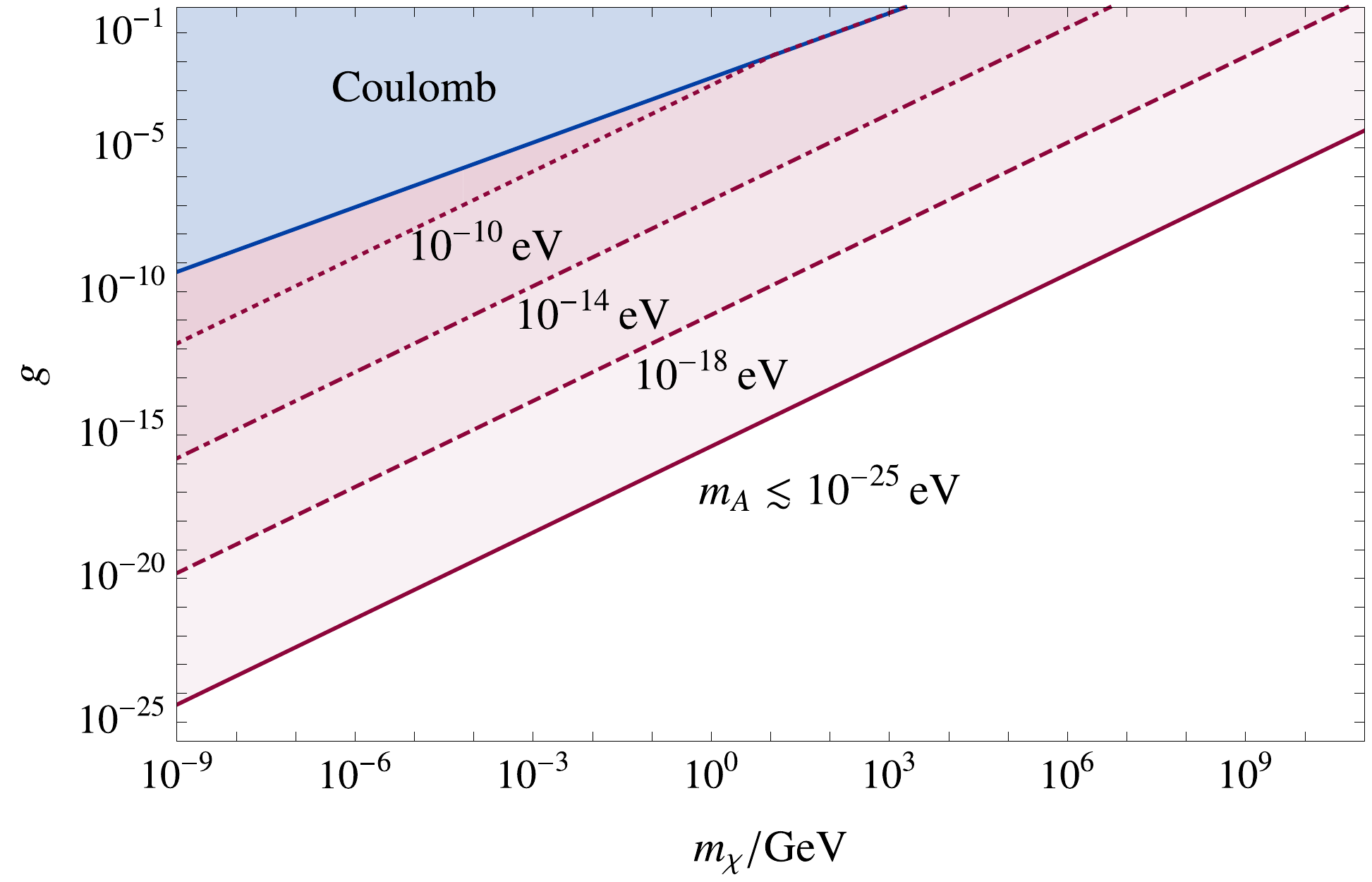}
		\caption{Estimated parameter space for which small perturbations
		would experience significant growth in astrophysical
		systems, for a vector mediator of mass $m_A$,
		coupling to DM particles of mass $m_\chi$
		with coupling $g$. 
		The shaded parameter space above the broken red lines, labelled
		with mediator mass $m_A$, is the space in which 
		perturbations would grow for that mediator mass.
		The region above the solid red line is the region
		in which perturbations would grow for sufficiently light
		mediators.
		The astrophysical situations we consider
		are cluster collisions such as 
		the Bullet Cluster, and dense DM subhalos (such as dwarf
		galaxies) moving through a larger halo (see Section~\ref{secastro}).
		The `Coulomb' region
		shows the parameter space in which $2 \rightarrow 2$ Coulomb
		collisions would have a significant impact in
		DM halos (see section~\ref{secexisting}).
		Note that, for DM masses $m_\chi \lesssim \eV$, the 
		DM occupation number in Galactic halos must be large,
		which may result in additional coherent
		scattering effects (Section~\ref{seclightbosonic}).}
\label{figmg}
	\end{center}
\end{figure}

As discussed in the Introduction, our goal in this paper
is not to fully calculate the consequences
of DM self-interactions. Instead, our aim
is to identify the parameter space in which
self-interactions are likely to be significant,
so that future work can investigate
this space more thoroughly, and in particular,
identify whether there are observational consequences.

To map out this parameter space, we want to identify
the astrophysical systems where, as we increase
the mediator coupling from zero, the first significant
effects arise. In this section, we identify different classes
of candidate systems --- cluster collisions, subhalos
within larger halos, and halos with
anisotropic velocity distributions --- and discuss their properties.

Cluster collisions, as observed in systems such as
the Bullet Cluster, are standard laboratories for
constraining DM self-interactions~\cite{Robertson_2016}.
The ionised gas in each cluster interacts during 
the collision, through SM plasma processes.
In constrast, the post-collision DM distribution,
as mapped out by gravitational lensing, 
is compatible
with no non-gravitational interactions~\cite{Clowe_2006,Robertson_2016}.

\begin{figure}[t]
	\begin{center}
\includegraphics[width = 0.8\textwidth]{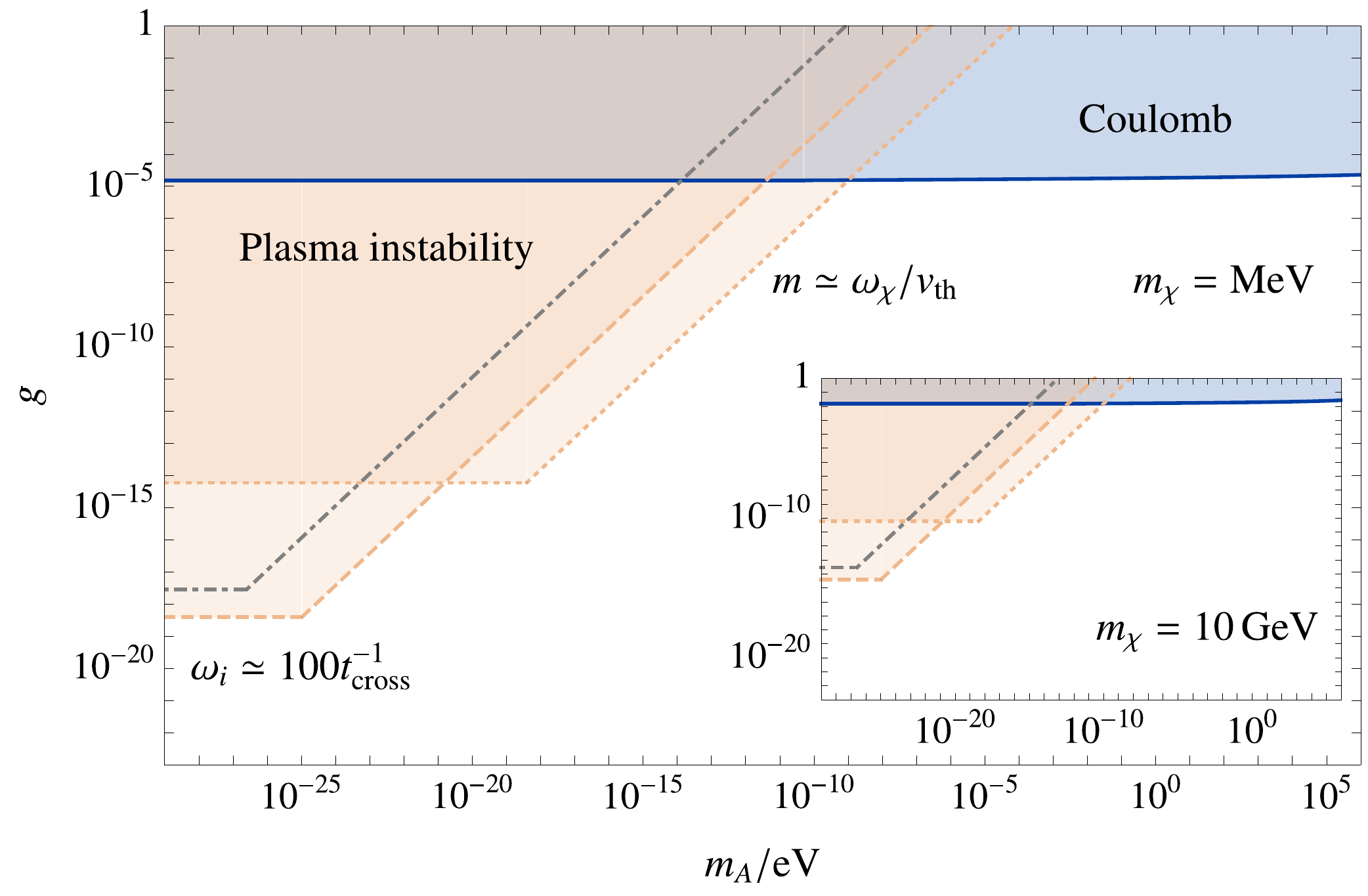}
		\caption{Estimated parameter space for which small
		perturbations would experience significant
		exponential growth in astrophysical systems,
		for a vector mediator of mass $m_A$, coupling 
		to DM particles of mass $m_\chi$ with coupling $g$.
		In this plot, we fix $m_\chi = \MeV$, and vary
		$m_A$ (the inset plot shows $m_\chi = 10 \GeV$).
		The dashed curve corresponds to the parameter
		space in which instabilities would grow
		in a Bullet Cluster-like collision, while the dotted
		curve corresponds to the parameter space in which 
		instabilities would grow for a dwarf-galaxy-like
		subhalo moving through a galactic halo,
		as discussed in Section~\ref{secastro}.
		The dot-dashed grey line shows the estimated instability
		region for magnetic instabilities, in halos with anisotropic
		velocity distributions.
		The horizontal parts of these curves are set
		by the growth rate of the instability being fast
		compared to the crossing time in the system,
		while the sloping parts are set by the mediator
		mass being small enough for instabilities to exist.
		 The `Coulomb' region
		shows the parameter space in which $2 \rightarrow 2$ Coulomb
		collisions would have a significant impact in
		DM halos (see section~\ref{secexisting}).
		The inset illustrates how these regions
		move to higher couplings as $m_\chi$ increases;
		the threshold $g$ values for the plasma instabilities
		increase $\propto m_\chi$, while the Coulomb
		threshold value increases approximately as $\propto m_\chi^{3/4}$.}
\label{figmag1}
	\end{center}
\end{figure}

To estimate when DM-DM plasma instabilities would
arise during such collisions, we need some model
for the DM distributions in the colliding clusters.
\cite{Paraficz_2016} estimates these distributions,
for the Bullet Cluster collision, using
gravitational lensing information. Assuming cored
DM density profiles\footnote{profiles with central
cusps, such as NFW, would lead to higher
central DM densities --- to be conservative, we consider cored profiles 
(gravitational lensing data does not have the spatial
resolution to distinguish these possibilities).}, their best-fit parameters have
core radii $\sim 100 \kpc$, and central
densities $\sim {\rm few} \times 0.1 \GeV \cm^{-3}$,
with inferred halo velocity dispersions of 
$\sigma \sim 800 {\rm km \, s^{-1}}$.
Simulations of the gas dynamics in the collision
suggest that the relative velocity of the halos
is $\sim 3000 \kms$\cite{Robertson_2016}.
These parameters give a central plasma frequency of
\begin{equation}
	\omega_p = \sqrt{\frac{g^2 \rho_\chi}{m_\chi^2}}
	\simeq 4 \times 10^{-7} \yr^{-1} \sqrt{\frac{\rho_\chi}{0.1 \GeV \cm^{-3}}} \frac{\GeV}{m_\chi} \frac{g}{10^{-17}}
\end{equation}
compared to a crossing time of $\sim 100 \kpc / (3000 \kms)
\sim 3 \times 10^7 {\rm \, yr}$.
If we take a conservative threshold of $\OO(100)$ $e$-folding 
times, we would expect plasma instabilities to grow
significantly for
\begin{equation}
	g \gtrsim 10^{-16} \frac{m_\chi}{\GeV}
	\label{eqg16}
\end{equation}
if the vector mediator is sufficiently light.
The latter condition should be met for 
\begin{equation}
	m \lesssim \frac{\omega_p}{\sigma}
	\simeq 3 \times 10^{-26} \eV
	\sqrt{\frac{\rho_\chi}{0.1 \GeV \cm^{-3}}}
	\frac{800 \kms}{\sigma}
\frac{\GeV}{m_\chi} \frac{g}{10^{-16}}
	\label{eqmcluster}
\end{equation}
The corresponding region in $m_\chi, g$ parameter space
is illustrated in Figure~\ref{figmg} (for various $m_A$),
and in $m_A, g$ parameter
space (for fixed $m_\chi$) in Figure~\ref{figmag1}.
Compared to constraints from $2 \rightarrow 2$
Coulomb scatterings, discussed in Section~\ref{secexisting},
plasma instabilities could be important at far smaller couplings,
for light enough mediators (though their observational
consequences need further investigation
--- see Section~\ref{secobscons}). 
From equations~\ref{eqg16} and~\ref{eqmcluster},
the threshold for the mediator mass to alter
the parameter space in which perturbations
grow is $m \gtrsim {\rm few} \times 10^{-26} \eV$, basically
independent of $m_\chi$.

For a two-stream instability,
the wavevector for the fastest-growing
mode (with a massless mediator) has $k \sim \frac{\omega_p}{v_0}
\sim {\rm few} \times \frac{\omega_i}{v_0}$.
For the threshold coupling we considered,
which has $\omega_i \sim \OO(100) / t_{\rm cross}
\sim \OO(100) v / R_{\rm cross}$,
this implies that
$k \sim \OO(100) / R_{\rm cross}$. Consequently,
the wavelength of the growing modes is small 
compared to the scale of the DM halo.
If the DM halo is smooth,
then our approximation of a spatially-uniform,
freely-falling plasma should be a reasonable one
(though more careful analysis, and potentially
simulations, would be necessary to confirm this).
For a massive mediator, the $k$ of the fastest-growing modes
is even higher.
On the other hand, if the DM halo has significant substructure, e.g.\
a dense central spike, or a large
number of `miniclusters', this may have important effects.

For more massive mediators, we are limited
by the $m \lesssim \omega_\chi/v_{\rm th}$ condition
on the existence of an instability.
To maximise the mass range we can cover,
we want DM systems with large $n_\chi$ and small $v_{\rm th}$.
Of the DM systems we have observational evidence
for, the most extreme values of both $n_\chi$ and $v_{\rm th}$ are obtained in small
subhalos of galactic halos, such as DM-dominated
dwarf galaxies. Even assuming cored dark matter
profiles, these can have central DM densities
of $\gtrsim 10 \GeV \cm^{-3}$ 
and velocity dispersion $\sim 10 \kms$ \cite{Walker_2009,Wolf_2010,Adams_2014}
(though it should be mentioned that some studies
claim that Milky Way dwarf galaxies
may contain no dark matter at all~\cite{Hammer_2019,Hammer_2020}).
These parameters correspond to a plasma frequency of 
\begin{equation}
	\omega_p  
	\simeq 4 \times 10^{-6} \yr^{-1} \sqrt{\frac{\rho_\chi}{10 \GeV \cm^{-3}}} \frac{\GeV}{m_\chi} \frac{g}{10^{-17}}
\end{equation}
and a maximum mediator mass
(see section~\ref{secbp}) of roughly
\begin{equation}
	m \lesssim \frac{\omega_p}{{\rm few} \times \sigma}
	\sim 10^{-23} \eV
	\sqrt{\frac{\rho_\chi}{10 \GeV \cm^{-3}}}
	\frac{10 \kms}{\sigma}
\frac{\GeV}{m_\chi} \frac{g}{10^{-16}}
\end{equation}
As illustrated in Figure~\ref{figmag1}, this means
that for large enough $g$, instabilities persist up to larger
$m$ than in the cluster collisions case.

The low density and high velocity dispersion of
the galactic halo in which the subhalo moves means
that we are in the (hot) beam-plasma scenario,
so the instability rate is suppressed by the density
contrast between the halos.
Taking a standard model for the Milky Way DM
halo (e.g.~\cite{Nesti_2013}), the density at
radii $\gg 10 \kpc$ is
\begin{equation}
	\rho_{\rm DM} \sim 0.1 \GeV \cm^{-3} \left(\frac{20 \kpc}{r}\right)^3 
\end{equation}
For typical dwarf galaxies at tens to hundreds of kpc from
the galactic centre, this gives a MW halo density of $\sim 10^{-2} - 10^{-3} \GeV \cm^{-3}$. Consequently, the maximum instability rate is
\begin{equation}
	\omega_i \sim 0.5 R \omega_p \sim 4 \times 10^{-6} {\rm yr^{-1}} \frac{R}{10^{-3}} \sqrt{\frac{\rho_\chi}{10 \GeV \cm^{-3}}} \frac{\GeV}{m_\chi} \frac{g}{10^{-14}}
	\label{eqwisub}
\end{equation}

In the case of cluster collisions, the available timescale
for perturbation growth is set by the crossing 
time of the DM halos (as noted above, further investigation
would be needed to understand the effect of spatial inhomogeneities
in these halos). For the case of a subhalo moving through a
larger halo, the crossing time (e.g.\ 
$\sim \kpc/10^{-3} \sim 4 \times 10^6 \yr$) is significantly smaller
than the timescale over which the subhalo's surroundings will
significantly change 
(e.g.\ $\sim 100 \kpc / 10^{-3} \sim 4 \times 10^8 \yr$).
Consequently, it is plausible that perturbations can build
up in the subhalo over timescales significantly longer than the crossing
time. This is compatible with the fact that the group velocity of growing electrostatic perturbations,
in the hot-beam case, is suppressed relative to the stream velocity,
since the real part of the frequency is set by the dispersion
relation for plasma oscillations .
Here, we leave a proper analysis to future work,
and conservatively restrict ourselves to perturbations
that grow substantially during a single crossing time.
With this assumption, equation~\ref{eqwisub} shows
that for very light mediators, cluster
collisions will be more prone to instabilities at smaller
couplings.

The unstable modes in a beam-plasma
setup have $k \sim \omega_p v_0$, and $\omega_i \sim R \omega_p$.
At the threshold we took,
$\omega_i \sim \OO(100) v_0 / R_{\rm cross}$,
so $k \sim R^{-1} \OO(100)/R_{\rm cross}$,
and the separation between the mode's wavelength
and the scale of the halo is larger than in the two-stream case.
For more massive mediators, the threshold coupling
is higher, and the fastest-growing modes have larger
$k$ (if a significant instability is present).

Cluster collisions and dense subhalos are
the most obvious examples of bimodal DM
velocity distributions, of the kind that
lead to electrostatic instabilities. 
For standard DM halos, the (virialised)
velocity distribution is expected to 
be roughly Maxwellian, so electrostatic
instabilities should not be present. However, 
as we reviewed in Section~\ref{secmagnetic},
for light mediators, magnetic instabilities
exist even for close-to-Maxwellian velocity distributions.
In fact,
numerical simulations indicate
that DM halos may, outside their central regions,
have quite anisotropic velocity distributions
\cite{Evans_2006,Wojtak_2008,Ludlow_2011,Lemze_2012,Sparre_2012,Wojtak_2013},
with the radial velocity dispersion
being $\sim 1.5$ times larger than the transverse ones.
This could lead to Weibel-type instabilities.

As per Section~\ref{secmagnetic}, the growth rates of magnetic instabilities
are velocity-suppressed, 
and the maximum mediator mass for which they exist is
$\sim \omega_p$ (for $\OO(1)$ velocity
anisotropies) rather
than $\sim \omega_p / v_{\rm th}$. 
Consequently, a basic estimate
of the parameter space in which they would arise
within galactic halos
gives a subset
of the electrostatic estimate,
as illustrated in Figure~\ref{figmag1}.
In addition, the evidence for
anisotropic DM distributions is somewhat indirect
\cite{Host_2008,Lemze_2011}, compared
to the evidence for bimodal velocity distributions
in subhalos and cluster collisions.
Nevertheless, the coupling threshold estimates for cluster collisions
and for anisotropic halos are quite close at 
small mediator masses, indicating that a more detailed
investigation is worthwhile.

Bound structures in the universe
most likely came about through a process
of hierarchial structure formation, in which smaller
structures formed first, and then merged
to form successively larger structures~\cite{Dodelson_2003}.
During this process, the dark matter velocity
distribution was often far from Maxwellian (e.g.\
during merger events), 
allowing instabilities to occur.
In some sense, the cluster mergers that we observe
are the late-time part of this process.
Generically, we might expect these later
events, which occur over longer timescales, to
allow more time for instabilities to grow,
and so to be the first that become important as we
increase the coupling from zero.
This is why we have focussed on such events
(along with the fact that we have observations
of them).
These arguments are, of course, schematic,
and would need to be checked more carefully.

Going further backwards in time, we can ask
whether DM-DM instabilities could be important
during the radiation era. We do not expect the DM
velocity distribution then to have been significantly bimodal,
but it could have been slightly anisotropic, raising the
possibility of magnetic instabilities. However, even being
optimistic, it does not seem likely that
these would be competitive with late-time
systems, in terms of possessing instabilities at small couplings.
Since $H \propto z^2$ then,
while $\omega_p \propto n_{\rm DM}^{1/2} \propto z^{3/2}$,
the increased instability rate possible in denser environments
is more than offset by the shorter instability time available,
and the most favourable ratios occur at later times.
Performing basic estimates at around the time of matter-radiation inequality,
it seems likely that the late-universe systems
we discussed above would allow instabilities
at smaller couplings. There is, of course, a separate
question of whether SM-DM couplings could lead to interesting
early-universe behaviour --- in particular,
the pressure in the baryon-photon fluid makes it behave
differently from DM, giving rise to a relative DM-baryon
bulk velocity~\cite{Dvorkin_2014}. We leave such investigations
to future work.

\subsection{Observational consequences}
\label{secobscons}

As we have emphasised, the instability regions
in Figure~\ref{figmag1} do not represent observational
constraints. Nevertheless, it is interesting to consider
how one might be able to obtain constraints,
at least in the close-to-threshold regime 
(that is, assuming that the relevant process is the first
time in the evolution of the system that collisions
become important, so that we can take the initial state
to be the same as for collisionless CDM).
For cluster collisions such as the Bullet Cluster,
if plasma instabilities lead to significant
momentum exchange, then we would expect
the DM density, as reconstructed by lensing, to be 
more similar to the collisional SM gas,
rather than the collisionless stars. Since, in reality,
the latter case is observed, we could place constraints on
such plasma instabilities.

A potential issue is
that gravitational lensing measurements
are compatible with $\OO(1)$ effects
on the DM halos, e.g.\ $\sim 30\%$ 
mass loss from the smaller halo \cite{Markevitch_2004}.
Consequently, the effects of plasma instabilities
would have to be not just $\OO(1)$, but
would have to introduce highly efficient momentum exchange,
in order to be constrained.
While our calculations have been purely at the level of small
perturbations, which can be treated using a linear
approximation, understanding the eventual consequences 
once perturbations become large would require a
more complete analysis. It should be noted that
the behaviour of SM matter in such collisions,
which does undergo significant momentum exchange due to plasma 
effects~\cite{2010Maxim}, and DM simulations such as \cite{Sepp:2016tfs},
suggest that the effects may be large enough to be visible.
However, establishing this carefully would require
significant extra work.

For the subhalo case, even if only a small fraction
of the stream kinetic energy and/or momentum is transferred
to the subhalo during one crossing, then the fact that the subhalo's lifetime 
of billions of years is much longer than the crossing time
(of only a few million years) means that there could be
large cumulative effects. Momentum transfer
would lead to an effective drag force on the DM;
this could modify the distribution of
SM matter bound to the halo~\cite{Kahlhoefer_2013,Kahlhoefer_2015},
and/or cause the subhalo's orbit to decay.
Kinetic energy transfer would heat up the subhalo,
expanding or eventually evaporating it~\cite{Kahlhoefer_2013}.
Due to these cumulative effects,
it may be the case that subhalos provide 
more robust signals of DM plasma instabilities (though again,
further work would be required to understand these).
While some of these effects have been considered in the context
of $2 \rightarrow 2$ DM scattering~\cite{Kahlhoefer_2013,Kahlhoefer_2015},
the coherent nature of plasma instabilities means
that, for light enough mediators, they may be important
at much smaller DM-mediator couplings.

It is possible that, if structure formation does
result in `cuspy' DM profiles with significantly higher
densities than we have considered here, then plasma
instabilities could be important
at even
smaller couplings. Relatedly, it is possible that the
first situations in which self-interactions are important
could occur earlier in time than our example scenarios,
and so change the initial conditions that we have been assuming
(though, as per the discussion of structure formation and the early
universe earlier in this section, there are reasons
to believe that our our scenarios might be the first
to become important as we increase the coupling from zero).
If one is attempting to identify the observational
signatures of a model, then such issues would
be need to be investigated carefully. 
Figure~\ref{figmag1} represents a conservative
estimate of the parameter space in which
such investigations may be warranted; as discussed above,
the true observationally-constrained region may reach
either smaller or larger couplings.

For couplings
significantly larger than the threshold value
where self-interaction effects first become significant, 
the DM's behaviour may be very complicated,
with self-interactions at earlier times affecting
the initial conditions for later processes.
In more complicated hidden sectors
than the simple model of symmetric DM that we have been
assuming, large enough self-interactions can be lead
to the formation of neutral bound states.
These `mirror atoms', in analogy to electron-nucleon
bound states in the SM, occur in a variety of models
(e.g.\ \cite{Hodges_1993,FOOT_2004,Kaplan_2010,Cyr_Racine_2013,Garc_a_Garc_a_2015}), and if the ionised fraction is small enough, would not display
the plasma behaviours we have investigated.
Of course, they would also have different behaviour to our simple model in other settings,
e.g.\ different scattering properties with the SM
in direct detection experiments, so they do not represent
a way to open up the same parameter space.


\section{Millicharged DM}
\label{secmillicharged}

So far, we have considered entirely hidden-sector
dynamics, with DM-SM interactions occurring only through
gravity. However, in some models, DM interacts 
directly with SM states. An important example 
is `millicharged' DM, which has a (small) charge
under SM electromagnetism.
In this case, the interaction dynamics are literally those of 
a usual EM plasma, with the addition
of a different particle species.
Millicharged particles can arise in a number of
ways~\cite{Davidson_2000,Foot_1993}
(it should be noted that models
incorporating an extra hidden-sector photon have
different behaviour, due to this extra mediator,
as we discuss in Section~\ref{secap}).

Coherent scattering effects for millicharged DM, in the form of 
DM particles scattering off large-scale coherent
EM fields, have been investigated extensively
in the literature~\cite{Chuzhoy_2009,McDermott_2011,Kadota:2016tqq,Hu_2017,Dunsky_2019,Stebbins:2019xjr}. These can be viewed as 
`one-many' scatterings, with a single DM
particle scattering off a field sourced by many SM particles, as opposed to the 
`many-many' scatterings we have been considering.
For example, \cite{Stebbins:2019xjr} analyses
scattering of millicharged DM with coherent magnetic fields
in the ISM, finding that if $q_{\rm DM} / m_{\rm DM} \gtrsim 10^{-13} \GeV^{-1}$,
the galactic disk would have been spun down by interactions
with the DM. \cite{Kadota:2016tqq} analyses the interactions
of DM
with magnetic fields in galaxy clusters, and finds a similar
coupling limit, above which the DM profile would be
strongly affected. While the plasma dynamics of these
scenarios are complicated, and such limits should be viewed as
plausible estimates rather than established constraints,
they illustrate that, even for rather small couplings,
one-many effects may be important.

An obvious question is whether, for even smaller
couplings, `many-many' scatterings of the kinds
we have been considering could be important.
For example, from Figure~\ref{figmag1}, 
for a massless hidden-sector mediator, and $m_\chi = \GeV$,
the electrostatic instability may be important in cluster collisions
for $g \gtrsim {\rm few} \times 10^{-16}$.
In a purely-DM environment, such effects would be
of interest. However, in most settings, the SM matter
cannot be neglected. In particular, since the DM charge
must be $\lll 1$ (for $m_\chi \ll 10^{13} \GeV$),
the relatively large electron charge $(e \simeq 0.3)$
means that SM matter will usually dominate the plasma
dynamics, even when the DM number density is higher.
As we discuss below, this can suppress plasma
instabilities (or makes them dominantly SM-sector,
so that DM scatters against SM-sourced fields
as per the `one-many' scenario considered above).

To take an example, we can consider
a beam-plasma scenario in which the 
high-velocity `beam' is DM-dominated,
and the denser `plasma' is SM-dominated.
This might arise when e.g.\ a bound structure
with DM and SM matter is passing through a DM
halo. From section~\ref{secbp}, the electrostatic instability
rate is $\omega_i \sim R \omega_p
\sim \frac{q^2 n_\chi}{m_\chi \omega_p}$, where
$q$ is the DM millicharge (the hot-beam limit
 is appropriate, as $R$ is very small).
Consequently, the instability rate is $\propto q^2$, 
rather than $\propto g$ as in the purely hidden-sector
case.\footnote{For magnetic instabilities, we are effectively
in the small-velocity-anisotropy regime, so the instability
rate should be $\propto R^{3/2}$.}
Physically, the SM matter makes the plasma `stiffer',
suppressing the instability.
The instability rate being suppressed by two
powers of the small DM charge means
that one-many scattering probes,
for which the momentum transfer rate 
is $\propto q$, are generally more powerful.

On the other hand, there may be other situations
in which collective DM effects are important. 
An obvious example would would be systems in which the SM
density is small enough. Another possibility is
that coherent SM fields provide a starting configuration
for which collective instabilities involving the DM
can arise.
\cite{Li_2020} investigates 
DM instabilities in supernova shock waves, finding 
that if the SM plasma is strongly magnetised, 
collective instabilities involving the DM can be important.
In order for this to happen, the effect of the initial
SM magnetic field on the DM trajectories must be significant,
and further investigation would be required to understand
whether such instabilities can occur in parameter
space regions which are not already constrained by one-many scattering
processes.


\section{Kinetically mixed mediator}
\label{secap}

Another important class of models are those
in which the hidden-sector mediator interacts
with the SM sector. Of these, the simplest
have the new vector interacting with 
a conserved SM current; interactions with a
non-conserved current lead to $({\rm energy}/\mbox{vector mass})^2$--enhanced rates for production of the vector's longitudinal mode~\cite{Schwartz_2013},
resulting in more complicated behaviour, and strong
constraints on light vectors~\cite{Dror_2017,Dror_2017_prl,Dror_2020}. The only
conserved currents in the SM are
$B-L$ (if neutrinos are Dirac) and EM. Couplings to $B-L$
result in a long-range fifth force between bulk matter,
so we restrict ourselves to an EM coupling.

If $\chi$ does not couple directly to the massless
SM photon (i.e.\ it does not have a millicharge),
then the Lagrangian for a `dark photon' mediator is
\begin{equation}
	\LL \supset - \frac{1}{4} F^2 - \frac{1}{4} F'^2 + \frac{1}{2} m^2 A'^2
	+ g A' J_\chi + J_{\rm EM} (A + \epsilon A')
	\label{eqdpl}
\end{equation}
This is equivalent to the standard `kinetic mixing' interaction
through the field redefinition $\hat A = A + \epsilon A'$,
giving (to leading order in $\epsilon \ll 1$)
\begin{equation}
	\LL \supset - \frac{1}{4} \hat F^2 - \frac{1}{4} F'^2 - \frac{1}{2} \epsilon \hat F F' + \frac{1}{2} m^2 A'^2
	+ g A' J_\chi + J_{\rm EM} \hat A
\end{equation}
(note the $\epsilon$ corrections simply modify the normalisations
of these terms, rather than introducing new ones).
Such an interaction can arise through integrating out
heavier matter that couples to both $A'$ and $\hat A$
\cite{Holdom_1986}.

\begin{figure}
	\begin{center}
\includegraphics[width = 0.7\textwidth]{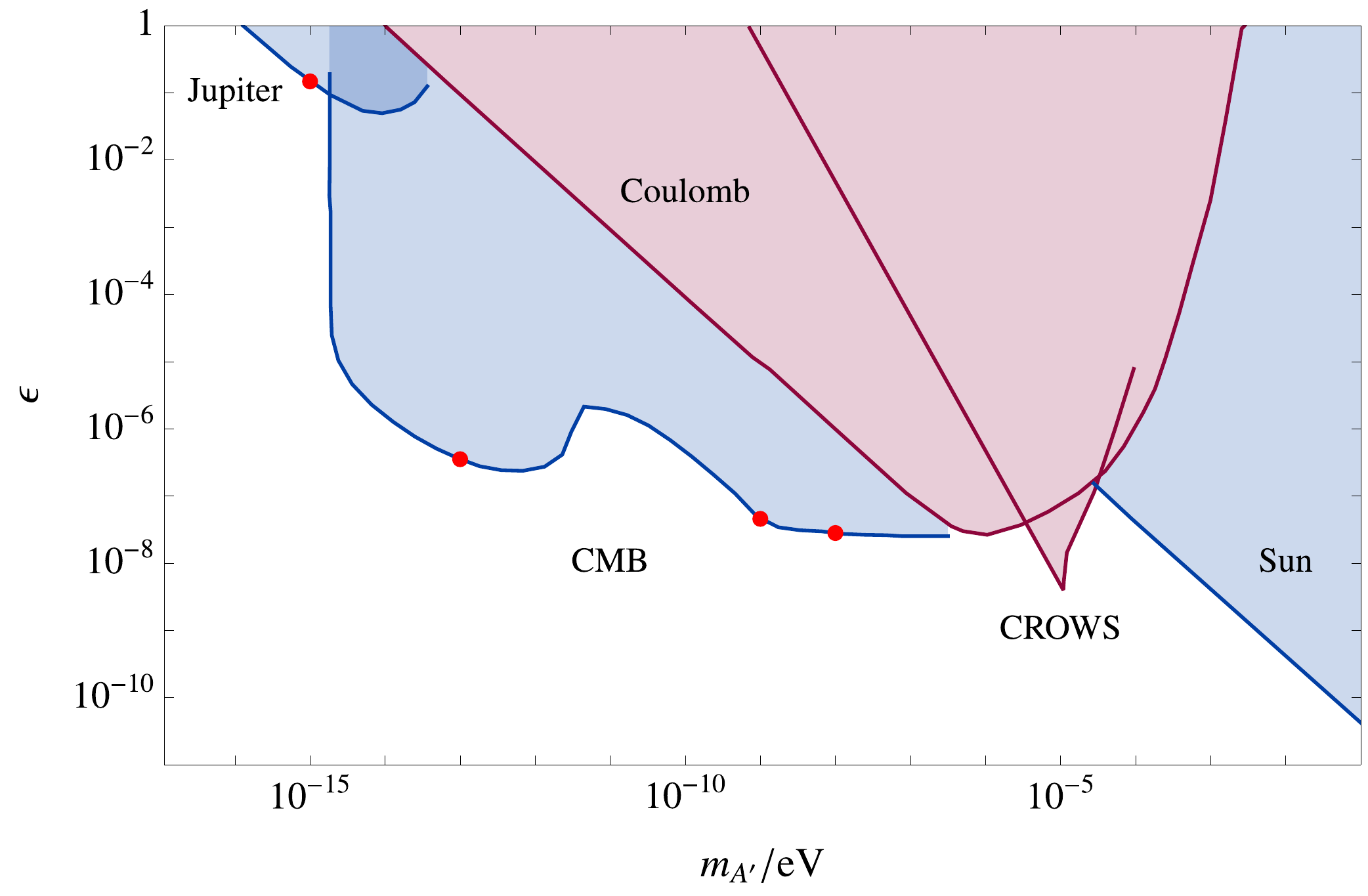}
\caption{Constraints on a dark photon of mass
		$m_{A'}$, possessing a kinetic mixing $\epsilon$
		with the SM photon. These constraints do not
		assume anything about the coupling of DM
		to the dark photon, relying only
		on SM processes (astrophysical constraints are plotted
		in blue, and laboratory constraints in red). 
		The constraints come from Jupiter's magnetic
		field~\cite{GOLDHABER_1971},
		CMB observations~\cite{Caputo_2020},
		laboratory tests of Coulomb's law~\cite{Williams_1971,Bartlett_1988},
		the CROWS microwave cavity experiment~\cite{Betz_2013},
		and stellar cooling bounds~\cite{Vinyoles_2015}.		The red dots correspond to the dark 
		photon masses illustrated in
		Figure~\ref{figqeff}.}
\label{figdp}
	\end{center}
\end{figure}

Another useful field redefinition is to define
the `active'/`sterile' basis, with $\tilde A = A + \epsilon A'$
the `active' field, and $\tilde A' = A' - \epsilon A$
the `sterile' field (again, we assume that $\epsilon \ll 1$,
and ignore normalisation changes). Then,
\begin{equation}
	\LL \supset - \frac{1}{4} F^2 - \frac{1}{4} F'^2 
	+ J_{\rm EM} \tilde A  
	+ g J_\chi (\tilde A' + \epsilon \tilde A)
	+ \frac{1}{2} m^2 \tilde A'^2 
	+ \epsilon m^2 \tilde A \tilde A'
	+ \frac{1}{2} \epsilon^2 m^2 \tilde A^2
	\label{eqactivesterile}
\end{equation}
If $m$ is small compared to the appropriate scales
of interest (e.g.\ the SM plasma mass, as we will discuss),
then the $m^2$ mass-mixing term is small,
and SM matter interacts only with $\tilde A$.
This illustrates why constraints on a dark photon
from SM processes decouple as $m \searrow 0$
(these constraints are shown in Figure~\ref{figdp}).

Kinetically mixed mediators are of considerable 
theoretical and phenomenological interest~\cite{1311review}. 
The kinetic mixing interaction can arise from arbitrarily
high scales~\cite{Holdom_1986}, and is generically present whenever
there is charged matter than interacts with the hidden
sector and SM.
Phenomenologically, the effectively medium-dependent
coupling of the dark photon to SM matter enables
models to access parameter space that would otherwise
be constrained~\cite{Knapen_2017}. This latter point will be important for
models targeted by low-threshold DM direct detection experiments,
as we discuss in section~\ref{secap}.

\subsection{Existing constraints}
\label{secexisting}

Some papers in the literature
(e.g.~\cite{Stebbins:2019xjr}) claim that the `one-many'
bounds from DM scattering off coherent SM fields,
as discussed in section~\ref{secmillicharged},
apply in the case of a `sufficiently light' massive dark photon.
However, unless the DM interactions directly
with the massless SM photon, the threshold
for `sufficiently light' is very small.
If the 
coupling is entirely through the kinetic
mixing term, then we can see from equation~\ref{eqdpl}
that $\chi$ interacts only with the massive $A'$
field, which decays exponentially at distances $\gtrsim m^{-1}$
from sources.
Consequently, if an SM magnetic field
is sourced by currents on a scale
$L \gg m^{-1}$, then the corresponding $B'$ 
field is suppressed by $\sim (m L)^{-1}$ compared
to $B$ (since it is dominantly sourced by currents 
within a range $\sim m^{-1}$).
For the galactic magnetic fields
considered in~\cite{Stebbins:2019xjr},
which have coherence length $\sim $ kpc, 
this suppression is $\sim 10^{-11} \epsilon \left(\frac{10^{-15} \eV}{m}\right)$,
where the nominal $m = 10^{-15} \eV$ value is small
enough to alleviate almost all constraints
on $\epsilon$ from SM processes (see Figure~\ref{figdp}).
This illustrates that `small' masses by particle physics
standards can nevertheless be high enough to very significantly
alleviate millicharge-type constraints.

The $2 \rightarrow 2$ DM-DM scattering constraints for a dark 
photon mediator can be estimated from modified versions
of the Coulomb scattering calculations
in~\cite{Ackerman:mha,Agrawal_2017}. 
Since we are most interested in these constraints
for $m \gtrsim \omega_p / v_{\rm th}$, where 
plasma instabilities are suppressed, 
the mediator's Compton wavelength will be smaller
than the Debye length, $\lambda_D \sim v_{\rm th} / \omega_p$. This means that the maximum impact parameter for 
Coulomb collisions is set by $m^{-1}$, rather 
than $\lambda_D^{-1}$, so the Coulomb logarithm is 
reduced compared to the massless mediator case.\footnote{In 
\cite{Agrawal_2017}, it is claimed that, for a 
two-species plasma with equal masses and opposite
charges, the appropriate maximum impact parameter
in the Coulomb logarithm is the interparticle
spacing $n^{-1/3}$, rather then $\lambda_D$ (in the case of a 
massless mediator), if $n^{-1/3} < \lambda_D$.
This does not agree with the results of standard
scattering integrals (e.g.~\cite{ll1981}), or analyses
of electron-positron plasmas~\cite{van_Erkelens_1981}.
Conversely, \cite{Ackerman:mha} does not take
into account collective effects at all, and takes
the maximum impact parameter to be set by
the Galactic radius. For a massless mediator,
the Coulomb logarithm will be set by $\lambda_D$,
between these two results.
}
There are a number of disagreements
between \cite{Ackerman:mha} and \cite{Agrawal_2017}
regarding the details of the Coulomb scattering
calculation; we will not attempt to resolve
these issues, but will use the (more conservative)
expressions from \cite{Agrawal_2017}, with a modified
Coulomb logarithm. This gives a relaxation
time, through Coulomb collisions, of 
\begin{equation}
	\tau_{\rm iso} = \frac{3}{16 \sqrt{\pi}}
	\frac{m_\chi^3 v^3}{\alpha_\chi^2 \rho_\chi} 
	\frac{1}{\log \Lambda_C}
	\label{eqtiso}
\end{equation}
where $\alpha_\chi \equiv g_\chi^2 / (4 \pi)$, and
$\Lambda_C \simeq \frac{m_\chi v^2}{\alpha_\chi m_{A'}}$
(for $\omega_\chi/v \lesssim m_{A'} \lesssim m_\chi/v$).
To avoid making DM too collisional, we need
$\tau_{\rm iso} \lesssim $ few Gyr in galaxies~\cite{Agrawal_2017},
which gives a bound
\begin{equation}
	g \lesssim 2 \times 10^{-5} \left(\frac{m_\chi}{\MeV}\right)^{3/4}
	\left(\frac{40}{\log \Lambda_C}\right)^{1/4}
	\label{eqcoulomb}
\end{equation}
where we have taken a typical value
of the Coulomb logarithm ($\Lambda_C$ will
depend on $g, m_\chi$ and $m_{A'}$, but
these effects will not be large).
As mentioned in Section~\ref{secastro}, 
it would require further analysis to determine
whether all couplings above this bound
are constrained, since strong enough 
self-interactions could lead to very complicated
behaviour. However, in the absence
of bound state formation (which, in
the symmetric DM model we are considering,
would lead to annihilation), it seems
likely that frequent collisions would
not be compatible with observations.

\subsection{Plasma instabilities}

In a situation with no charged SM matter, 
the growth of plasma instabilties would
be exactly as in the massive-mediator
case treated in Section~\ref{secinst}. To understand
how the presence of SM matter affects this,
we can analyse the coupled equations of motion for Fourier modes
of the vector fields,
\begin{equation}
	\left(
	\begin{pmatrix}
		\omega^2 - k^2 & 0 \\
		0 && \omega^2 - k^2 - m^2
	\end{pmatrix}
	- 
	\begin{pmatrix}
		\Pi && \epsilon \Pi \\
		\epsilon \Pi && \epsilon^2 \Pi
	\end{pmatrix}
	- 
	\begin{pmatrix}
		0 & 0 \\
		0 & \Pi_\chi
	\end{pmatrix}
	\right)
	\begin{pmatrix}
		A \\ A'
	\end{pmatrix}
	= 0
	\label{eqmat2}
\end{equation}
Here, $\Pi$ is the self-energy contribution from
charged SM matter, while $\Pi_\chi$ is from the DM
(for electrostatic oscillations, we can reduce
this six-dimensional system to a two-dimensional one, so that
the equations below become scalar equations).
For a null eigenvector $(A, A')$, and writing
$K^2 \equiv \omega^2 - k^2$, we have
$(K^2 - \Pi) A = \epsilon \Pi A'$; using
this in the $A'$
equation, we have
\begin{equation}
	(K^2 - \Pi) (K^2 - m^2 - \epsilon^2 \Pi - \Pi_\chi) A' = \epsilon^2 \Pi^2 A'
	\label{eqk2}
\end{equation}
where $K^2 \equiv \omega^2 - k^2$. In the DM-only case,
we have $(K^2 - m^2 - \Pi_\chi)A' = 0$. 
Using the ansatz $(K^2 - m^2 - \Pi_\chi) A' = 
(\epsilon^2 \Pi + G)A'$, equation~\ref{eqk2} becomes
\begin{equation}
	(K^2 - \Pi) G A' = \epsilon^2 \Pi^2 A'
	\quad \Rightarrow \quad G A' = (K^2 -\Pi)^{-1} \epsilon^2 \Pi^2 A'
	\label{eqg1}
\end{equation}
If the SM charge density is appreciable,
and the DM-mediator coupling is $g \ll 1$, then the SM plasma
self-energy may be much larger than the DM contribution $\Pi_\chi$.
If our solution is a small perturbation
of the DM-only case, so $K^2 A' \simeq (m^2 + \Pi_\chi)A'$,
then $\Pi$ being much larger than $m^2 + \Pi_\chi$
allows us to expand the $(K^2 - \Pi)^{-1}$
term in equation~\ref{eqg1} by treating $K^2$ as small, giving
\begin{equation}
	(K^2 - m^2 - \Pi_\chi)A' \simeq - \epsilon^2
	\left(
	K^2 + K^4 \Pi^{-1} + \dots
	\right) A'
	\label{eqexp1}
\end{equation}
The physical intuition behind this
is that the SM matter is `stiff', due to its
strong coupling to the photon, and so decouples from
the hidden sector dynamics.

\begin{figure}
	\begin{center}
\includegraphics[width = 0.6\textwidth]{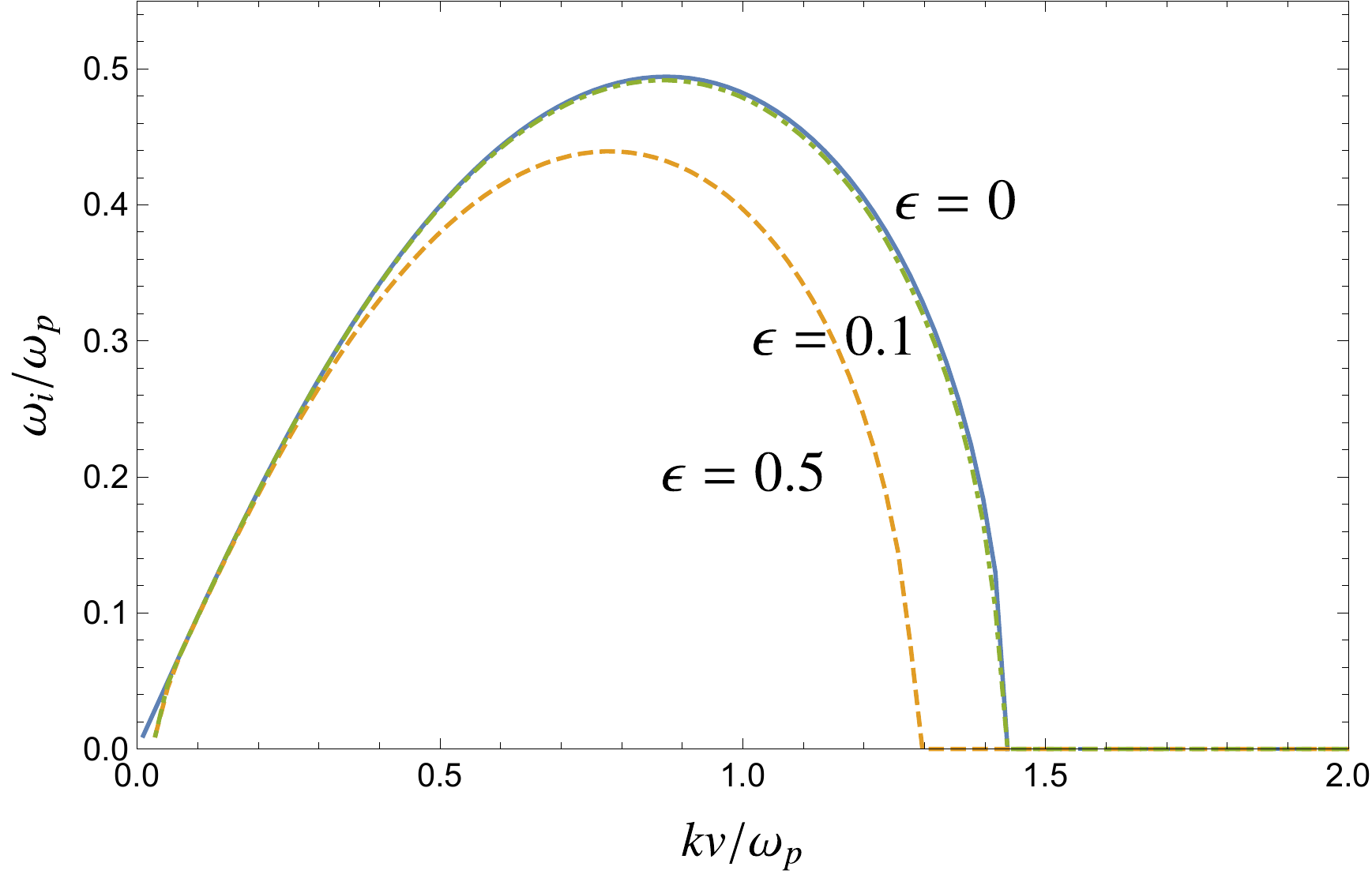}
	\caption{Longitudinal instability rate for the two-stream 
		velocity distribution from Figure~\ref{fig2s1}, for DM
		interacting via a light mediator that is kinetically
		mixed with the SM photon (Section~\ref{secap}). The SM plasma distribution
		is taken to be cold (for simplicity), with a plasma
		frequency 100 times that of the DM plasma,
		and a collision rate of $\sim 10 \omega_p$ (where
		$\omega_p$ is the plasma frequency of the DM plasma).
		The different curves show the instability rate for
		different kinetic mixings. As these illustrate, the
		effect of the SM plasma is small, despite its high
		density and large collision rate.}
\label{figtwomed}
	\end{center}
\end{figure}

In the opposite limit, where $\Pi$ can be treated
as small compared to $K^2$ in equation~\ref{eqg1},
we obtain
\begin{equation}
	(K^2 - m^2 - \Pi_\chi)A' \simeq \epsilon^2
	\left(
	\Pi + \Pi^2 / K^2 + \dots
	\right) A'
\end{equation}
Consequently, if $\Pi$ is small compared
to $m^2 + \Pi_\chi$, then we again obtain behaviour
close to the DM-only case.
Thus, in both of these scenarios, there
are modes whose behaviour is close to the DM-only case.
Even when the SM and DM plasma effects are comparable,
we can solve for the behaviour of the combined
system (equation~\ref{eqmat2}).
The upshot of this analysis is that,
in most scenarios,
DM-DM plasma instabilities are important
in the same parts of parameter
space as they would be if the mediator was
not kinetically mixed.
Figure~\ref{figtwomed} demonstrates this in a particular
situation, by showing the instability rate for
a two-stream DM velocity distribution, in the
presence of a denser, collisional SM plasma.
This illustrates that, even for fairly large $\epsilon$,
the main effect is to slightly rescale the single-mediator
instability behaviour, as per equation~\ref{eqexp1}.

\begin{figure}
\includegraphics[width = 0.5\textwidth]{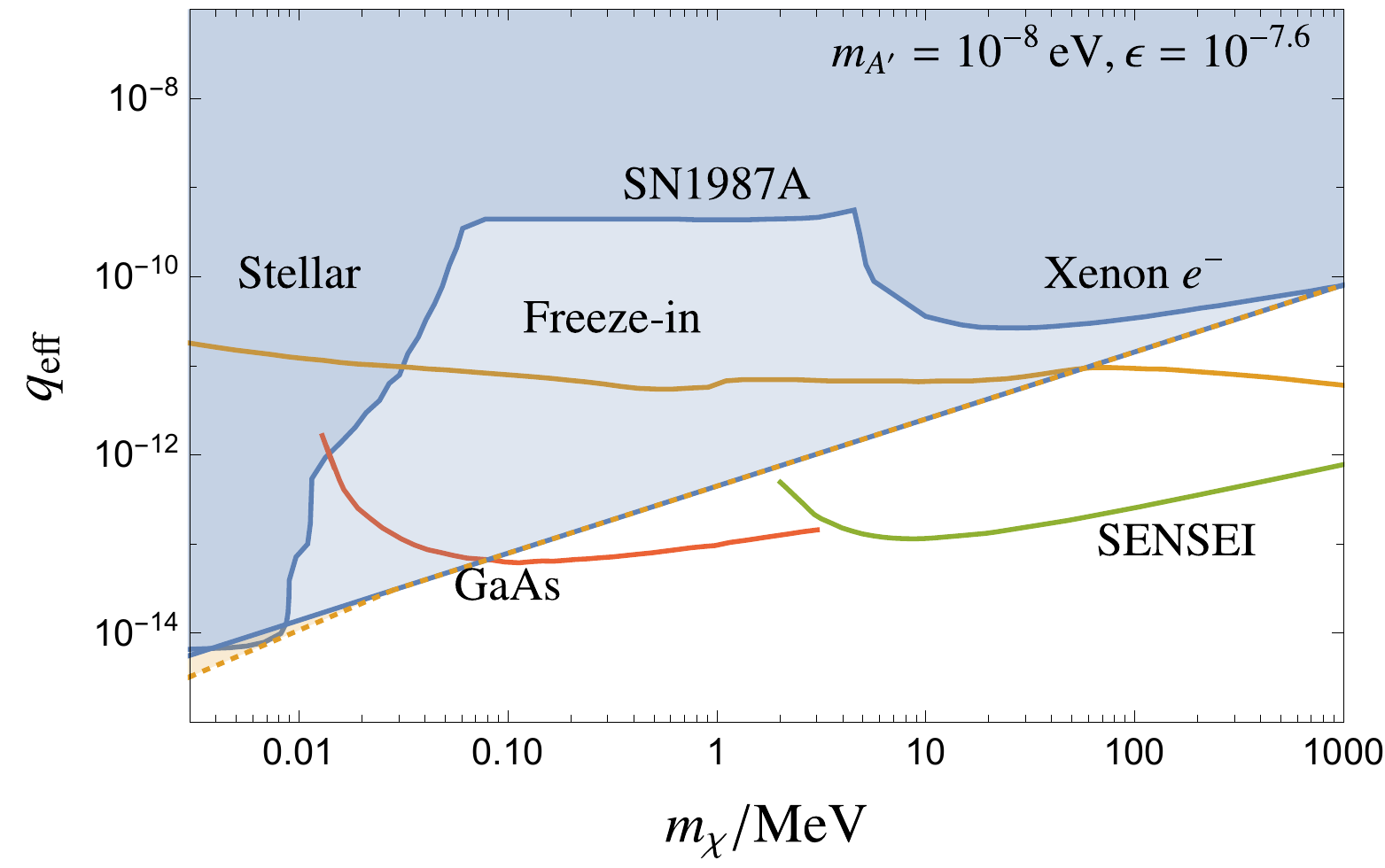}
\includegraphics[width = 0.5\textwidth]{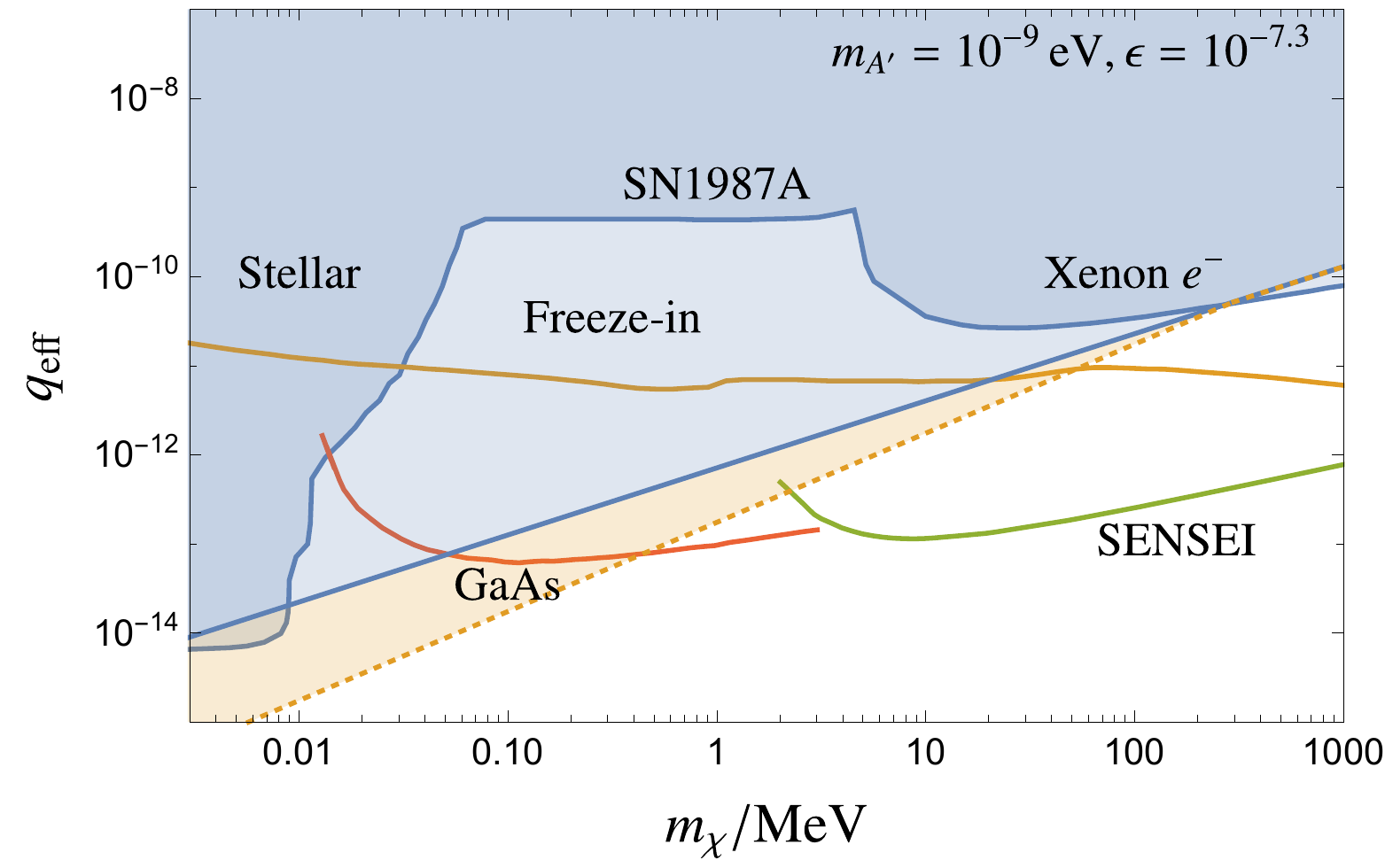}
\includegraphics[width = 0.5\textwidth]{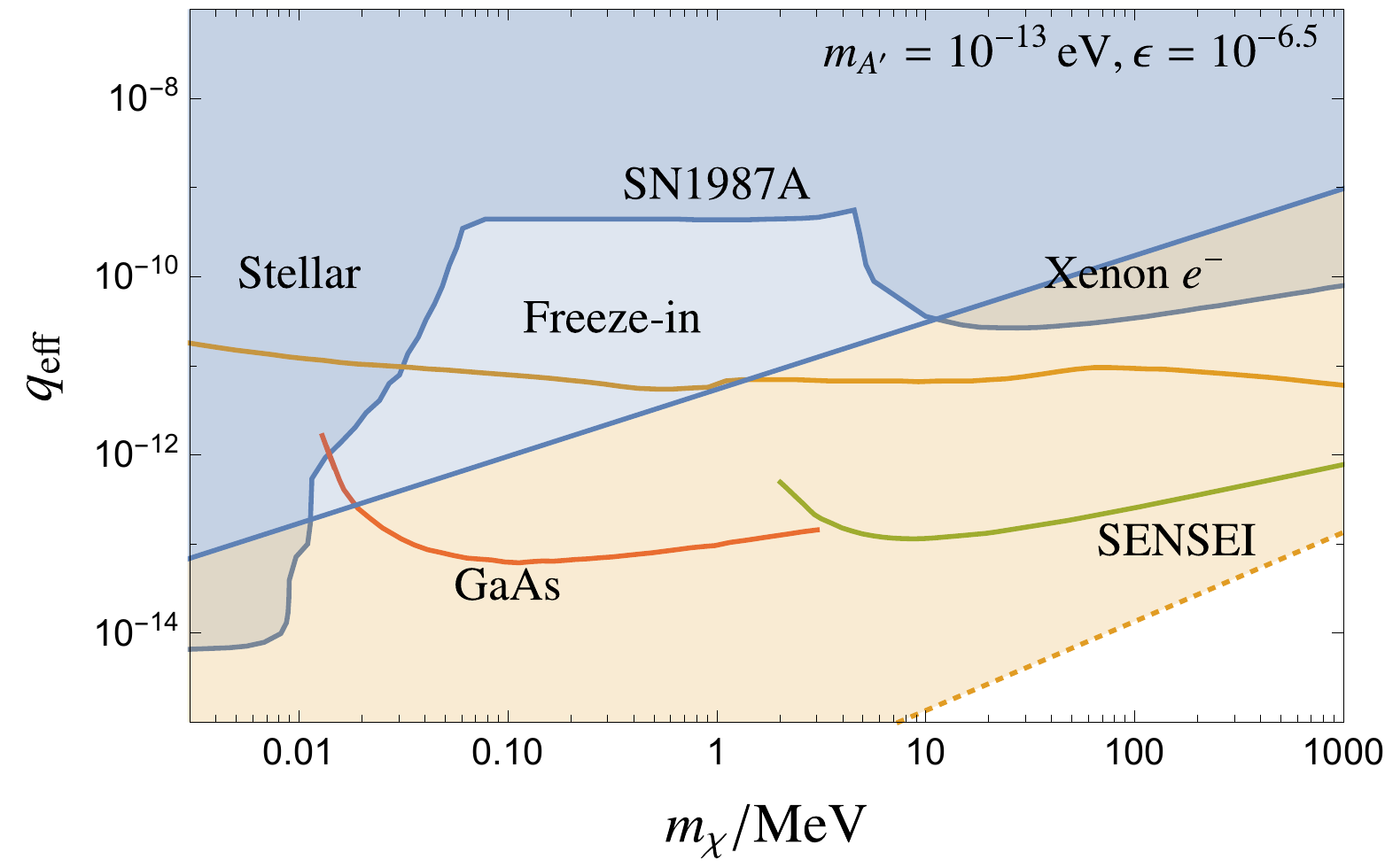}
\includegraphics[width = 0.5\textwidth]{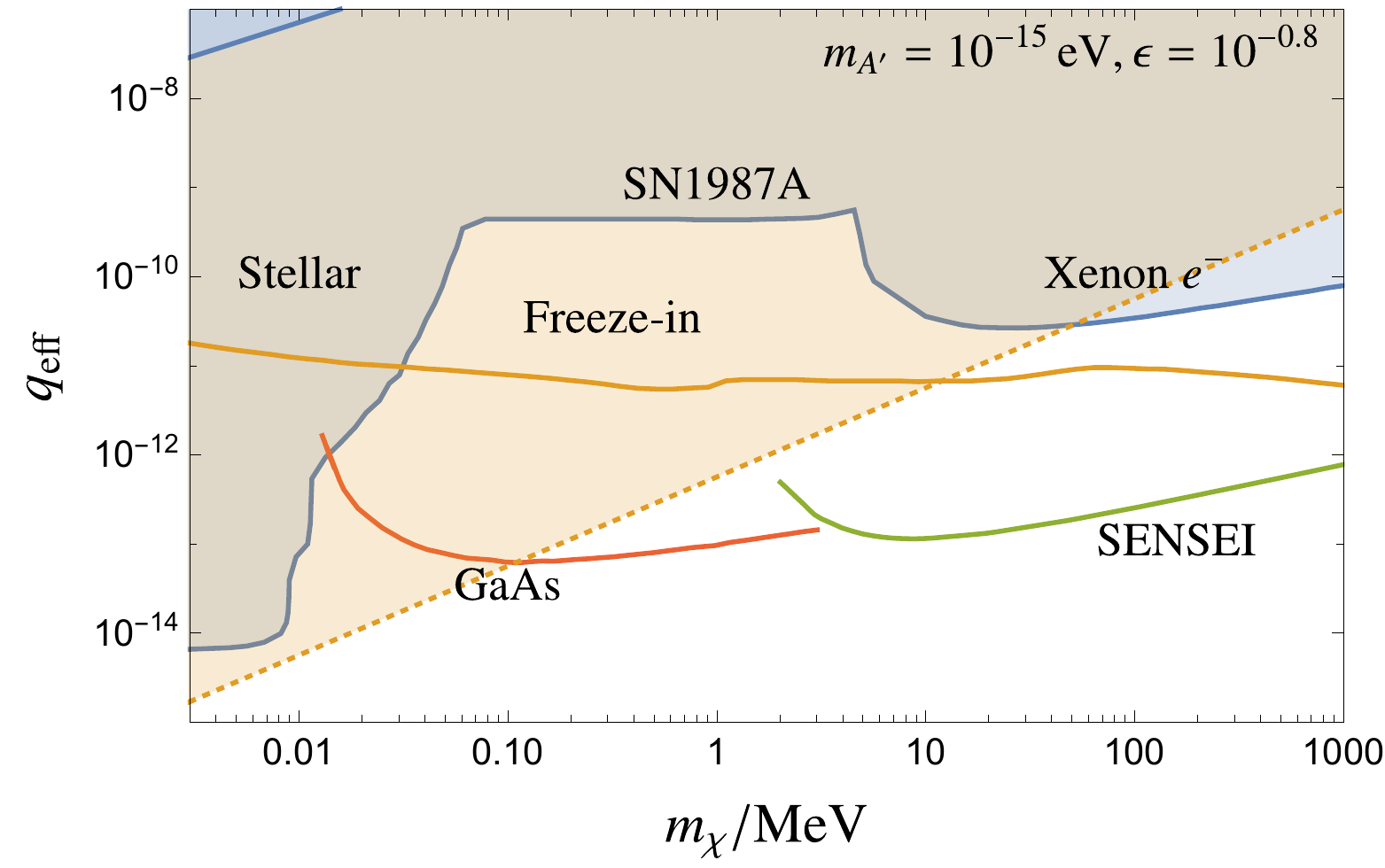}
	\caption{Constraints on the effective charge
	$q_{\rm eff} \equiv \epsilon g$ of a dark matter
	particle $\chi$ with mass $m_\chi$, coupling to a dark photon of mass
	$m_{A'}$ with coupling $g$, where the dark photon
	has a kinetic mixing $\epsilon$ with the SM photon.
	The dark blue shaded region corresponds to existing constraints
	from stellar cooling~\cite{Vogel_2014}, from SN1987A
	\cite{Chang_2018}, and from the
	Xenon10 experiment~\cite{Essig_2012,Essig_2017}.
	The `freeze-in' curve corresponds to
	the $q_{\rm eff}$ required for early-universe
	production from the SM radiation
	bath to produce the correct DM abundance~\cite{Dvorkin_2019,Essig_2012a}.
	The SENSEI~\cite{Crisler_2018,Abramoff_2019} and GaAs~\cite{Knapen_2018,Griffin_2018} lines correspond
	to projected sensitivites
	for future dark matter
	direct detection experiments.
	The light blue shaded regions correspond
	to parameter space in which $2 \rightarrow 2$ Coulomb
	scattering would have a significant effect in galactic
	halos (Section~\ref{secexisting}).
	The orange regions correspond to 
	parameter space in which
	dark-sector plasma instabilities would
	undergo significant exponential growth
	in astrophysical systems, as discussed in Sections~\ref{secastro}
	and~\ref{secap}. The different panels correspond to different
	dark photon masses; for each dark photon
	mass, the largest kinetic mixing parameter $\epsilon$
	consistent with SM constraints is assumed (illustrated
	in Figure~\ref{figdp}), allowing the smallest $g$ for a given $q_{\rm eff}$. We can see that, while the smaller $g$ permitted by smaller $m_{A'}$
	relaxes the $2 \rightarrow 2$ scattering constraints,
	the lighter mediator mass promotes plasma instabilities.}
\label{figqeff}
\end{figure}

\subsection{Dark matter parameter space}

For a direct detection experiment, if $m \ll \omega_p$
in the target material (or $m \ll L^{-1}$, where $L$ is the scale of
the conductive shielding around the experiment, etc),
then the interactions of a DM particle with the target can
be treated as occurring via the `active' $\tilde A$
state (equation~\ref{eqactivesterile}). Consequently,
the DM scatters as if it had an effective
charge $q_{\rm eff} = \epsilon g$. Processes such as the production of
$\bar \chi \chi$ pairs in stars, and in the early
universe, also scale with $\epsilon g$, so 
the effective DM parameter space, for given $m_{A'}$, 
is the $(m_\chi, q_{\rm eff})$ plane,
illustrated in the plots of Figure~\ref{figqeff}.

For a given scattering cross section
(set by $q_{\rm eff}$), we can decrease
the required $g$ by increasing $\epsilon$.
A major difference between a kinetically mixed mediator
and others (such as $B-L$ mediator,
or a scalar), is that we can
increase the SM coupling $\epsilon$ all the way up to $\OO(1)$ by
taking $m_{A'}$ small enough; as discussed above,
the constraints on $\epsilon$ from SM processes
decouple as $m \searrow 0$. These constraints are shown
in Figure~\ref{figdp}; 
for $m_{A'} \lesssim 10^{-16} \eV$, 
$\epsilon \sim \OO(1)$ is allowed.
Consequently, the $2 \rightarrow 2$
DM-DM scattering constraints discussed in Section~\ref{secexisting},
which are only weakly dependent on $m_{A'}$,
can be significantly relaxed, to the point where
they do not constrain parameter space. This
is illustrated quantitatively in Figure~\ref{figqeff},
where the different panels shows the effect
of decreasing $m_{A'}$ and increasing $\epsilon$.
As noted in Section~\ref{secexisting}, while the $m_{A'}$ required are small on
particle physics scales, they are still significantly
larger than the inverse length scales of astrophysical magnetic
fields, so such models evade millicharge-type constraints.
For other light mediators, strong constraints on the SM
coupling from e.g.\ stellar cooling, or fifth force
tests, significantly reduce the maximum
allowed scattering cross-section in laboratory experiments.
This generally results in near-term experiments having no
realistic prospects of detecting such DM models,
at least for DM masses $\lesssim 10 \MeV$~\cite{Knapen_2017}.

However, since we must go to rather small $m_{A'}$
to open up this parameter space, this raises
the possibility
that coherent DM-DM scattering processes become
important. Figure~\ref{figqeff} also shows
the parameter space regions in which plasma
instabilities will start growing in astrophysical systems.
Unlike the $2 \rightarrow 2$ scattering bounds, 
the enhancement of instabilities due to
decreased $m_{A'}$ compensates for the increased
$\epsilon$ that is possible, and much of the low-mass
target range is within the plasma instability
region, even for $\epsilon \sim \OO(1)$.
In particular, the `freeze-in' cross section~\cite{Dvorkin_2019,Essig_2012a},
for which thermal production of $\chi$ from the SM
radiation bath is sufficient to produce the whole
DM abundance, is within this regime up to $\OO(10) \MeV$.
Conversely, when $m_{A'}$ is large enough that plasma instabilities
would be unimportant, then $\epsilon$ is constrained to be
small, and 
the Coulomb scattering bounds constrain
much of the target parameter space for proposed
light DM detection experiments.

As we discussed in section~\ref{secastro}, these plasma instability regions
should not be viewed as observationally constrained,
since it is logically possible that the growth of perturbations
does not lead to observable effects. However, our calculations
do motivate more careful investigations into
whether these regions of parameter space
are astrophysically viable. Of course, even if
it turns out that such particles cannot be all
of the dark matter, it is possible that
$\OO(10\%)$ of the DM mass density is strongly collisional~\cite{Knapen_2017}.
To predict the signals of such subcomponent models at DM detection experiments,
the effects of this collisionality on the DM
distribution at Earth would need to be taken into account.


\section{Light bosonic DM}
\label{seclightbosonic}

In our discussions so far, we have assumed that dark
matter can be treated as point particles,
with some interparticle separation
$\sim n^{-1/3}$. However, when the de Broglie wavelength
of of the DM becomes comparable to $n^{-1/3}$,
i.e.\ the occupation number becomes $\gtrsim 1$,
this picture breaks down. For DM in the vicinity of the Earth,
this occurs at $m_{\rm DM} \lesssim \eV$.
While most experiments searching for DM scattering do
not cover such small masses, some types of
experiments~\cite{Berlin_2020}
and astrophysical observations~\cite{Caputo_2019} may have sensitivity there,
and in any case, understanding
the behaviour is of theoretical interest.

In the $m_{\rm DM} \ll \eV$ regime, we can treat
the (necessarily bosonic) DM as a large occupation number
classical field. If, before structure formation,
the particle and antiparticle fields have some independent
perturbations\footnote{for example, as would arise from
inflationary isocurvature fluctuations, or from small-scale
fluctuations arising from a phase transition production mechanism.
We leave the question of how large such fluctuations
need to be for future work.}, then (ignoring non-gravitational
interactions) the virialization process will
result in them becoming effectively independent Gaussian
random fields in halos, with characteristic wavelength 
$\lambda_B \sim 1/(m_{\rm DM} v)$.
If we consider a particular spatial point,
then the amplitude of the particle field
will generically be $\OO(1)$ different from that
of the antiparticle field. Supposing that the amplitude
of the particle field is larger, then 
there is a net positive charge density over a spatial
scale $\sim \lambda_B$. In charge density terms,
the particle and antiparticle fields result in
an array of charged `blobs', each of typical size
$\sim \lambda_B$, and typical charge 
$\sim \pm n_{\rm DM} \lambda_B^3$.

If $\lambda_B$ is significantly smaller than the
wavelength of perturbations that would grow
in a uniform plasma, then the smooth approximations that we used above
will be valid (c.f.\ the discussion around
equation~\ref{eqn20}). If not, then a modified
treatment would be required. For a two-stream velocity distribution,
and a light mediator,
the wavenumber of the most unstable perturbations
is $k \sim \omega_p / v$, so
at the threshold coupling from 
equation~\ref{eqg16}, and taking the nominal cluster
collision parameters from
Section~\ref{secastro}, 
\begin{align}
	&\lambda_B \ll k^{-1} 
	\quad \Leftrightarrow \quad \\
	&m_{\rm DM} \gg \frac{(g/m_{\rm DM}) \rho^{1/2}}{v^2}
	\sim 10^{-24} \eV \left(\frac{\rho}{0.1 \GeV \cm^{-3}}\right)^{1/2}
	\left(\frac{g/m_{\rm DM}}{10^{-16} \GeV^{-1}}\right)
	\left(\frac{3000 \kms}{v}\right)^2 \nonumber
\end{align}
Bearing in mind that the DM mass must be 
$\gtrsim 10^{-21} \eV$ to be compatible with small scale
structure observations~\cite{Hui_2017}, we see that for a light enough mediator,
significantly above-threshold couplings
would be required to affect the plasma instability behaviour.
For a heavier mediator, the perturbation wavenumber near 
the coupling threshold is $k \sim m_{A'}$, so
we need $m_{\rm DM} \gg m_{A'} / v$ there for the smooth plasma description
to be accurate.

For a heavy enough mediator and small enough coupling, plasma instabilities will be
suppressed. However, `collisional' effects on the plasma evolution
will still exist, and will no longer be well modelled as
$2 \rightarrow 2$ Coulomb collisions. Instead, 
`blob-blob' collisions will occur coherently (this is in direct
analogy to the picture of gravitational relaxation for light bosonic
DM~\cite{Hui_2017,Levkov_2018,Bar_Or_2019}). Treating scattering off a blob via standard
scattering theory~\cite{Sakurai}, the scattering rate
(for a single DM particle) will be parametrically set by
\begin{equation}
	\Gamma \sim \sigma v n_B
	\sim \frac{g^4 N_B^2}{16 \pi m_{\rm DM}^2 v^4 } v n_B\log \Lambda_C
	\sim \frac{\log \Lambda_C}{16 \pi} \frac{g^4 \rho^2}{m_{\rm DM}^7 v^6}
	\label{eqbc1}
\end{equation}
where $N_B \sim n_{\rm DM} \lambda_B^3$ is the number of particles in a blob,
$n_B \sim \lambda_B^{-3}$ is the number density of blobs, and
$\log \Lambda_C$ indicates the appropriate
Coulomb logarithm.
This form will be valid if $m_{A'} \lesssim m_{\rm DM} v$, so
that the mediator is long-range compared to the blob size.\footnote{
If the mediator is short-range compared to the de Broglie
wavelength, then the behaviour will be as per a contact interaction,
giving
\begin{equation}
	\Gamma 
	\sim \frac{1}{16 \pi} \frac{g^4 \rho^2}{m_{A'}^4 m^3 v^2}
\end{equation}
(the constant factor here should be not taken seriously).
	Similar parametric behaviour would occur for e.g.\ a scalar
	of mass $m$
	interacting via a quartic interaction $\lambda \varphi^4$, which
	would give
\begin{equation}
	\Gamma \sim 
	\frac{1}{16\pi} \frac{\lambda^2 \rho^2}{m^7 v^2}
	\sim 
	\frac{1}{16\pi} \frac{\rho^2}{f^4 m^3 v^2}
\end{equation}
where we have used the usual scaling of the quartic
coupling $\lambda \sim m^2/f^2$
in terms of the symmetry breaking scale $f$,
for a potential arising from high-scale
breaking of a shift symmetry.
This matches the parametric relaxation timescale
from~\cite{Semikoz_1995,Sikivie_2009}.
	}
Equation~\ref{eqbc1} has the same density and velocity dependence
as for gravitational relaxation of light bosonic DM, but 
differs from the $2 \rightarrow 2$ Coulomb collisions
rate (equation~\ref{eqtiso}), which may result in stronger constraints
than the latter at small $m_{\rm DM}$.

As an example, we can consider parameters motivated by
the experiment proposed in~\cite{Berlin_2020}. If we take
$m_{\rm DM} = 10^{-4} \eV$, so that the de Broglie wavelength is not
larger than the detector, and $m_{A'} = 10^{-7} \eV$, giving 
a mediator range on the same scale, then for
$g \lesssim 5 \times 10^{-14}$, astrophysical plasma
instabilities are suppressed (according to our estimates
in Section~\ref{secastro}).
If we naively applied the $2 \rightarrow 2$ Coulomb constraints
from Section~\ref{secexisting}, these would give 
$g \lesssim 6 \times 10^{-13}$. However, if we take the
estimate of the coherent scattering rate from equation
\ref{eqbc1}, and set $\Gamma \sim 1/{\rm Gyr}$ as a rough constraint,
this would give 
\begin{equation}
	g \lesssim 4 \times 10^{-17} \left(\frac{\GeV \cm^{-3}}{\rho}\right)^{1/2}
	\left(\frac{v}{10^{-3}}\right)^{3/2}
\end{equation}
Consequently, the coherence-enhanced scattering rate is 
more significant than plasma instabilities in this regime,
and potentially more constraining than a naive application of the $2 \rightarrow 2$
Coulomb scattering bounds.

Despite these rates, it is unclear whether coherent scattering
processes of these kinds will provide observational constraints.
One issue is that, if the scatterings drive
the particle and antiparticle fields towards identical spatial
profiles, then we will no longer have charged blobs.
Consequently, it is not obvious whether the
scattering process will saturate before it has
observable effects on halo shapes. 

For light enough mediators ($m_{A'} \lesssim m_{\rm DM} v^2$), it would also be necessary to check
whether coherent emission of the mediator
is ever important. While particle-antiparticle
annihilation to the mediator is possible when $2 m_{\rm DM} \ge m_{A'}$,
the rate of this process is $\propto g^4$,
so would likely be unimportant (there will be a rate $\propto g^2$
from `off-shell' parts of the DM field's wavefunction,
due to it being bound in a gravitational potential,
but these will be extremely small in a typical halo).

These caveats illustrate that more investigation would
be needed to understand the observational consequences
of light bosonic DM interacting via a light
vector mediator. However, our estimates in this section
do suggest that coherent effects may be important in
new regions of parameter space.


\section{Conclusions}

As we have emphasised throughout, the goal of
our paper has not been to constrain models of
self-interacting DM, but to map out parameter
space regions in which long-range coherent
self-interactions may be important. The obvious
follow-on would be to more carefully investigate the
physics of these models, and to determine whether
they are observationally constrained.
One of the most
important models is the kinetically mixed
mediator scenario, for which it would be very interesting
to determine whether the models
being targeted by proposed DM direct detection
experiments~\cite{Hochberg_2016,Hochberg_2017,Hochberg_2018,Crisler_2018,Knapen_2018,Berlin_2020} are astrophysically viable.

Even for purely hidden-sector interactions,
such investigations may be quite complicated.
In principle, one would like to start from 
the very early universe (e.g.\ from when the DM
was produced) and track the DM's evolution until today.
Without this, one would not know the initial DM distribution
in scenarios such as cluster mergers, so unless
one could demonstrate that all observationally-acceptable
initial conditions behave in a certain way,
relating the model to observations would be difficult.
Of course, in practice, it would hopefully be possible
to make significant approximations, e.g.\ that self-interactions
have negligible effects before a certain point, or conversely,
that they effectively thermalise the distribution.
SM plasma behaviour may provide some guide, at least in the
regime of a light mediator and strong coupling.

As well as understanding the constraints that could be obtained
from plasma instabilities, it would be useful to understand
the observational signatures that could arise,
in models with a coupling that is just large enough.
The discussion in Section~\ref{secobscons} represents 
a first attempt in this direction.
In particular, understanding the relationships between 
observational signatures in different astrophysical systems
would be important.
For DM that is not super-heavy, such threshold-level couplings
would need to be extremely small --- however, there
are potential theoretical reasons why such couplings
might arise~\cite{Craig_2019}.

There are many dark matter models
which feature long-range vector mediators,
beyond the simple examples we have considered.
The most obvious examples are models
with additional species, having different charges
and/or masses~\cite{Feng_2009,Garc_a_Garc_a_2015} (such as the proton and electron
in the SM). 
If most of the present-day dark matter is ionized (i.e.\ consists
of isolated charged particles), then analogues of the plasma
calculations we have carried out here will apply,
though the presence of different-mass species
will change the behaviour in potentially-interesting
ways (c.f.\ ion effects in SM plasmas).
On the other hand, if dark matter is dominantly in the form
of overall-neutral bound states (`dark atoms'),
then the interactions between these will be effectively
short-ranged, and will not give rise to plasma
instabilities.

As mentioned in Section~\ref{secastro}, it may be observationally
viable for a $\sim \OO(10\%)$ subcomponent of DM
to be strongly self-interacting~\cite{Knapen_2017}. At the threshold
of viability, there may be interesting observational effects
of long-range interactions in such a subcomponent.
In section~\ref{secprev}, we briefly reviewed
\cite{Heikinheimo:2015kra,Sepp:2016tfs,Heikinheimo:2017meg},
which simulated the dynamics of a DM subcomponent interacting
via a massless hidden-sector mediator.
Other questions, including the DM distribution at
Earth of a subcomponent~\cite{Clarke_2016}, would also be interesting
to consider, and relevant to the direct detection
phenomenology of such a subcomponent.


\begin{acknowledgments}
	We acknowledge helpful conversations with
	Prateek Agrawal,
	Asimina Arvanitaki,
	Daniel Egana-Ugrinovic,
	Peter Graham,
	Tongyan Lin, and
	Julian Munoz, and thank
	Jung-Tsung Li and Tongyan Lin for
	comments on a draft of this paper.
	RL's research is supported in part by the National Science Foundation under Grant No.~PHYS-1720397, and the Gordon and Betty Moore Foundation Grant GBMF7946.
\end{acknowledgments}

\appendix

\section{Maxwellian velocity distribution}
\label{appMaxwell}

Virialised DM halos are expected to have approximately
Maxwellian velocity dispersions,
and the dispersion relations for these
can be treated analytically.
For a Maxwellian velocity distribution 
\begin{equation}
	f(v) = \frac{1}{(2 \pi \sigma^2)^{3/2}} e^{-v^2 / (2 \sigma^2)}
\end{equation}
we have
\begin{equation}
	\Pi A = \begin{pmatrix}
		- \omega_p^2 \frac{\omega}{\sqrt{2} k \sigma} Z\left(
		\frac{\omega}{\sqrt{2} k \sigma}\right) & 0 \\ 
		0 & - (\omega^2 - k^2) \frac{\omega_p^2}{k^2 \sigma^2}
		\frac{\omega}{\sqrt{2} \sigma k} \left( Z\left(
		\frac{\omega}{\sqrt{2} k \sigma}\right) +
		\sqrt{2} \frac{k \sigma}{\omega}\right)
	\end{pmatrix}
	\begin{pmatrix}
		A_\perp \\ A_\parallel
	\end{pmatrix}
	\label{eqmaxwell1}
\end{equation}
where $Z$ is the plasma dispersion function~\cite{fitzpatrick2015plasma}.
This gives the usual small-$k$ expansion of the non-relativistic
dispersion relations~\cite{Raffelt_1996}.
To find the response function for a velocity distribution
with non-zero mean velocity,
we can we perform the appropriate Lorentz boost.

Another relevant limit of the Maxwellian
self-energy is the $\omega \rightarrow 0$ limit, 
for which we have 
\begin{equation}
	\Pi A \rightarrow \begin{pmatrix}
		0 & 0 \\ 0 & \omega_p^2/\sigma^2
	\end{pmatrix} \begin{pmatrix} A_\perp \\ 
		A_\parallel
	\end{pmatrix}
\end{equation}
This shows that static magnetic fields are not screened,
while electric fields are screened over the Debye length
scale $\lambda_D \sim (\omega_p/\sigma)^{-1}$.


\section{Penrose stability criteria}
\label{appPenrose}

In this appendix, we review the Penrose
stability criteria~\cite{Penrose_1960} for electrostatic plasma
oscillations, and extend them to the case of a massive
vector mediator.

From section~\ref{secelectro}, the dispersion
relation for an electrostatic perturbation is
\begin{equation}
	(c^2 k^2 + c^4 m^2) \simeq c^2 \sum_s \frac{q_s^2}{m_s} \int d^3 v
	\frac{k \cdot \partial_v f_s}{k \cdot v - \omega}
\end{equation}
Considering a single species for simplicity (or equivalently,
symmetric positive and negative charges),
we can define $v_k = \hat k \cdot v$, and
$g(v_k) = \int d^2 v_\perp f(v)/n$, where
$v_\perp$ is the velocity component perpendicular to
$k$ (so $\int dv_k g(v_k) = 1$). Then, we have
\begin{equation}
	(k^2 + m^2 c^2) \simeq \omega_p^2 \int dv_k
	\frac{g'(v_k)}{v_k - \omega/k}
\end{equation}
where $\omega_p^2 = \sum_s q_s^2 n_s / m_s$.
If we define $G(u) \equiv \int dv_k \frac{g'(v_k)}{v_k - u}$,
then the dispersion relation is $G(\omega/k) = (k^2 + m^2 c^2) / \omega_p^2$.
If there is a solution where $\omega$ has positive imaginary
part, then that mode will be unstable to growth.
For a massless mediator, it can be shown that the existence of such a growing mode 
is equivalent to there existing some real $u$ for which
$G(u)$ is real and $\ge 0$ (effectively, if there are growing
modes, then there must be a marginally
stable mode).

\begin{figure}
	\begin{center}
\includegraphics[width = 0.6\textwidth]{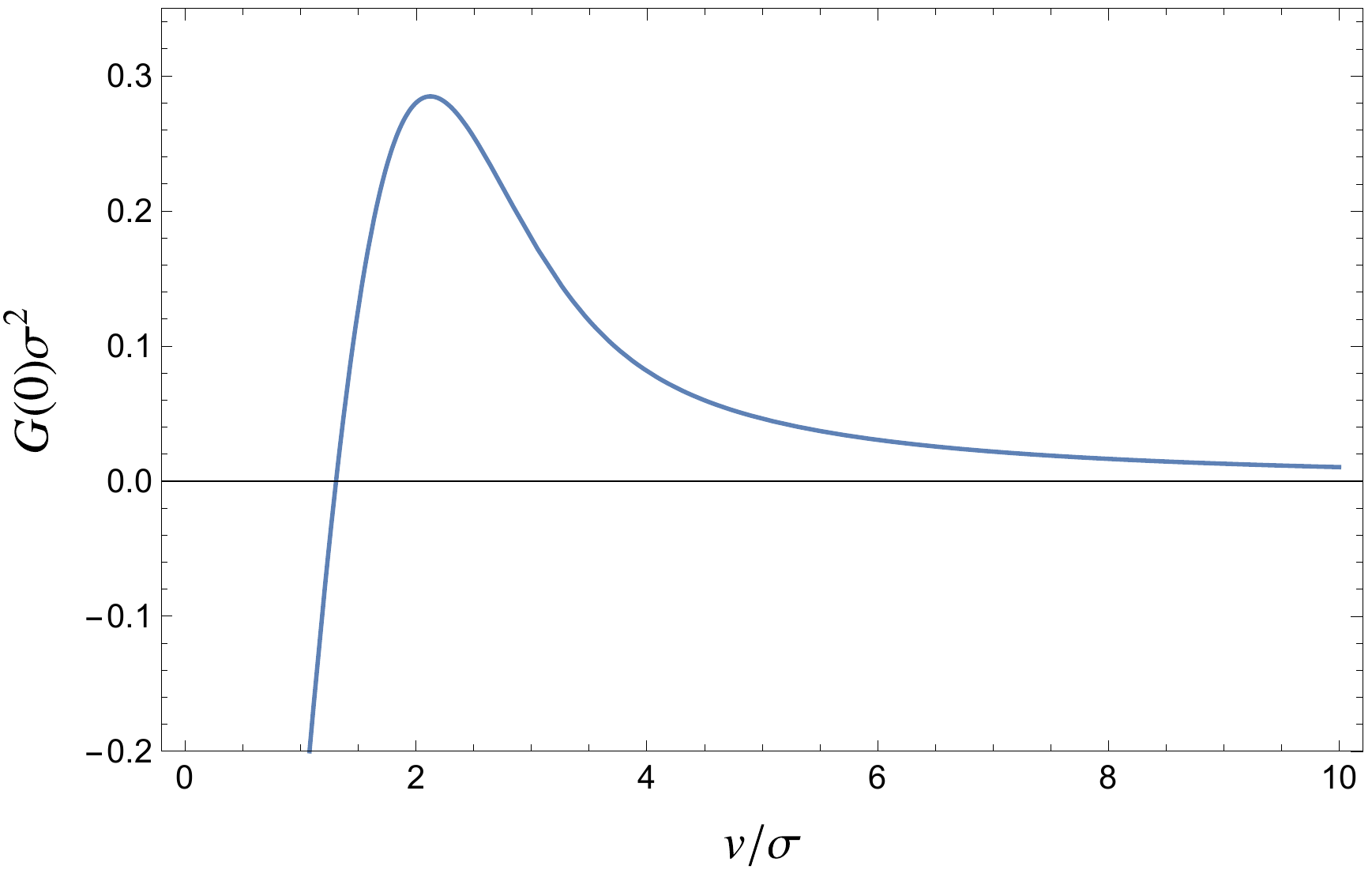}
	\caption{Normalised longitudinal self-energy for a 
		two-Gaussian-stream velocity distribution,
		evaluated at zero phase velocity (see equation~\ref{eqdawson}). As discussed 
		in appendix~\ref{appPenrose}, this function is related to
		the existence of unstable longitudinal modes.}
\label{figdawson}
	\end{center}
\end{figure}

Since
\begin{equation}
	G(u + i 0_+) = P \int dv_k \frac{g'(v_k)}{v_k - u} + i \pi g'(u)
\end{equation}
to have $G(u) \ge 0$, we need $g'(u) = 0$. 
For the example of a symmetric `two-stream' velocity distribution,
where each stream is Maxwellian with velocity distribution
$\sigma$, and the closing velocity is $2 v$, we have
\begin{equation}
	G(0 + i0_+) = 
	\frac{-1 + 2 \frac{v}{\sqrt{2} \sigma}
	D \left(\frac{v}{\sqrt{2} \sigma}\right)}{\sigma^2}
	\label{eqdawson}
\end{equation}
where $D$ is the Dawson integral~\cite{abramowitz_stegun}.
 This function is
is plotted in Figure~\ref{figdawson}. For a massless
mediator, an unstable mode exists
if $G(0 + i 0_+) \ge 0$, which is true for $v \gtrsim 1.3 \sigma$.

For a massive mediator, the dispersion relation
is
\begin{equation}
	\frac{k^2 + c^2 m^2}{\omega_p^2} = G(\omega/k)
\end{equation}
By the same logic as in the massless case,
an instability exists if there is some real
$u$ for which $G(u) - c^2 m^2 / \omega_p^2$ is real
and non-negative. From Figure~\ref{figdawson},
we can see that this will be true only
for some bounded range of $v/\sigma$, if at all.
If $m^2 c^2 / \omega_p^2 
\ge \max G(0 + i 0_+) \simeq 0.3/\sigma^2$, then 
there are no solutions, and consequently no instability.
In terms of the notation in Section~\ref{secelectro},
where $\omega_p$ labelled the plasma frequency for each
stream, rather than for both streams, 
we need $m^2 c^2 / (2 \omega_p^2) \ge G(0 + i 0_+)$.

\begin{figure}
	\begin{center}
\includegraphics[width = 0.6\textwidth]{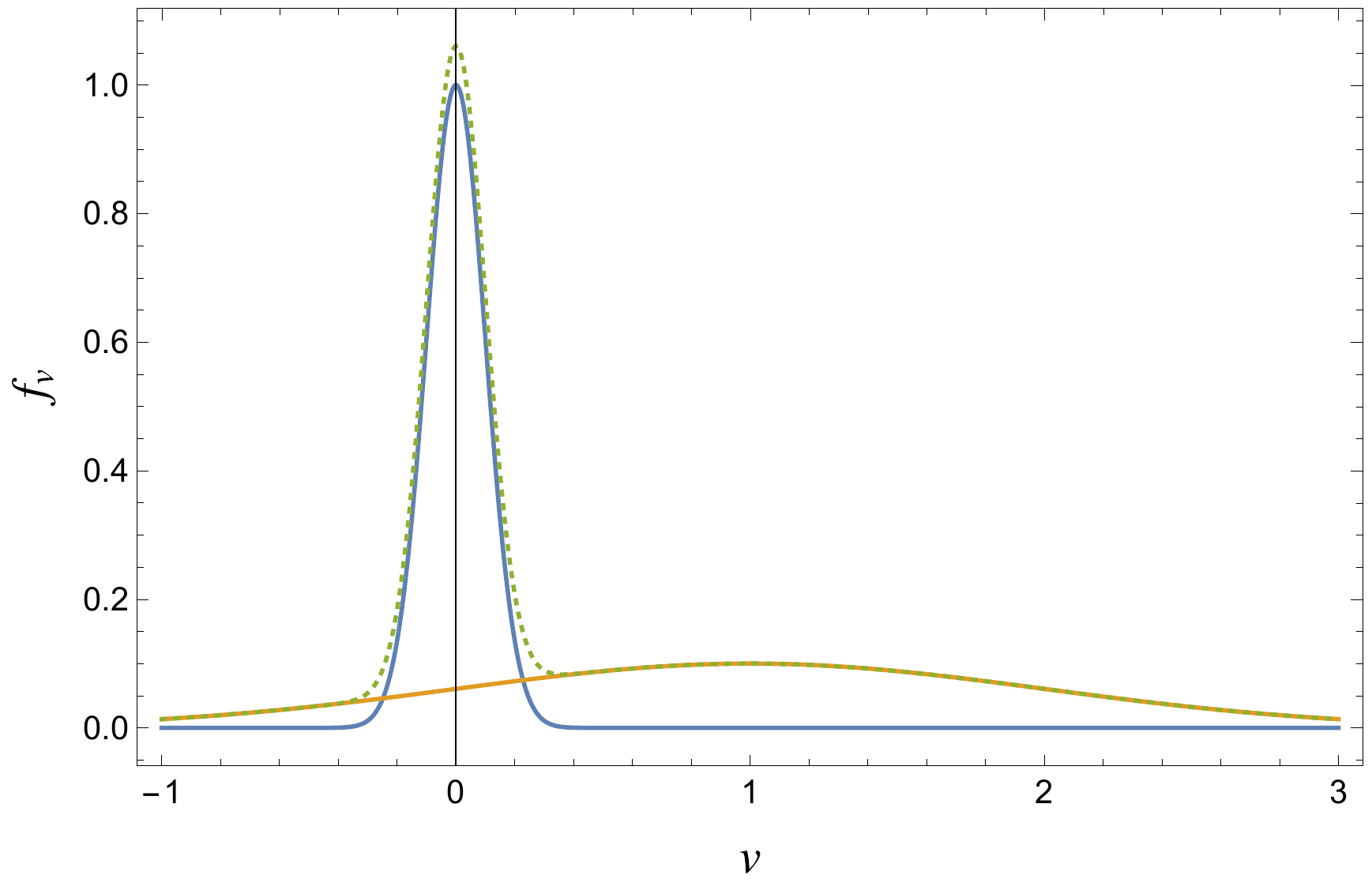}
	\caption{Velocity distribution for a `beam-plasma'
		scenario, in which the beam velocity dispersion
		is $\sigma_s = 1$, the beam velocity is $v_s = 1$,
		the plasma velocity dispersion is $\sigma_p = 0.1$,
		and the density ratio is $R = 0.1$. The blue curve
		shows the velocity distribution of the `plasma'
		stream,
		the orange curve that of the `beam', and the dotted
		curve the combined velocity distribution.}
\label{figveldist}
	\end{center}
\end{figure}

Less symmetric velocity distributions are also
of interest. An example is the case of a dense,
low-velocity-dispersion stream passing through 
a lower-density, higher-velocity-dispersion
stream --- the `beam-plasma' scenario discussed 
in Section~\ref{secbp}.
This kind of velocity distribution is illustrated
in Figure~\ref{figveldist}.
If both of the streams have Maxwellian velocity distributions,
and we take the denser stream to be at rest, so
\begin{equation}
	g(v_k) \simeq \frac{1}{\sqrt{2 \pi \sigma_p^2}}
	e^{-v_k^2/(2 \sigma_p^2)}
	+ \frac{R}{\sqrt{2\pi\sigma_s^2}} e^{-(v_k-v_s)^2/(2 \sigma_s^2)}
\end{equation}
with $R \ll 1$, then the $u$ for which $g'(u) = 0$ 
satisfies
\begin{equation}
	u^2 \simeq 2 \sigma_p^2 \log \frac{u \sigma_s^3}{v_s \sigma_p^3 R} + v_s^2 \frac{
		\sigma_p^2}{\sigma_s^2}
\end{equation}
(where we have assumed that $v_s \gg v$, which is true if
$\sigma_s \gg \sigma_p$). If the beam velocity distribution
is broad, so $v_s \sim \sigma_s$, then we have
$u \sim C \sigma_p$, where $C$ is at most logarithmically large.
The integral for $G(u + i 0_+)$ is then dominated by the
`plasma' peak near the origin, which (via integration by
parts) is
\begin{equation}
	G(u + i 0_+)
	= P \int dv_k \frac{g'(v_k)}{v_k - u}
	= \int dv_k \frac{g(v_k) - g(u)}{(v_k - u)^2}
	\simeq \frac{1}{u^2} \sim \frac{1}{C^2 \sigma_p^2}
\end{equation}
Consequently, for a massive mediator, an instability exists 
if 
\begin{equation}
	G(u + i 0_+) \ge \frac{m^2 c^2}{\omega_p^2}
	\quad \Leftrightarrow \quad
	m^2 \lesssim \frac{\omega_p^2}{C^2 \sigma_p^2}
\end{equation}
for some $C \sim \OO(10)$. 


\bibliography{dmplasma}

\providecommand{\href}[2]{#2}\begingroup\raggedright\begin{thebibliography}{100}

\bibitem{Clowe_2006}
D.~Clowe, M.~Brada{\v{c}}, A.~H. Gonzalez, M.~Markevitch, S.~W. Randall,
  C.~Jones et~al., \emph{A direct empirical proof of the existence of dark
  matter}, \href{http://dx.doi.org/10.1086/508162}{\emph{The Astrophysical
  Journal} {\bf 648} (Aug, 2006) L109--L113}.

\bibitem{Robertson_2016}
A.~Robertson, R.~Massey and V.~Eke, \emph{What does the bullet cluster tell us
  about self-interacting dark matter?},
  \href{http://dx.doi.org/10.1093/mnras/stw2670}{\emph{Monthly Notices of the
  Royal Astronomical Society} {\bf 465} (Oct, 2016) 569--587}.

\bibitem{Kahlhoefer_2013}
F.~Kahlhoefer, K.~Schmidt-Hoberg, M.~T. Frandsen and S.~Sarkar, \emph{Colliding
  clusters and dark matter self-interactions},
  \href{http://dx.doi.org/10.1093/mnras/stt2097}{\emph{Monthly Notices of the
  Royal Astronomical Society} {\bf 437} (Nov, 2013) 2865--2881}.

\bibitem{Chacko_2006}
Z.~Chacko, H.-S. Goh and R.~Harnik, \emph{Natural electroweak breaking from a
  mirror symmetry},
  \href{http://dx.doi.org/10.1103/physrevlett.96.231802}{\emph{Physical Review
  Letters} {\bf 96} (Jun, 2006) }.

\bibitem{Garc_a_Garc_a_2015}
I.~G. Garc{\'{\i}}a, R.~Lasenby and J.~March-Russell, \emph{Twin higgs
  asymmetric dark matter},
  \href{http://dx.doi.org/10.1103/physrevlett.115.121801}{\emph{Physical Review
  Letters} {\bf 115} (Sep, 2015) }.

\bibitem{Damour_1990}
T.~Damour, G.~W. Gibbons and C.~Gundlach, \emph{Dark matter, time-{varyingG},
  and a dilaton field},
  \href{http://dx.doi.org/10.1103/physrevlett.64.123}{\emph{Physical Review
  Letters} {\bf 64} (jan, 1990) 123--126}.

\bibitem{Gradwohl_1992}
B.-A. Gradwohl and J.~A. Frieman, \emph{Dark matter, long-range forces, and
  large-scale structure}, \href{http://dx.doi.org/10.1086/171865}{\emph{The
  Astrophysical Journal} {\bf 398} (Oct, 1992) 407}.

\bibitem{Nusser_2005}
A.~Nusser, S.~S. Gubser and P.~J.~E. Peebles, \emph{Structure formation with a
  long-range scalar dark matter interaction},
  \href{http://dx.doi.org/10.1103/physrevd.71.083505}{\emph{Physical Review D}
  {\bf 71} (Apr, 2005) }.

\bibitem{Kesden_2006}
M.~Kesden and M.~Kamionkowski, \emph{Tidal tails test the equivalence principle
  in the dark-matter sector},
  \href{http://dx.doi.org/10.1103/physrevd.74.083007}{\emph{Physical Review D}
  {\bf 74} (Oct, 2006) }.

\bibitem{Hellwing_2009}
W.~A. Hellwing and R.~Juszkiewicz, \emph{Dark matter gravitational clustering
  with a long-range scalar interaction},
  \href{http://dx.doi.org/10.1103/physrevd.80.083522}{\emph{Physical Review D}
  {\bf 80} (Oct, 2009) }.

\bibitem{Comelli_2012}
D.~Comelli, M.~Crisostomi and L.~Pilo, \emph{Perturbations in massive gravity
  cosmology}, \href{http://dx.doi.org/10.1007/jhep06(2012)085}{\emph{Journal of
  High Energy Physics} {\bf 2012} (jun, 2012) }.

\bibitem{Chiuderi_2015}
C.~Chiuderi and M.~Velli, \emph{Basics of Plasma Astrophysics}.
\newblock Springer Milan, 2015,
  \href{http://dx.doi.org/10.1007/978-88-470-5280-2}{10.1007/978-88-470-5280-2}.

\bibitem{Ackerman:mha}
L.~Ackerman, M.~R. Buckley, S.~M. Carroll and M.~Kamionkowski, \emph{{Dark
  Matter and Dark Radiation}},
  \href{http://dx.doi.org/10.1103/PhysRevD.79.023519,
  10.1142/9789814293792_0021}{\emph{Phys. Rev.} {\bf D79} (2009) 023519},
  [\href{http://arxiv.org/abs/0810.5126}{{\tt 0810.5126}}].

\bibitem{Chen_2015}
F.~Chen, \emph{Introduction to Plasma Physics and Controlled Fusion}.
\newblock Springer International Publishing, 2015.

\bibitem{Briggs_71}
R.~J. {Briggs}, \emph{{Two-Stream Instabilities}}, {\emph{Advances in Plasma
  Physics} {\bf 4} (Jan., 1971) 43}.

\bibitem{Paraficz_2016}
D.~Paraficz, J.-P. Kneib, J.~Richard, A.~Morandi, M.~Limousin, E.~Jullo et~al.,
  \emph{The bullet cluster at its best: weighing stars, gas, and dark matter},
  \href{http://dx.doi.org/10.1051/0004-6361/201527959}{\emph{Astronomy {\&}
  Astrophysics} {\bf 594} (Oct, 2016) A121}.

\bibitem{Mardon}
B.~Feldstein, J.~Mardon and L.~O'Silva, ``A (nearly) weaker-than-gravity bound
  on dark matter electromagnetism.''

\bibitem{Heikinheimo:2015kra}
M.~Heikinheimo, M.~Raidal, C.~Spethmann and H.~Veermäe, \emph{{Dark matter
  self-interactions via collisionless shocks in cluster mergers}},
  \href{http://dx.doi.org/10.1016/j.physletb.2015.08.012}{\emph{Phys. Lett.}
  {\bf B749} (2015) 236--241}, [\href{http://arxiv.org/abs/1504.04371}{{\tt
  1504.04371}}].

\bibitem{Sepp:2016tfs}
C.~Spethmann, H.~Veermäe, T.~Sepp, M.~Heikinheimo, B.~Deshev, A.~Hektor
  et~al., \emph{{Simulations of Galaxy Cluster Collisions with a Dark Plasma
  Component}},
  \href{http://dx.doi.org/10.1051/0004-6361/201731299}{\emph{Astron.
  Astrophys.} {\bf 608} (2017) A125},
  [\href{http://arxiv.org/abs/1603.07324}{{\tt 1603.07324}}].

\bibitem{Stebbins:2019xjr}
A.~Stebbins and G.~Krnjaic, \emph{{New Limits on Charged Dark Matter from
  Large-Scale Coherent Magnetic Fields}},
  \href{http://dx.doi.org/10.1088/1475-7516/2019/12/003}{\emph{JCAP} {\bf 1912}
  (2019) 003}, [\href{http://arxiv.org/abs/1908.05275}{{\tt 1908.05275}}].

\bibitem{Li_2020}
J.-T. Li and T.~Lin, \emph{Dynamics of millicharged dark matter in supernova
  remnants},
  \href{http://dx.doi.org/10.1103/physrevd.101.103034}{\emph{Physical Review D}
  {\bf 101} (May, 2020) }.

\bibitem{Dai_2009}
D.-C. Dai, K.~Freese and D.~Stojkovic, \emph{Constraints on dark matter
  particles charged under a hidden gauge group from primordial black holes},
  \href{http://dx.doi.org/10.1088/1475-7516/2009/06/023}{\emph{Journal of
  Cosmology and Astroparticle Physics} {\bf 2009} (Jun, 2009) 023--023}.

\bibitem{Feng_2009}
J.~L. Feng, M.~Kaplinghat, H.~Tu and H.-B. Yu, \emph{Hidden charged dark
  matter}, \href{http://dx.doi.org/10.1088/1475-7516/2009/07/004}{\emph{Journal
  of Cosmology and Astroparticle Physics} {\bf 2009} (Jul, 2009) 004--004}.

\bibitem{Agrawal_2017}
P.~Agrawal, F.-Y. Cyr-Racine, L.~Randall and J.~Scholtz, \emph{Make dark matter
  charged again},
  \href{http://dx.doi.org/10.1088/1475-7516/2017/05/022}{\emph{Journal of
  Cosmology and Astroparticle Physics} {\bf 2017} (May, 2017) 022--022}.

\bibitem{Garny_2019}
M.~Garny, A.~Palessandro, M.~Sandora and M.~S. Sloth, \emph{Charged planckian
  interacting dark matter},
  \href{http://dx.doi.org/10.1088/1475-7516/2019/01/021}{\emph{Journal of
  Cosmology and Astroparticle Physics} {\bf 2019} (Jan, 2019) 021--021}.

\bibitem{Clarke_2016}
J.~Clarke and R.~Foot, \emph{Plasma dark matter direct detection},
  \href{http://dx.doi.org/10.1088/1475-7516/2016/01/029}{\emph{Journal of
  Cosmology and Astroparticle Physics} {\bf 2016} (Jan, 2016) 029--029}.

\bibitem{BT}
J.~Binney and S.~Tremaine, \emph{Galactic Dynamics: (Second Edition)}.
\newblock Princeton Series in Astrophysics. Princeton University Press, 2008.

\bibitem{Lynden_Bell_1968}
D.~Lynden-Bell, R.~Wood and A.~Royal, \emph{The gravo-thermal catastrophe in
  isothermal spheres and the onset of red-giant structure for stellar systems},
  \href{http://dx.doi.org/10.1093/mnras/138.4.495}{\emph{Monthly Notices of the
  Royal Astronomical Society} {\bf 138} (feb, 1968) 495--525}.

\bibitem{fitzpatrick2015plasma}
R.~Fitzpatrick, \emph{Plasma physics : an introduction}.
\newblock CRC Press, Taylor \& Francis Group, 2015.

\bibitem{Le_Bellac_1996}
M.~L. Bellac, \emph{Thermal Field Theory}.
\newblock Cambridge University Press, Aug, 1996,
  \href{http://dx.doi.org/10.1017/cbo9780511721700}{10.1017/cbo9780511721700}.

\bibitem{Raffelt_1996}
G.~Raffelt, \emph{{Stars as laboratories for fundamental physics}: {The
  astrophysics of neutrinos, axions, and other weakly interacting particles}}.
\newblock 5, 1996.

\bibitem{Penrose_1960}
O.~Penrose, \emph{Electrostatic instabilities of a uniform non-maxwellian
  plasma}, \href{http://dx.doi.org/10.1063/1.1706024}{\emph{Physics of Fluids}
  {\bf 3} (1960) 258}.

\bibitem{Weibel_1959}
E.~S. Weibel, \emph{Spontaneously growing transverse waves in a plasma due to
  an anisotropic velocity distribution},
  \href{http://dx.doi.org/10.1103/physrevlett.2.83}{\emph{Physical Review
  Letters} {\bf 2} (Feb, 1959) 83--84}.

\bibitem{Fried_1959}
B.~D. Fried, \emph{Mechanism for instability of transverse plasma waves},
  \href{http://dx.doi.org/10.1063/1.1705933}{\emph{Physics of Fluids} {\bf 2}
  (1959) 337}.

\bibitem{Walker_2009}
M.~G. Walker, M.~Mateo, E.~W. Olszewski, J.~Pe{\~{n}}arrubia, N.~W. Evans and
  G.~Gilmore, \emph{A {UNIVERSAL} {MASS} {PROFILE} {FOR} {DWARF} {SPHEROIDAL}
  {GALAXIES}?}, \href{http://dx.doi.org/10.1088/0004-637x/704/2/1274}{\emph{The
  Astrophysical Journal} {\bf 704} (Oct, 2009) 1274--1287}.

\bibitem{Wolf_2010}
J.~Wolf, G.~D. Martinez, J.~S. Bullock, M.~Kaplinghat, M.~Geha, R.~R.
  Mu{\~{n}}oz et~al., \emph{Accurate masses for dispersion-supported galaxies},
  \href{http://dx.doi.org/10.1111/j.1365-2966.2010.16753.x}{\emph{Monthly
  Notices of the Royal Astronomical Society} (May, 2010) no--no}.

\bibitem{Adams_2014}
J.~J. Adams, J.~D. Simon, M.~H. Fabricius, R.~C.~E. van~den Bosch, J.~C.
  Barentine, R.~Bender et~al., \emph{{DWARF} {GALAXY} {DARK} {MATTER} {DENSITY}
  {PROFILES} {INFERRED} {FROM} {STELLAR} {AND} {GAS} {KINEMATICS}},
  \href{http://dx.doi.org/10.1088/0004-637x/789/1/63}{\emph{The Astrophysical
  Journal} {\bf 789} (Jun, 2014) 63}.

\bibitem{Hammer_2019}
F.~Hammer, Y.~Yang, J.~Wang, F.~Arenou, M.~Puech, H.~Flores et~al., \emph{On
  the absence of dark matter in dwarf galaxies surrounding the milky way},
  \href{http://dx.doi.org/10.3847/1538-4357/ab36b6}{\emph{The Astrophysical
  Journal} {\bf 883} (Oct, 2019) 171}.

\bibitem{Hammer_2020}
F.~Hammer, Y.~Yang, F.~Arenou, J.~Wang, H.~Li, P.~Bonifacio et~al.,
  \emph{Orbital evidences for dark-matter-free milky way dwarf spheroidal
  galaxies}, \href{http://dx.doi.org/10.3847/1538-4357/ab77be}{\emph{The
  Astrophysical Journal} {\bf 892} (Mar, 2020) 3}.

\bibitem{Nesti_2013}
F.~Nesti and P.~Salucci, \emph{The dark matter halo of the milky way, {AD}
  2013}, \href{http://dx.doi.org/10.1088/1475-7516/2013/07/016}{\emph{Journal
  of Cosmology and Astroparticle Physics} {\bf 2013} (Jul, 2013) 016--016}.

\bibitem{Evans_2006}
N.~W. Evans and J.~H. An, \emph{Distribution function of dark matter},
  \href{http://dx.doi.org/10.1103/physrevd.73.023524}{\emph{Physical Review D}
  {\bf 73} (Jan, 2006) }.

\bibitem{Wojtak_2008}
R.~Wojtak, E.~L. {\L}okas, G.~A. Mamon, S.~Gottlöber, A.~Klypin and
  Y.~Hoffman, \emph{The distribution function of dark matter in massive
  haloes},
  \href{http://dx.doi.org/10.1111/j.1365-2966.2008.13441.x}{\emph{Monthly
  Notices of the Royal Astronomical Society} {\bf 388} (Aug, 2008) 815--828}.

\bibitem{Ludlow_2011}
A.~D. Ludlow, J.~F. Navarro, S.~D.~M. White, M.~Boylan-Kolchin, V.~Springel,
  A.~Jenkins et~al., \emph{The density and pseudo-phase-space density profiles
  of cold dark matter haloes},
  \href{http://dx.doi.org/10.1111/j.1365-2966.2011.19008.x}{\emph{Monthly
  Notices of the Royal Astronomical Society} {\bf 415} (Jun, 2011) 3895--3902}.

\bibitem{Lemze_2012}
D.~Lemze, R.~Wagner, Y.~Rephaeli, S.~Sadeh, M.~L. Norman, R.~Barkana et~al.,
  \emph{{PROFILES} {OF} {DARK} {MATTER} {VELOCITY} {ANISOTROPY} {IN}
  {SIMULATED} {CLUSTERS}},
  \href{http://dx.doi.org/10.1088/0004-637x/752/2/141}{\emph{The Astrophysical
  Journal} {\bf 752} (Jun, 2012) 141}.

\bibitem{Sparre_2012}
M.~Sparre and S.~H. Hansen, \emph{The behaviour of shape and velocity
  anisotropy in dark matter haloes},
  \href{http://dx.doi.org/10.1088/1475-7516/2012/10/049}{\emph{Journal of
  Cosmology and Astroparticle Physics} {\bf 2012} (Oct, 2012) 049--049}.

\bibitem{Wojtak_2013}
R.~Wojtak, S.~Gottlöber and A.~Klypin, \emph{Orbital anisotropy in
  cosmological haloes revisited},
  \href{http://dx.doi.org/10.1093/mnras/stt1113}{\emph{Monthly Notices of the
  Royal Astronomical Society} {\bf 434} (Jul, 2013) 1576--1585}.

\bibitem{Host_2008}
O.~Host, S.~H. Hansen, R.~Piffaretti, A.~Morandi, S.~Ettori, S.~T. Kay et~al.,
  \emph{{MEASUREMENT} {OF} {THE} {DARK} {MATTER} {VELOCITY} {ANISOTROPY} {IN}
  {GALAXY} {CLUSTERS}},
  \href{http://dx.doi.org/10.1088/0004-637x/690/1/358}{\emph{The Astrophysical
  Journal} {\bf 690} (Dec, 2008) 358--366}.

\bibitem{Lemze_2011}
D.~Lemze, Y.~Rephaeli, R.~Barkana, T.~Broadhurst, R.~Wagner and M.~L. Norman,
  \emph{{QUANTIFYING} {THE} {COLLISIONLESS} {NATURE} {OF} {DARK} {MATTER} {AND}
  {GALAXIES} {IN} a1689},
  \href{http://dx.doi.org/10.1088/0004-637x/728/1/40}{\emph{The Astrophysical
  Journal} {\bf 728} (Jan, 2011) 40}.

\bibitem{Dodelson_2003}
S.~Dodelson, \emph{{Modern Cosmology}}.
\newblock Academic Press, Elsevier Science, 2003.

\bibitem{Dvorkin_2014}
C.~Dvorkin, K.~Blum and M.~Kamionkowski, \emph{Constraining dark matter-baryon
  scattering with linear cosmology},
  \href{http://dx.doi.org/10.1103/physrevd.89.023519}{\emph{Physical Review D}
  {\bf 89} (jan, 2014) }.

\bibitem{Markevitch_2004}
M.~Markevitch, A.~H. Gonzalez, D.~Clowe, A.~Vikhlinin, W.~Forman, C.~Jones
  et~al., \emph{Direct constraints on the dark matter self-interaction cross
  section from the merging galaxy cluster 1e 0657-56},
  \href{http://dx.doi.org/10.1086/383178}{\emph{The Astrophysical Journal} {\bf
  606} (May, 2004) 819--824}.

\bibitem{2010Maxim}
M.~Markevitch, \emph{Intergalactic shock fronts}, {\emph{arxiv:1010.3660}
  (2010) }.

\bibitem{Kahlhoefer_2015}
F.~Kahlhoefer, K.~Schmidt-Hoberg, J.~Kummer and S.~Sarkar, \emph{On the
  interpretation of dark matter self-interactions in abell 3827},
  \href{http://dx.doi.org/10.1093/mnrasl/slv088}{\emph{Monthly Notices of the
  Royal Astronomical Society: Letters} {\bf 452} (jul, 2015) L54--L58}.

\bibitem{Hodges_1993}
H.~M. Hodges, \emph{Mirror baryons as the dark matter},
  \href{http://dx.doi.org/10.1103/physrevd.47.456}{\emph{Physical Review D}
  {\bf 47} (Jan, 1993) 456--459}.

\bibitem{FOOT_2004}
R.~FOOT, \emph{{MIRROR} {MATTER}-{TYPE} {DARK} {MATTER}},
  \href{http://dx.doi.org/10.1142/s0218271804006449}{\emph{International
  Journal of Modern Physics D} {\bf 13} (Dec, 2004) 2161--2192}.

\bibitem{Kaplan_2010}
D.~E. Kaplan, G.~Z. Krnjaic, K.~R. Rehermann and C.~M. Wells, \emph{Atomic dark
  matter}, \href{http://dx.doi.org/10.1088/1475-7516/2010/05/021}{\emph{Journal
  of Cosmology and Astroparticle Physics} {\bf 2010} (May, 2010) 021--021}.

\bibitem{Cyr_Racine_2013}
F.-Y. Cyr-Racine and K.~Sigurdson, \emph{Cosmology of atomic dark matter},
  \href{http://dx.doi.org/10.1103/physrevd.87.103515}{\emph{Physical Review D}
  {\bf 87} (May, 2013) }.

\bibitem{Davidson_2000}
S.~Davidson, S.~Hannestad and G.~Raffelt, \emph{Updated bounds on milli-charged
  particles},
  \href{http://dx.doi.org/10.1088/1126-6708/2000/05/003}{\emph{Journal of High
  Energy Physics} {\bf 2000} (May, 2000) 003--003}.

\bibitem{Foot_1993}
R.~Foot, H.~Lew and R.~R. Volkas, \emph{Electric-charge quantization},
  \href{http://dx.doi.org/10.1088/0954-3899/19/3/005}{\emph{Journal of Physics
  G: Nuclear and Particle Physics} {\bf 19} (Mar, 1993) 361--372}.

\bibitem{Chuzhoy_2009}
L.~Chuzhoy and E.~W. Kolb, \emph{Reopening the window on charged dark matter},
  \href{http://dx.doi.org/10.1088/1475-7516/2009/07/014}{\emph{Journal of
  Cosmology and Astroparticle Physics} {\bf 2009} (Jul, 2009) 014--014}.

\bibitem{McDermott_2011}
S.~D. McDermott, H.-B. Yu and K.~M. Zurek, \emph{Turning off the lights: How
  dark is dark matter?},
  \href{http://dx.doi.org/10.1103/physrevd.83.063509}{\emph{Physical Review D}
  {\bf 83} (Mar, 2011) }.

\bibitem{Kadota:2016tqq}
K.~Kadota, T.~Sekiguchi and H.~Tashiro, \emph{{A new constraint on millicharged
  dark matter from galaxy clusters}},
  \href{http://arxiv.org/abs/1602.04009}{{\tt 1602.04009}}.

\bibitem{Hu_2017}
P.-K. Hu, A.~Kusenko and V.~Takhistov, \emph{Dark cosmic rays},
  \href{http://dx.doi.org/10.1016/j.physletb.2017.02.035}{\emph{Physics Letters
  B} {\bf 768} (May, 2017) 18--22}.

\bibitem{Dunsky_2019}
D.~Dunsky, L.~J. Hall and K.~Harigaya, \emph{{CHAMP} cosmic rays},
  \href{http://dx.doi.org/10.1088/1475-7516/2019/07/015}{\emph{Journal of
  Cosmology and Astroparticle Physics} {\bf 2019} (Jul, 2019) 015--015}.

\bibitem{Schwartz_2013}
M.~D. Schwartz, \emph{{Quantum Field Theory and the Standard Model}}.
\newblock Cambridge University Press, 3, 2014.

\bibitem{Dror_2017}
J.~A. Dror, R.~Lasenby and M.~Pospelov, \emph{Dark forces coupled to
  nonconserved currents},
  \href{http://dx.doi.org/10.1103/physrevd.96.075036}{\emph{Physical Review D}
  {\bf 96} (Oct, 2017) }.

\bibitem{Dror_2017_prl}
J.~A. Dror, R.~Lasenby and M.~Pospelov, \emph{New constraints on light vectors
  coupled to anomalous currents},
  \href{http://dx.doi.org/10.1103/physrevlett.119.141803}{\emph{Physical Review
  Letters} {\bf 119} (oct, 2017) }.

\bibitem{Dror_2020}
J.~A. Dror, \emph{Discovering leptonic forces using nonconserved currents},
  \href{http://dx.doi.org/10.1103/physrevd.101.095013}{\emph{Physical Review D}
  {\bf 101} (May, 2020) }.

\bibitem{Holdom_1986}
B.~Holdom, \emph{{Two U(1)'s and {$\epsilon$} charge shifts}},
  \href{http://dx.doi.org/10.1016/0370-2693(86)91377-8}{\emph{Physics Letters
  B} {\bf 166} (Jan, 1986) 196--198}.

\bibitem{1311review}
R.~Essig, J.~A. Jaros, W.~Wester, P.~H. Adrian, S.~Andreas, T.~Averett et~al.,
  \emph{Dark sectors and new, light, weakly-coupled particles},
  {\emph{arxiv:1311.0029} (2013) }.

\bibitem{Knapen_2017}
S.~Knapen, T.~Lin and K.~M. Zurek, \emph{Light dark matter: Models and
  constraints},
  \href{http://dx.doi.org/10.1103/physrevd.96.115021}{\emph{Physical Review D}
  {\bf 96} (Dec, 2017) }.

\bibitem{GOLDHABER_1971}
A.~S. GOLDHABER and M.~M. NIETO, \emph{Terrestrial and extraterrestrial limits
  on the photon mass},
  \href{http://dx.doi.org/10.1103/revmodphys.43.277}{\emph{Reviews of Modern
  Physics} {\bf 43} (Jul, 1971) 277--296}.

\bibitem{Caputo_2020}
A.~Caputo, H.~Liu, S.~Mishra-Sharma and J.~T. Ruderman, \emph{Dark photon
  oscillations in our inhomogeneous universe}, {\emph{arxiv:2002.05165} (2020)
  }.

\bibitem{Williams_1971}
E.~R. Williams, J.~E. Faller and H.~A. Hill, \emph{New experimental test of
  coulomb's law: A laboratory upper limit on the photon rest mass},
  \href{http://dx.doi.org/10.1103/physrevlett.26.721}{\emph{Physical Review
  Letters} {\bf 26} (Mar, 1971) 721--724}.

\bibitem{Bartlett_1988}
D.~F. Bartlett and S.~Lögl, \emph{Limits on an electromagnetic fifth force},
  \href{http://dx.doi.org/10.1103/physrevlett.61.2285}{\emph{Physical Review
  Letters} {\bf 61} (Nov, 1988) 2285--2287}.

\bibitem{Betz_2013}
M.~Betz, F.~Caspers, M.~Gasior, M.~Thumm and S.~W. Rieger, \emph{First results
  of the {CERN} resonant weakly interacting sub-{eV} particle search
  ({CROWS})},
  \href{http://dx.doi.org/10.1103/physrevd.88.075014}{\emph{Physical Review D}
  {\bf 88} (Oct, 2013) }.

\bibitem{Vinyoles_2015}
N.~Vinyoles, A.~Serenelli, F.~Villante, S.~Basu, J.~Redondo and J.~Isern,
  \emph{New axion and hidden photon constraints from a solar data global fit},
  \href{http://dx.doi.org/10.1088/1475-7516/2015/10/015}{\emph{Journal of
  Cosmology and Astroparticle Physics} {\bf 2015} (Oct, 2015) 015--015}.

\bibitem{ll1981}
L.~D. Landau and E.~M. Lifshitz, \emph{Physical Kinetics}.
\newblock 1981.

\bibitem{van_Erkelens_1981}
H.~van Erkelens, \emph{Relativistic boltzmann theory for a plasma},
  \href{http://dx.doi.org/10.1016/0378-4371(81)90023-6}{\emph{Physica A:
  Statistical Mechanics and its Applications} {\bf 107} (May, 1981) 48--70}.

\bibitem{Vogel_2014}
H.~Vogel and J.~Redondo, \emph{Dark radiation constraints on minicharged
  particles in models with a hidden photon},
  \href{http://dx.doi.org/10.1088/1475-7516/2014/02/029}{\emph{Journal of
  Cosmology and Astroparticle Physics} {\bf 2014} (Feb, 2014) 029--029}.

\bibitem{Chang_2018}
J.~H. Chang, R.~Essig and S.~D. McDermott, \emph{Supernova 1987a constraints on
  sub-{GeV} dark sectors, millicharged particles, the {QCD} axion, and an
  axion-like particle},
  \href{http://dx.doi.org/10.1007/jhep09(2018)051}{\emph{Journal of High Energy
  Physics} {\bf 2018} (Sep, 2018) }.

\bibitem{Essig_2012}
R.~Essig, A.~Manalaysay, J.~Mardon, P.~Sorensen and T.~Volansky, \emph{First
  direct detection limits on sub-{GeV} dark matter from {XENON}10},
  \href{http://dx.doi.org/10.1103/physrevlett.109.021301}{\emph{Physical Review
  Letters} {\bf 109} (Jul, 2012) }.

\bibitem{Essig_2017}
R.~Essig, T.~Volansky and T.-T. Yu, \emph{New constraints and prospects for
  sub-{GeV} dark matter scattering off electrons in xenon},
  \href{http://dx.doi.org/10.1103/physrevd.96.043017}{\emph{Physical Review D}
  {\bf 96} (Aug, 2017) }.

\bibitem{Dvorkin_2019}
C.~Dvorkin, T.~Lin and K.~Schutz, \emph{Making dark matter out of light:
  Freeze-in from plasma effects},
  \href{http://dx.doi.org/10.1103/physrevd.99.115009}{\emph{Physical Review D}
  {\bf 99} (Jun, 2019) }.

\bibitem{Essig_2012a}
R.~Essig, J.~Mardon and T.~Volansky, \emph{Direct detection of sub-{GeV} dark
  matter}, \href{http://dx.doi.org/10.1103/physrevd.85.076007}{\emph{Physical
  Review D} {\bf 85} (apr, 2012) }.

\bibitem{Crisler_2018}
M.~Crisler, R.~Essig, J.~Estrada, G.~Fernandez, J.~Tiffenberg, M.~S. Haro
  et~al., \emph{{SENSEI}: First direct-detection constraints on sub-{GeV} dark
  matter from a surface run},
  \href{http://dx.doi.org/10.1103/physrevlett.121.061803}{\emph{Physical Review
  Letters} {\bf 121} (Aug, 2018) }.

\bibitem{Abramoff_2019}
O.~Abramoff, L.~Barak, I.~M. Bloch, L.~Chaplinsky, M.~Crisler, Dawa et~al.,
  \emph{{SENSEI}: Direct-detection constraints on sub-{GeV} dark matter from a
  shallow underground run using a prototype skipper {CCD}},
  \href{http://dx.doi.org/10.1103/physrevlett.122.161801}{\emph{Physical Review
  Letters} {\bf 122} (Apr, 2019) }.

\bibitem{Knapen_2018}
S.~Knapen, T.~Lin, M.~Pyle and K.~M. Zurek, \emph{Detection of light dark
  matter with optical phonons in polar materials},
  \href{http://dx.doi.org/10.1016/j.physletb.2018.08.064}{\emph{Physics Letters
  B} {\bf 785} (Oct, 2018) 386--390}.

\bibitem{Griffin_2018}
S.~Griffin, S.~Knapen, T.~Lin and K.~M. Zurek, \emph{Directional detection of
  light dark matter with polar materials},
  \href{http://dx.doi.org/10.1103/physrevd.98.115034}{\emph{Physical Review D}
  {\bf 98} (Dec, 2018) }.

\bibitem{Berlin_2020}
A.~Berlin, R.~T. D'Agnolo, S.~A. Ellis, P.~Schuster and N.~Toro, \emph{Directly
  deflecting particle dark matter},
  \href{http://dx.doi.org/10.1103/physrevlett.124.011801}{\emph{Physical Review
  Letters} {\bf 124} (Jan, 2020) }.

\bibitem{Caputo_2019}
A.~Caputo, L.~Sberna, M.~Fr{\'{\i}}as, D.~Blas, P.~Pani, L.~Shao et~al.,
  \emph{Constraints on millicharged dark matter and axionlike particles from
  timing of radio waves},
  \href{http://dx.doi.org/10.1103/physrevd.100.063515}{\emph{Physical Review D}
  {\bf 100} (sep, 2019) }.

\bibitem{Hui_2017}
L.~Hui, J.~P. Ostriker, S.~Tremaine and E.~Witten, \emph{Ultralight scalars as
  cosmological dark matter},
  \href{http://dx.doi.org/10.1103/physrevd.95.043541}{\emph{Physical Review D}
  {\bf 95} (Feb, 2017) }.

\bibitem{Levkov_2018}
D.~Levkov, A.~Panin and I.~Tkachev, \emph{Gravitational bose-einstein
  condensation in the kinetic regime},
  \href{http://dx.doi.org/10.1103/physrevlett.121.151301}{\emph{Physical Review
  Letters} {\bf 121} (Oct, 2018) }.

\bibitem{Bar_Or_2019}
B.~Bar-Or, J.-B. Fouvry and S.~Tremaine, \emph{Relaxation in a fuzzy dark
  matter halo}, \href{http://dx.doi.org/10.3847/1538-4357/aaf28c}{\emph{The
  Astrophysical Journal} {\bf 871} (Jan, 2019) 28}.

\bibitem{Sakurai}
J.~J. Sakurai, \emph{Modern Quantum Mechanics (Revised Edition)}.
\newblock Addison Wesley, 1993.

\bibitem{Semikoz_1995}
D.~V. Semikoz and I.~I. Tkachev, \emph{Kinetics of bose condensation},
  \href{http://dx.doi.org/10.1103/physrevlett.74.3093}{\emph{Physical Review
  Letters} {\bf 74} (Apr, 1995) 3093--3097}.

\bibitem{Sikivie_2009}
P.~Sikivie and Q.~Yang, \emph{Bose-einstein condensation of dark matter
  axions},
  \href{http://dx.doi.org/10.1103/physrevlett.103.111301}{\emph{Physical Review
  Letters} {\bf 103} (Sep, 2009) }.

\bibitem{Hochberg_2016}
Y.~Hochberg, M.~Pyle, Y.~Zhao and K.~M. Zurek, \emph{Detecting superlight dark
  matter with fermi-degenerate materials},
  \href{http://dx.doi.org/10.1007/jhep08(2016)057}{\emph{Journal of High Energy
  Physics} {\bf 2016} (Aug, 2016) }.

\bibitem{Hochberg_2017}
Y.~Hochberg, Y.~Kahn, M.~Lisanti, C.~G. Tully and K.~M. Zurek,
  \emph{Directional detection of dark matter with two-dimensional targets},
  \href{http://dx.doi.org/10.1016/j.physletb.2017.06.051}{\emph{Physics Letters
  B} {\bf 772} (Sep, 2017) 239--246}.

\bibitem{Hochberg_2018}
Y.~Hochberg, Y.~Kahn, M.~Lisanti, K.~M. Zurek, A.~G. Grushin, R.~Ilan et~al.,
  \emph{Detection of sub-{MeV} dark matter with three-dimensional dirac
  materials},
  \href{http://dx.doi.org/10.1103/physrevd.97.015004}{\emph{Physical Review D}
  {\bf 97} (Jan, 2018) }.

\bibitem{Craig_2019}
N.~Craig, I.~G. Garcia and S.~Koren, \emph{The weak scale from weak gravity},
  \href{http://dx.doi.org/10.1007/jhep09(2019)081}{\emph{Journal of High Energy
  Physics} {\bf 2019} (Sep, 2019) }.

\bibitem{Heikinheimo:2017meg}
M.~Heikinheimo, M.~Raidal, C.~Spethmann and H.~Veermae, \emph{{Collisionless
  shocks in self-interacting dark matter}},
  \href{http://dx.doi.org/10.1088/1361-6587/aa7f48}{\emph{Plasma Phys. Control.
  Fusion} {\bf 60} (2017) 014011}, [\href{http://arxiv.org/abs/1707.03662}{{\tt
  1707.03662}}].

\bibitem{abramowitz_stegun}
M.~Abramowitz and I.~A. Stegun, \emph{Handbook of Mathematical Functions, With
  Formulas, Graphs, and Mathematical Tables,}.
\newblock Dover Publications, Inc., USA, 1974.

\end{thebibliography}\endgroup
\bibliographystyle{JHEP}

\end{document}